\documentclass[aps,pra,twocolumn,superscriptaddress,notitlepage,showpacs,showkeys]{revtex4-1}
\usepackage{graphicx,amsmath,amsfonts,amssymb,upgreek,txfonts,color}
\usepackage[colorlinks,linkcolor=blue,citecolor=blue,urlcolor=blue,breaklinks=true]{hyperref}
\usepackage{color, colortbl}
\definecolor{lightgreen}{rgb}{0.88,1,1}
\usepackage{float}

\newcommand{\be}{\begin{equation}}
\newcommand{\ee}{\end{equation}}
\newcommand{\bae}{\begin{eqnarray}} \newcommand{\eae}{\end{eqnarray}}

\begin{document}

\title{Squeezed-Light-Enhanced Multiparameter Quantum Estimation in Cavity Magnonics}

\author{Hamza Harraf}
\affiliation{LPHE-Modeling and Simulation, Faculty of Sciences, Mohammed V University in Rabat, Rabat, Morocco}	
\author{Mohamed Amazioug} 
\thanks{m.amazioug@uiz.ac.ma}
\affiliation{LPTHE-Department of Physics, Faculty of sciences, Ibnou Zohr University, Agadir, Morocco}	
\author{Rachid Ahl Laamara}
\affiliation{LPHE-Modeling and Simulation, Faculty of Sciences, Mohammed V University in Rabat, Rabat, Morocco}	
\affiliation{Centre of Physics and Mathematics, CPM, Faculty of Sciences, Mohammed V University in Rabat, Rabat, Morocco}

\date{\today}

\begin{abstract}

Improving multiparameter quantum estimation in magnonic systems via quantum noise suppression is a well-established and critical research objective. In this work, we propose an experimentally realistic scheme to improve the precision of simultaneously estimating different parameters in a cavity-magnon system by utilizing a degenerate optical parametric amplifier (OPA). The OPA enhances the estimation precision by decreasing the most informative quantum Cram\'er-Rao bound, calculated employing the symmetric logarithmic derivative (SLD) and the right logarithmic derivative (RLD). We show that when nonlinearity is introduced into the system, quantum noise is significantly suppressed. Our results show how different physical parameters influence multiparameter estimation precision and provide a detailed discussion of the associated physical mechanisms in the steady state. Our results focus on exploring practical Gaussian measurement schemes that can be realized experimentally. Besides, we further analyze the system's dynamics, comparing both the SLD quantum Fisher information (QFI) and the classical Fisher information (CFI) for both homodyne and heterodyne detection. This approach provides a robust foundation for multiparameter quantum estimation, offering significant potential for application in hybrid magnomechanical and optomechanical systems.

\end{abstract}

\maketitle

\section{\label{sec1}Introduction}

Gaussian states within continuous-variable (CV) frameworks \cite{Kim02,Paris05,Van05} have become prominent in quantum information for two main reasons: their theoretical description is simplified by the use of only first and second moments, and they are readily generated and controlled in experimental setups. These states can be implemented in various protocols, ranging from quantum optomechanics \cite{Vitali07, Mauro07, Groblasher09, Tian10, Teufel11, Chan11} and teleportation channels \cite{Paris03} to recent advances in magnomechanics \cite{Zhang16, Li18}. Additionally, the Gaussian framework is justified by its successful application in key experimental milestones, such as studies on Bose-Einstein condensates \cite{Kevrekidis03, Gross11, Wade16} and the high-precision detection of gravitational waves \cite{Aasi13}.\\

Within the rapidly developing field of condensed matter physics, cavity magnonics has emerged to study the coherent coupling between photons and quantized collective spin excitations, known as magnons \cite{Wagle2024,Tabuchi2014,Zhang2014,Andrade2025,Ebrahimi21,Peng2025,Xu2023}. Due to their quantized nature in magnetically ordered systems, magnons offer efficient interaction with external magnetic fields and provide the advantage of highly tunable operational frequencies. Furthermore, magnons offer a powerful framework for implementing hybrid quantum systems with enhanced controllability, owing to their coherent coupling across a wide range of platforms. Their ability to serve as high-fidelity carriers for quantum information processing has further established their prominence in the field \cite{yuan}. The adoption of yttrium iron garnet (YIG) spheres has facilitated the study of magnonic interactions within a high-precision regime. The combination of dense spin populations and low damping \cite{Tabuchi2014,Zhang2014,Goryachev2014,19,20,21,22,23,Flaig17} positions these spheres as excellent candidates for implementing quantum information and metrological protocols \cite{Tabuchi2014,25}. Experimental progress has established magnon-based sensing as a high-precision tool, utilizing dispersive coupling with superconducting qubits \cite{magnon sensing}. Additionally, experiments in cavity magnonic systems have validated the magnetic dipole-mediated coupling between magnons and microwave cavity modes \cite{Tabuchi2014,Zhang2014,Goryachev2014,Highcooperativity,Nori}. Notably, such interactions are not limited by thermal constraints, as they can be experimentally realized in YIG systems operating at both cryogenic \cite{Tabuchi2014,Goryachev2014,Highcooperativity,Nori} and room temperatures \cite{Zhang2014,Nori}. Additionally, the versatility of this coupling is demonstrated by the observation of the Purcell effect and magnetically induced transparency \cite{Zhang2014}. Such interactions are particularly crucial in cavity magnomechanical systems \cite{Zhang16, Dynamical Backaction Magnomechanics}; here, magnons couple to microwave photons through magnetic dipoles while simultaneously interacting with phonons via magnetostriction. Furthermore, various quantum protocols have been implemented within this framework, including microwave-to-optical transduction \cite{30}, the engineering of quantum entanglement \cite{Li18,32,33,34,35,36,last,electromagnonics-optomechanics, foroudcrystalentanglement,hussian2022,Amazioug2023}, magnon blockade \cite{Deng24,Amazioug2025} and the generation of squeezed magnonic and photonic states \cite{37, microwace field squeezing}. Additional applications encompass phononic lasing \cite{38}, quantum sensing specifically thermometry \cite{39} and magnetometry \cite{one,phasemodulatedmagnetometry,8} as well as magnon blockade \cite{magnonblockade1,magnonblockade2}, Spin cat states \cite{Sharma}, quantum illumination \cite{quantumillumination}, and state-transfer processes such as photon–phonon conversion and quantum memory \cite{photon-phononconversion, Cavity magnomechanical storage and retrieval of quantum states,Zhu2025}. \\

The Quantum Fisher Information Matrix (QFIM) originated from the work of Helstrom and Holevo \cite{Holevo11, Helstrom67, Helstrom76, Bures69} and was later refined and popularized by Braunstein and Caves \cite{Braunstein94}. The QFIM aims to describe the limits of distinguishability for quantum states characterized by infinitesimally different parameters. Within the field of multiparameter quantum metrology, this matrix represents a vital component, given that its inverse sets the boundary for the maximum precision attainable in joint parameter estimation. Investigating the enhancement of the QFIM is thus crucial for the advancement of multiparameter quantum metrology. However, the ultimate precision dictated by the quantum Cram\'er-Rao bound (QCRB) \cite{paris2009} is not universally saturable, often hindered by experimental realities or inherent mathematical constraints in the multiparameter regime \cite{Szczykulska2016}. Despite these limitations, the QCRB provides a benchmark to evaluate whether a proposed quantum detector is viable and if it can surpass the precision limits of current technological standards \cite{Giovannetti04, Zwierz10, Giovannetti11, Demkowicz12}. Unlike multiparameter scenarios, the QCRB is always achievable for single-parameter estimation. This theoretical certainty underpins numerous advancements, ranging from gravitational wave sensing \cite{Abbott16} and quantum thermometry \cite{Monras11,Hofer17,Correa15,Mirkhalaf24, Khan22} to the precise measurement of phase, squeezing parameters, time, and frequency \cite{Ballester04, Gaiba09, Zhang13, Frowis14}. Conversely, the simultaneous estimation of multiple parameters often fails to saturate the QCRB. This is primarily due to the non-commutativity of the optimal observables, which creates an incompatibility between the measurement schemes required to achieve maximum precision for different parameters \cite{Matsumoto02,Vaneph13,Vidrighin14,Crowley14,Ragy16}. To address these challenges, substantial effort has been devoted to multiparameter quantum metrology. The goal is to generalize the criteria under which the QCRB can be saturated, thereby enabling the simultaneous estimation of multiple variables with maximum achievable accuracy. The QFIM is also closely connected to the quantification of key quantum properties, such as coherence and entanglement \cite{Seveso19, Hauke16, Zhang13, Liu17, Naimy2025, Asjad2023}. These potential applications underscore the importance of developing precise theoretical techniques for computing the QFIM. This work addresses that need by introducing an analytical method to obtain the QFIM for systems described by Gaussian-type bosonic continuous variables.\\

In this work, we theoretically investigate the enhancement of the so-called the most informative quantum Cram\'er-Rao bound ($B_{\rm MI}$) within a feasible experimental scheme utilizing an optical parametric amplifier (OPA). Specifically, we investigate the simultaneous estimation of the cavity--magnon coupling strength $g_{mc}$ and the cavity decay rate $\gamma_c$, which are jointly encoded in the same Gaussian state, and characterize their joint precision through the SLD and RLD QFIMs. The system under study consists of a cavity-magnonic system where magnons, acting as quantum excitations of spin waves, are coupled to microwave cavity photons via magnetic dipole interactions. Driven by an external classical field, the microwave cavity mode is analyzed using a set of feasible experimental parameters. Furthermore, we ensure that all reported results fall within the stable regime of the system's dynamics. Specifically, we show that the introduction of an OPA leads to an enhanced estimation regime where the $B_{\rm MI}$ is minimized, allowing for higher precision. In the steady state, we explore the dependence of the estimated precisions of $\rm g_{mc}$ and $\gamma_{\rm c}$ on the system temperature, the magnon-photon coupling strength, the driving power, and the cavity and magnon detuning. In addition, we analyze the dynamics of the quantum Fisher information (QFI) using the diagonal elements of the SLD QFIM and the classical Fisher information (CFI) obtained via homodyne and heterodyne detection to evaluate the estimation precision limits and measurement strategies for $g_{\text{mc}}$ and $\gamma_c$ in a cavity-magnonic system. By evaluating these quantities, we establish the ultimate precision bounds for $\rm g_{mc}$ estimation and pinpoint the most effective measurement protocols. Consequently, we focus on exploring practical Gaussian measurement schemes that can be realized experimentally. Both homodyne and heterodyne detection \cite{Meystre2007,Scully1999} represent fundamental Gaussian measurement strategies with significant applications in continuous-variable signal processing. Gaussian measurements on Gaussian states produce Gaussian-distributed results, allowing the parameters of interest to be determined from the statistical moments of the quadrature operators. Moreover, we investigate the comparative performance of homodyne and heterodyne detection for the estimation of $\rm g_{mc}$ and $\gamma_{\rm c}$, using the ultimate precision limit of optimal measurement settings as a performance benchmark.\\

This article is organized as follows. In Sec. \ref{sec2}, we investigate the system dynamics using the quantum Langevin equations (QLEs). Section \ref{sec3} focuses on the linearization of these equations and the derivation of the covariance matrix (CM) for both steady-state and dynamical regimes. Section \ref{sec4} gives the theoretical framework for multiparameter quantum estimation of Gaussian states; here, we derive the QFIM via the SLD and RLD operators, compare individual and simultaneous estimation strategies, and evaluate the classical Fisher information for homodyne and heterodyne detection. The numerical results are discussed in Sec. \ref{sec5}. In Section~\ref{sec6}, we show that the proposed setup is feasible by applying realistic, experimentally accessible parameters and configurations. Finally, Sec. \ref{sec7} provides a summary of our findings along with concluding remarks and a future outlook.%

\section{\label{sec2} The Model}

\begin{figure}
\includegraphics[width=9cm]{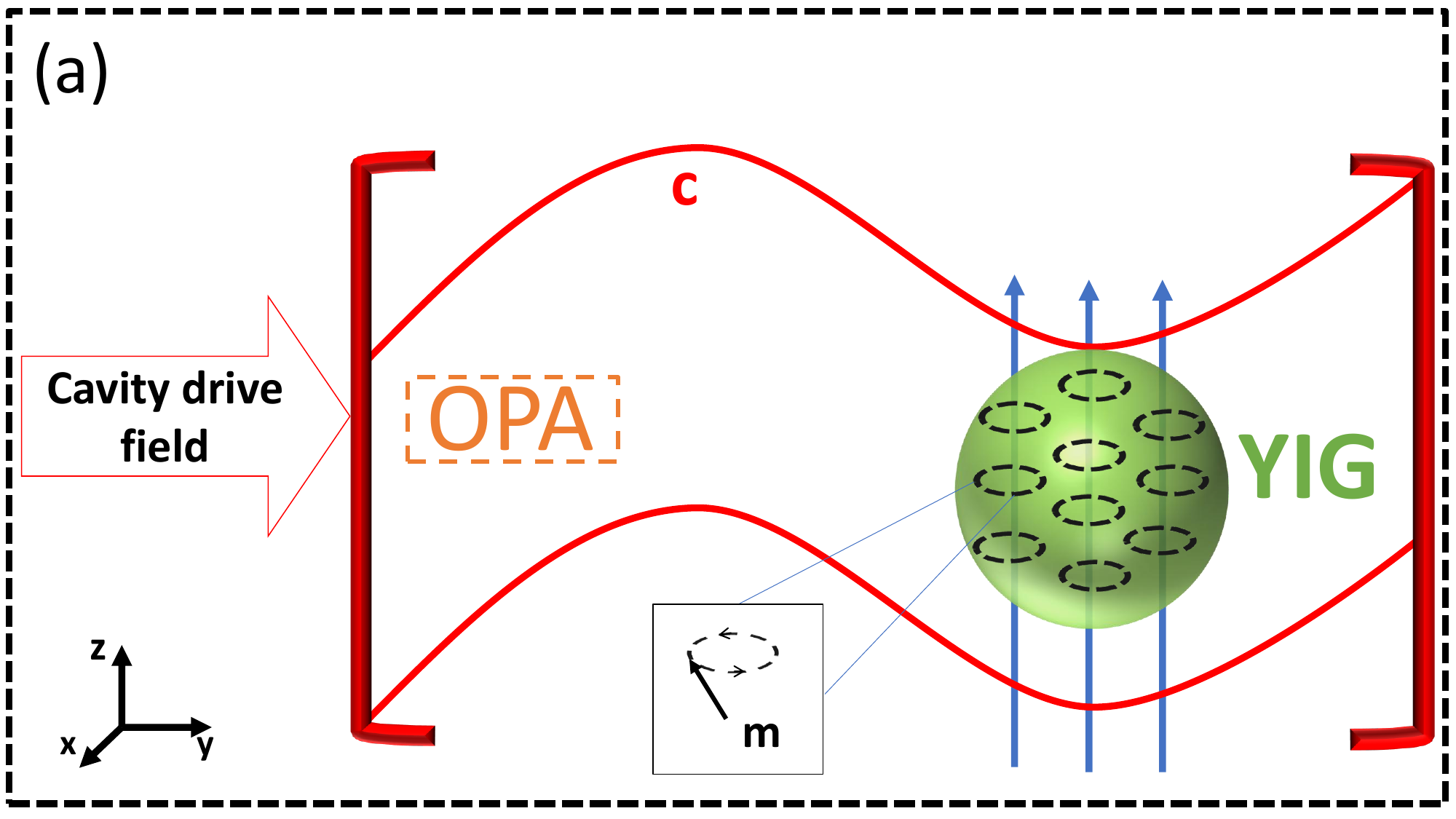}\\
\includegraphics[width=7cm]{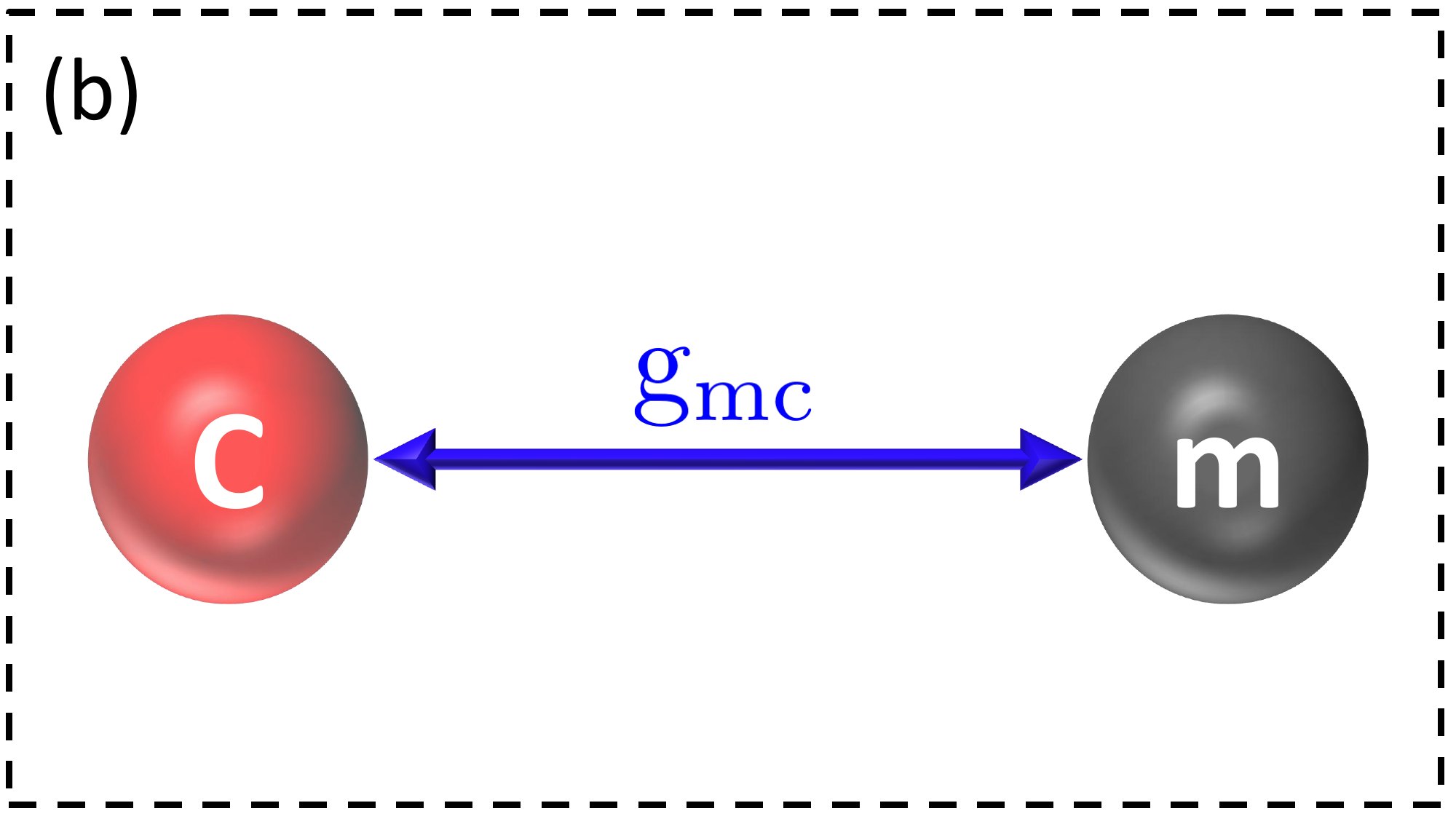}
\includegraphics[width=7cm]{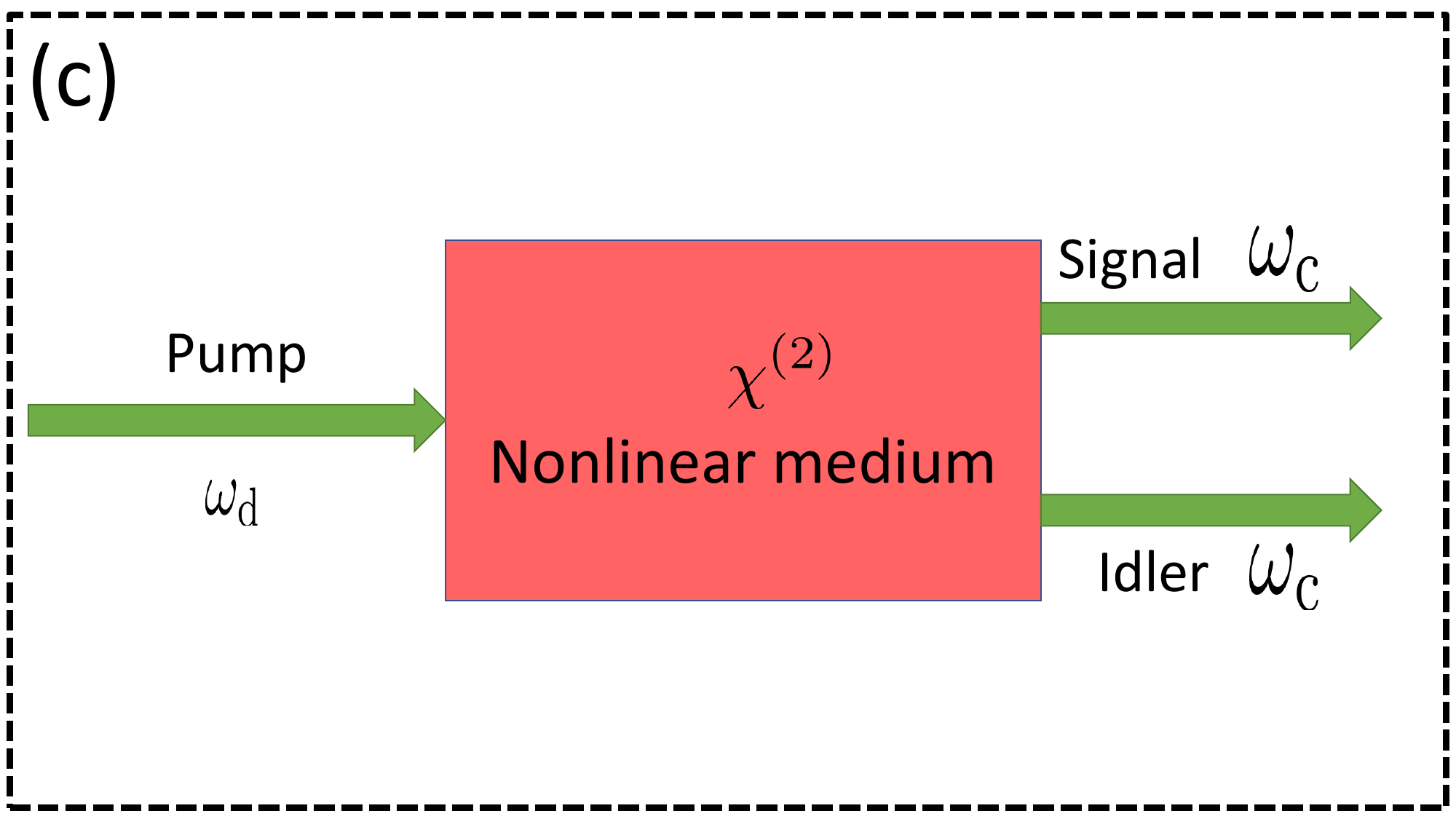}
\caption{(a) Schematic representation of the cavity-magnon system, with a YIG sphere positioned at the magnetic field antinode of the microwave cavity mode to maximize coupling. A uniform magnetic field $\mathbf{B}$ is applied to bias the YIG sphere, with its orientation including at least one component parallel to the $z$-axis. Moreover, an external electromagnetic field drives the cavity mode. (b) The interaction between the photon mode and the magnon mode presents by cavity-magnon coupling $\rm g_{ mc}$. (c) The system utilizes a degenerate optical parametric amplifier (OPA) process where, mediated by second-order nonlinearity $\chi^{(2)}$, a single pump photon of frequency $\omega_{\rm d}$ is converted into two cavity photons of frequency $\omega_{\rm c}$ (signal and Idler with same frequency $\omega_{\rm c}$). This process is governed by the resonance relation $\omega_{\rm d}=2\omega_{\rm c}$.}
\label{Fig1}
\end{figure}

Fig.~\ref{Fig1}(a) depicts the proposed system, which consists of a second-order nonlinear $\chi^{(2)}$ cavity (e.g., aluminum nitride~\cite{Wolf17, Szabados20, Wang20}) and a YIG sphere. A static magnetic field $H_{z}$ is applied to the YIG sphere to induce magnon excitations, which couple to the microwave cavity mode at a rate $\rm g_{mc}$~\cite{Tabuchi2014,Zhang2014,27,28} as shown in Fig.~\ref{Fig1}(b). In addition, the system incorporates a degenerate optical parametric amplifier (OPA) that pumps the cavity mode with a two-photon drive. Following the scheme shown in Fig.~\ref{Fig1}(c), the driving frequency is set to $\omega_{\rm d} = 2\omega_{\rm c}$, where $\omega_{\rm c}$ represents the cavity's resonance frequency and $\theta$ denotes the pump phase. By considering an external electromagnetic drive applied to the cavity mode, we can write the system Hamiltonian as follows
\begin{eqnarray}\label{1}
{\rm H}&=& \hbar \omega_{\rm c} \hat{\rm c}^{\dagger}\hat{\rm c}+\hbar \omega_{\rm m} \hat{\rm m}^{\dagger}\hat{\rm m}+\hbar {\rm g_{mc}} (\hat{\rm c}+\hat{\rm c}^{\dagger})(\hat{\rm m}+\hat{\rm m}^{\dagger})\nonumber\\
&&+i\hbar \epsilon_L( \hat{\rm c}^{\dagger} e^{-i\omega_0 t}-\hat{\rm c}e^{i\omega_0 t})+i\lambda\hbar(e^{-i\omega_{\rm d} t}e^{i\theta}\hat{\rm c}^{\dagger 2}-e^{i\omega_{\rm d} t}e^{-i\theta}\hat{\rm c}^{2}).\
\end{eqnarray}

The free-energy contribution of the system is captured by the first two terms of the Hamiltonian. The photonic and magnonic modes are described by the bosonic operators $\rm \hat{c}$ and $\rm \hat{m}$, which satisfy the fundamental commutation relation $\rm [\hat{o},\hat{ o}^{\dagger}] = 1$ (where $\rm \hat{o} \in \{\rm \hat{c}, \rm \hat{m}\}$), with $\hat{\rm c}^{\dagger}$ and $\hat{\rm m}^{\dagger}$ being the corresponding creation operators. Moreover, $\omega_{\rm c}$ ($\omega_{\rm m}$) represents the resonance frequency of the cavity (magnon) mode. Specifically, the magnonic resonance frequency $\omega_{\rm m} = \gamma \mathbf{B}$ is governed by the external magnetic field $\mathbf{B}$. Given the gyromagnetic ratio $\gamma/2\pi = 28~\mathrm{GHz/T}$, the magnon frequency can be readily tuned by adjusting the strength of the bias field. Moreover, the Kerr nonlinear effect in the magnons arising from magnetocrystalline anisotropy is considered sufficiently small to be ignored. When the magnon-photon coupling strength $\rm g_{mc}$ dominates the respective decay rates $\gamma_{\rm c}$ and $\gamma_{\rm m}$, the system reaches the strong-coupling regime (${\rm g_{ mc}} > \gamma_{\rm c}, \gamma_{\rm m}$). This condition has been extensively explored in recent literature \cite{1,2,3,4}. The third term of the Hamiltonian describes the microwave photon-magnon coupling originating from the magnetic dipole interaction. In the rotating wave approximation (RWA), the interaction Hamiltonian ${\rm g_{ mc}}(\hat{\rm c} + \hat{\rm c}^{\dagger})(\hat{\rm m} + \hat{\rm m}^{\dagger})$ simplifies to the beam-splitter-like form ${\rm g_{ mc}}(\hat{\rm c}\hat{\rm m}^{\dagger} + \hat{\rm c}^{\dagger}\hat{\rm m})$. This approximation remains valid in the regime where the resonance frequencies significantly exceed the coupling and decay rates ($\omega_{\rm c}, \omega_{\rm m} \gg {\rm g_{ mc}}, \gamma_{\rm c}, \gamma_{\rm m}$), a condition readily satisfied in current experimental setups \cite{Li18}.

The external drive contribution is captured by the fourth Hamiltonian term, where the cavity is driven by a microwave field of frequency $\omega_0$ and power $\rm P$. The resulting driving rate is defined as $\epsilon_L = \sqrt{2{\rm P}\gamma_{\rm c} / \hbar\omega_0}$, with $\gamma_{\rm c}$ representing the cavity decay rate. In the last Hamiltonian term, the OPA process is parameterized by the gain $\lambda$ and its drive phase $\theta$. By adopting a frame rotating at the drive frequency $\omega_0$ through the unitary transformation $\textbf{U}(t) = \exp[-i \omega_0 (\hat{\rm c}^{\dagger}\hat{\rm c} + \hat{\rm m}^{\dagger}\hat{\rm m})t]$, the total Hamiltonian is expressed as
\begin{eqnarray}\label{4X}
\hat{\rm H}&=& \textbf{U}(t) {\rm H} \textbf{U}^{\dagger}(t) - i\hbar \textbf{U}(t) \frac{\partial \textbf{U}^{\dagger}(t)}{\partial t} \nonumber\\
&=&\hbar\Delta_{\rm c}\hat{\rm c}^{\dagger}\hat{\rm c}+\hbar\Delta_{\rm m}\hat{\rm m}^{\dagger}\hat{\rm m}+\hbar {\rm g}_{\rm{mc}}(\hat{\rm c}\hat{\rm m}^{\dagger} + \hat{\rm c}^{\dagger}\hat{\rm m})\nonumber\\
&&+i\hbar\epsilon_L(\hat{\rm c}^{\dagger}-\hat{\rm c})+i\lambda\hbar(e^{i\theta}\hat{\rm c}^{\dagger 2}-e^{-i\theta}\hat{\rm c}^{2}),\
\end{eqnarray}
with the cavity and magnonic detunings defined as $\Delta_{\rm c} = \omega_{\rm c} - \omega_0$ and $\Delta_{\rm m} = \omega_{\rm m} - \omega_0$, respectively. We introduce the Bogoliubov transformation according to $\rm c = \cosh r\,a-i e^{i\theta}\sinh r\,a^\dagger,$ and the squeezing parameter $r$ is chosen such that $\tanh(2r) = \frac{2\lambda}{\Delta_c}.$ The Hamiltonian became
\begin{align}
\hat {\rm H}
\approx &
\hbar\Delta_{\rm a} \hat {\rm a}^\dag \hat {\rm a} + \hbar\Delta_{\rm m} \hat {\rm m}^\dag \hat {\rm m} + \hbar\rm g_1
\left(
\hat{\rm a} \hat{\rm m}^\dagger+\hat{\rm a}^\dagger \hat{\rm m}
\right)
\nonumber\\
&+
i\hbar\rm g_2
\left(
e^{-i\theta}\hat{\rm a}\hat{\rm m}
-
e^{i\theta}\hat{\rm a}^\dagger \hat{\rm m}^\dagger \right)
 + i\hbar(\tilde{\epsilon}_L \hat{\rm a}^\dag - \tilde{\epsilon}_L^* \hat{\rm a}),
\end{align}
where ${\Delta}_{\rm a}=\sqrt{\Delta_c^2-4\lambda^2}$, $\rm g_1 = \rm g_{mc}\cosh r$, $\rm g_2 = \rm g_{mc}\sinh r$ and $\tilde{\epsilon}_L = \cosh r + ie^{-i\theta} \sinh r$.

\section{\label{sec3} COVARIANCE MATRIX}

The system dynamics, accounting for both dissipation and quantum fluctuations, are governed by the quantum Langevin equations (QLEs) for the cavity ($\hat{\rm c}$) and magnon ($\hat{\rm m}$) operators. The quantum Langevin equations for the magnon and cavity modes, including their damping rates ($\gamma_{\rm m, c}$) and associated noise operators ($\hat{\rm m}^{\rm in}, \hat{\rm c}^{\rm in}$), are obtained from the Hamiltonian [Eq.~(\ref{4X})] as
\begin{align} 
\frac{d\hat{\rm c}}{dt} &= -(i\Delta_{\rm c}+\gamma_{\rm c})\hat{\rm c}-i{\rm g}_{\rm mc}\hat{\rm m}+2\lambda e^{i\theta}\hat{\rm c}^{\dagger}+\epsilon_L+\sqrt{2\gamma_{\rm c}}\hat{\rm c}^{\rm in}, \label{3}\\
\frac{d\hat{\rm m}}{dt} &=-(i\Delta_{\rm m}+\gamma_{\rm m})\hat{\rm m}-i{\rm g}_{\rm mc}\hat{\rm c}+\sqrt{2\gamma_{\rm m}}\hat{\rm m}^{\rm in}. \label{4}
\end{align}
The input quantum noise operators $\hat{\rm c}^{\rm in}$ and $\hat{\rm m}^{\rm in}$ are characterized by the correlation functions
\begin{eqnarray}
\langle \hat{\rm o}^{\mathrm{in}}(t)\hat{\rm o}^{\mathrm{in} \dagger}(t^\prime)\rangle &=&(1+\bar{n}_{\rm o})\delta(t-t^{\prime}), \label{9} \\
\langle \hat{\rm o}^{\mathrm{in} \dagger}(t)\hat{\rm o}^{\mathrm{in} }(t^\prime)\rangle &=&\bar{n}_{\rm o}\delta(t-t^{\prime}), \label{10}
\end{eqnarray}
with $\bar{n}_{\rm o}$ (for $\rm o = \rm c, \rm m$) representing the equilibrium thermal occupancy of the respective modes, defined by the Bose-Einstein distribution $\bar{n}_{\rm o} = [\exp(\hbar \omega_{\rm o} / k_B {\rm T}) - 1]^{-1}$, with $k_B$ being the Boltzmann constant. Under the condition $\omega_{\rm c} \simeq \omega_{\rm m}$, which is typical for resonant microwave-magnon coupling, we adopt the approximation $\bar{n}_{\rm c} \simeq \bar{n}_{\rm m} \equiv \bar{n}$ for the mean thermal excitation numbers \cite{Zhang2014}. From Eqs. (\ref{3}-\ref{4}) one can decompose each operator $\hat{\rm o}$ into its steady-state mean value $\langle \hat{\rm o} \rangle$ and a small quantum fluctuation $\delta\hat{\rm o}$, i.e., $\hat{\rm o} = \langle \hat{\rm o} \rangle + \delta\hat{\rm o}$, while discarding second-order fluctuation terms.  The steady-state values of the system operators are written as
\begin{widetext}
\begin{align}
\langle \hat{\rm c} \rangle &= \frac{\epsilon_L \left(\left(\Delta_{\rm m}^2+\gamma_{\rm m}^2\right)\left(\gamma_{\rm c}+2\lambda\cos(\theta)\right)+{\rm g}_{\rm mc}^2\gamma_{\rm m}\right)+i\epsilon_L \left(\left(\gamma_{\rm m}^2+\Delta_{\rm m}^2\right)\left(2\lambda\sin(\theta)-\Delta_{\rm c}\right)+{\rm g}_{\rm mc}^2\Delta_{\rm m}\right)}{\gamma_{\rm m}\left(2{\rm g}_{\rm mc}^2\gamma_{\rm c}+\gamma_{\rm m}\left(\Delta_{\rm c}^2-4\lambda\right)+\gamma_{\rm c}^2\gamma_{\rm m}\right)-2{\rm g}_{\rm mc}^2\Delta_{\rm c}\Delta_{\rm m}+\Delta_{\rm m}^2\left(\Delta_{\rm c}^2+\gamma_{\rm c}^2-4\lambda\right)+{\rm g}_{\rm mc}^4}, \\
\langle \hat{\rm m} \rangle &= \frac{{\rm g}_{\rm mc}\epsilon_L \left(\gamma_{\rm m}\left(\Delta_{\rm c}-2\lambda\sin(\theta)\right)+\Delta_{\rm m}\left(\gamma_{\rm c}+2\lambda\cos(\theta)\right)\right)+i{\rm g}_{\rm mc}\epsilon_L \left(\gamma_{\rm m}\left(\gamma_{\rm c}+2\lambda\cos(\theta)\right)+\Delta_{\rm m}\left(2\lambda\sin(\theta)-\Delta_{\rm c}\right)+{\rm g}_{\rm mc}^2\right)}{\gamma_{\rm m}\left(2{\rm g}_{\rm mc}^2\gamma_{\rm c}+\gamma_{\rm m}\left(\Delta_{\rm c}^2-4\lambda\right)+\gamma_{\rm c}^2\gamma_{\rm m}\right)-2{\rm g}_{\rm mc}^2\Delta_{\rm c}\Delta_{\rm m}+\Delta_{\rm m}^2\left(\Delta_{\rm c}^2+\gamma_{\rm c}^2-4\lambda\right)+{\rm g}_{\rm mc}^4}.
\end{align}
\end{widetext}
To analyze the fluctuation dynamics, we introduce the quadrature operators $\delta \hat{X}_{\rm c} = (\delta \hat{\rm c} + \delta \hat{\rm c}^\dagger)/\sqrt{2}$, $\delta \hat{P}_{\rm c} = i(\delta \hat{\rm c}^\dagger - \delta \hat{\rm c})/\sqrt{2}$, and similarly for the magnon mode. The resulting linearized QLEs for the quadrature vector $\mathbf{u}(t) = \left[\delta \hat{X}_{\rm c}, \delta \hat{P}_{\rm c}, \delta \hat{X}_{\rm m}, \delta \hat{P}_{\rm m}\right]^\top$ are written compactly as 
\begin{align}
\dot{\mathbf{u}}(t) = \mathcal{A}\mathbf{u}(t) + \mathbf{n}(t),
\end{align}
where $\mathbf{n}(t)$ is the input noise vector given by $\mathbf{n}(t) = \left[\sqrt{2\gamma_{\rm c}}\hat{X}_{\rm c}^{\rm in}, \sqrt{2\gamma_{\rm c}}\hat{P}_{\rm c}^{\rm in}, \sqrt{2\gamma_{\rm m}}\hat{X}_{\rm m}^{\rm in}, \sqrt{2\gamma_{\rm m}}\hat{P}_{\rm m}^{\rm in}\right]^\top$ and the drift matrix $\mathcal{A}$ takes the form
\begin{align}
\mathcal{A} = \begin{bmatrix}
-\gamma_{\rm c}+2\lambda\cos(\theta) & \Delta_{\rm c}+2\lambda\sin(\theta) & 0 & {\rm g}_{\rm mc} \\
-\Delta_{\rm c}+2\lambda\sin(\theta) & -\gamma_{\rm c}-2\lambda\cos(\theta) & -{\rm g}_{\rm mc} & 0 \\
0 & {\rm g}_{\rm mc} & -\gamma_{\rm m} & \Delta_{\rm m} \\
-{\rm g}_{\rm mc} & 0 & -\Delta_{\rm m}  & -\gamma_{\rm m} \\
\end{bmatrix}.
\end{align}
Taking into account the characteristic equation $\det(\mathcal{A}-\Lambda\mathbb{I})=\Lambda^4+f_1\Lambda^3+f_2\Lambda^2+f_3\Lambda+f_4$. Based on the Routh-Hurwitz stability criterion \cite{Dejesus1987}, the system is stable if and only if all Hurwitz determinants are positive. 

The linearized dynamics and Gaussian nature of the noise terms ensure that the steady-state fluctuations form a Gaussian state, characterized by the $4 \times 4$ covariance matrix $\mathcal{C}$
\begin{align}
\mathcal{C}_{ij} = \frac{1}{2} \langle \mathbf{u}_i(t)\mathbf{u}_j(t') + \mathbf{u}_j(t')\mathbf{u}_i(t) \rangle \quad (i, j = 1, 2, \dots, 4).
\end{align}
The dynamics-state values of the covariance matrix are governed by the following Lyapunov equation \cite{Vitali07, ref:31}
\begin{align} \label{CM}
\partial_t\mathcal{C} = \mathcal{A}\mathcal{C} + \mathcal{C}\mathcal{A}^{\mathsf{T}} + \mathcal{D},
\end{align}
where the diffusion matrix, determined by the noise correlations $\frac{1}{2}\langle \mathbf{n}_i(t)\mathbf{n}_j(t') + \mathbf{n}_j(t')\mathbf{n}_i(t) \rangle = \mathcal{D}_{ij}\delta(t-t')$, is given by $\mathcal{D} = \text{diag}[\gamma_{\rm c}(2\bar{n}_{\rm c}+1), \gamma_{\rm c}(2\bar{n}_{\rm c}+1), \gamma_{\rm m}(2\bar{n}_{\rm m}+1), \gamma_{\rm m}(2\bar{n}_{\rm m}+1)]$. Here, the system is taken to be initially in the vacuum state. The covariance matrix $\mathcal{C}$  is expressed as
\begin{equation} \label{eq:Sigmamm}
\mathcal{C} =
\begin{pmatrix}
	\mathcal{C}_{\rm c} & \mathcal{C}_{\rm cm} \\
     \mathcal{C}_{\rm cm}^{\top} &\mathcal{C}_{\rm m}
\end{pmatrix},
\end{equation} 
where $\mathcal{C}_{\rm c}$ and $\mathcal{C}_{\rm m}$ represent the $2 \times 2$ sub-matrices corresponding to each mode, while the $2 \times 2$ matrix $\mathcal{C}_{\rm cm}$ captures the cross-correlations between the two modes.

\section{\label{sec4}QUANTUM ENTANGLEMENT}%

The quantum entanglement shared between the cavity and magnon subsystems is measured via the logarithmic negativity, $\mathcal{E}_{\rm mc}$ \cite{adesso, simon}. In CV systems, the bipartite entanglement can be quantified by the logarithmic negativity $\mathcal{E}_{\rm mc}$, given by \cite{adesso, simon, Vidal, Plenio} 
\begin{equation} \label{eq:37}
\mathcal{E}_{\rm mc}=\max\left(0,-\log(2\mu^-)\right),
\end{equation}
where $\mu^-$ represents the minimum symplectic eigenvalue of the partially transposed covariance matrix $\mathcal{C}$ of the two modes
\begin{equation} \label{eq:38}
\mu^-=2^{\frac{-1}{2}} \left(\xi - \sqrt{\xi^2-4\det\mathcal{C}}\right)^{\frac{1}{2}},
\end{equation}
where $\xi = \det \mathcal{C}_{\rm c} + \det \mathcal{C}_{\rm m} - 2\det \mathcal{C}_{\rm cm}$. The two modes exhibit quantum entanglement if and only if the condition $\mu^- < 1/2$ ($\mathcal{E}_{\rm mc} > 0$) is satisfied.

\section{\label{sec5}QUANTUM AND CLASSICAL FISHER INFORMATION}
\subsection{Quantum multiparameter estimation \label{sec4.1}}

Multiparameter estimation constitutes a foundational problem within quantum estimation theory, extending the principles of single-parameter sensing to more complex systems. In a typical quantum estimation scheme, a probe is initialized and evolved through a channel encoded with the target parameters. By measuring the output state and applying an appropriate estimator, the values of the parameters are extracted. We consider a parameter-dependent family of quantum states $\rho_{\boldsymbol{\epsilon}}$, where $\boldsymbol{\epsilon} = (\epsilon_1, \epsilon_2, \dots, \epsilon_d)^\top$ denotes a vector of $d$ physical parameters to be estimated. Moreover, the objective of quantum estimation is to determine $\boldsymbol{\epsilon}$ through a measurement scheme $\{\mathsf{M}_{\boldsymbol{z}}\}$ and a corresponding estimator $\boldsymbol{\hat\epsilon}(\boldsymbol{z})$. Here, the measurement is characterized by a set of POVM elements $\mathsf{M}_{\boldsymbol{z}}$, satisfying the completeness relation $\sum_{z}\mathsf{M}^{\dagger}_z\mathsf{M}_z=\mathbb{I}$. The likelihood of obtaining a specific outcome $\boldsymbol{z}$ for a given parameter set $\boldsymbol{\epsilon}$ is determined by the Born rule, $p(\boldsymbol{z}|\boldsymbol{\epsilon}) = \text{Tr}[\rho_{\boldsymbol{\epsilon}}\mathsf{M}_{\boldsymbol{z}}]$. To quantify the sensitivity of the state to the parameters, we introduce the Symmetric Logarithmic Derivative (SLD) and Right Logarithmic Derivative (RLD) operators for each parameter $\epsilon_\alpha$, defined respectively by the following relations \cite{EPJD2014, hmz:9, Safranek2018, Bakmou2020, Razhin2017, hmz:27}
\begin{align}
\partial_{\boldsymbol{\epsilon}_\alpha}\rho_{\boldsymbol{\epsilon}}  &= \bigg({\rm L}_{\boldsymbol{\epsilon}_\alpha}^{(S)} \rho_{\boldsymbol{\epsilon}} + \rho_{\boldsymbol{\epsilon}} {\rm L}_{\boldsymbol{\epsilon}_\alpha}^{(S)}\bigg)/2 \quad {(\rm SLD)}.  \label{eq:SLD} \\
\partial_{\boldsymbol{\epsilon}_\alpha}\rho_{\boldsymbol{\epsilon}} &=\rho_{\boldsymbol{\epsilon}} {\rm L}_{\boldsymbol{\epsilon}_\alpha}^{(R)}  \quad {(\rm RLD)}, \label{eq:RLD}
\end{align}
where $\partial_{\boldsymbol{\epsilon}_\alpha}=\frac{\partial}{\partial \boldsymbol{\epsilon}_\alpha}$ represents the partial derivative. Using the SLD and RLD operators, we define the corresponding quantum Fisher information matrices as follows
\begin{align}
{\bf H}_{\boldsymbol{\epsilon}_\alpha\boldsymbol{\epsilon}_\beta} &=  {\rm Tr} \left[ \rho_{\boldsymbol{\epsilon}}\left(\frac{ {\rm L}^{(S)}_{\boldsymbol{\epsilon}_\alpha} {\rm L}^{(S)}_{\boldsymbol{\epsilon}_\beta} + {\rm L}^{(S)}_{\boldsymbol{\epsilon}_\beta} {\rm L}^{(S)}_{\boldsymbol{\epsilon}_\alpha}}{2}\right) \right ]\:,  \\
{\bf J}_{\boldsymbol{\epsilon}_\alpha\boldsymbol{\epsilon}_\beta} &= {\rm Tr}[\rho_{\boldsymbol{\epsilon}} {\rm L}_{\boldsymbol{\epsilon}_\beta}^{(R)} {\rm L}_{\boldsymbol{\epsilon}_\alpha}^{(R)\dag}],
\end{align}
where ${\rm Tr}[\mathbf{A}$] represents the trace operation acting on the finite-dimensional matrix $\mathbf{A}$. The precision of the estimation is bounded by two different Cram\'er-Rao inequalities. The corresponding QCRB is given by \cite{hmz:27,Bakmou2020}
\begin{align}
\textbf{Cov}(\boldsymbol{\epsilon})
&\geq \frac{{\bf \mathbf{H}}^{-1}}{\mathcal{M}},
\label{M} \\
\textbf{Cov}(\boldsymbol{\epsilon})
&\geq \frac{{\rm Re}\left({\bf J}^{-1}\right)
+ \left|{\rm Im}\left({\bf J}^{-1}\right)\right|}{\mathcal{M}},
\label{S}
\end{align}
where the covariance matrix $\textbf{Cov}(\epsilon_\alpha, \epsilon_\beta)$ writes as $\textbf{Cov}(\epsilon_\alpha, \epsilon_\beta) = {\rm E} [\epsilon_\alpha \epsilon_\beta] - {\rm E} [\epsilon_\alpha]{\rm E} [\epsilon_\beta]$. Moreover, we have decomposed the inverse of the Quantum Fisher Information Matrix into its real and imaginary parts, $\mathbf{J}^{-1} \equiv {\rm Re}(\mathbf{J}^{-1}) + i{\rm Im}(\mathbf{J}^{-1})$, and utilized the operator absolute value $|\mathbf{A}| \equiv \sqrt{\mathbf{A}\mathbf{A}^{\dagger}}$. $\mathcal{M}$ represents the number of measurement repetitions (or equivalently, the sample size), these inequalities simplify to lower bounds on the sum of the variances of the estimated parameters. Specifically, for the individual estimation strategy, the off-diagonal contributions vanish ($\mathbf{H}_{\epsilon_{\alpha}\epsilon_{\beta}} = \mathbf{J}_{\epsilon_{\alpha}\epsilon_{\beta}} = 0$ for $\alpha \neq \beta$). This implies that each parameter is independently characterized by its variance, and therefore Eqs.~(\ref{M}) and (\ref{S}) reduce to
\begin{align}
\mathrm{Var}[\epsilon_\alpha]
&\geq \frac{{\bf \mathbf{H}}^{-1}_{\epsilon_{\alpha}\epsilon_{\alpha}}}{\mathcal{M}},
\label{M1} \\
\mathrm{Var}[\epsilon_\alpha]
&\geq \frac{
\mathrm{Re}\!\left({\bf \mathbf{J}}^{-1}_{\epsilon_{\alpha}\epsilon_{\alpha}}\right)
+
\left|
\mathrm{Im}\!\left({\bf \mathbf{J}}^{-1}_{\epsilon_{\alpha}\epsilon_{\alpha}}\right)
\right|
}{\mathcal{M}}.
\label{S1}
\end{align}
Equations (\ref{M1}) and (\ref{S1}) are always saturable. This saturation corresponds to an optimal measurement of the parameter, where the optimal states are given by projections onto the eigenbasis of the symmetric logarithmic derivative (SLD). Furthermore, applying the trace operator to the two inequalities in (\ref{M}) and (\ref{S}) shows that they reduce to the sum of the variances of the estimated parameters.
\begin{align}
\label{BS} \sum_\alpha^{d} {\rm Var} [\epsilon_\alpha] &\geq { B}_{ S}/\mathcal{M} := \frac{{\rm Tr}\left[{\bf H}^{-1}\right]}{\mathcal{M}} \:,\\
\label{RD}\sum_\alpha ^{d}{\rm Var} [\epsilon_\alpha] &\geq { B}_{ R}/\mathcal{M} := \frac{{\rm Tr}\left[{\rm Re}\left({\bf J}^{-1}\right)\right] + {\rm Tr}\left[\left|{\rm Im}\left({\bf J}^{-1}\right)\right|\right]}{\mathcal{M}} \:,
\end{align}
the matrices $\mathbf{H}$ and $\mathbf{J}$ are referred to as the SLD \cite{hmz:11} and RLD \cite{hmz:12, hmz:13, hmz:14} quantum Fisher information matrices, respectively. The attainability of the SLD and RLD bounds is generally restricted \cite{hmz:15}. Due to quantum non-commutative, it is often impossible to reach the precision limits for all parameters simultaneously, as the optimization of one measurement typically disturbs the estimation of the others. Furthermore, the optimal estimator associated with the RLD bound does not always correspond to a physically realizable POVM. Conversely, even when the optimal measurements for individual parameters are non-commuting, it may still be possible to saturate both bounds through the implementation of a joint measurement strategy. While much of the work in this field centers on the SLD bound \cite{hmz:16, hmz:17, hmz:18, hmz:19, Monras11, hmz:21, Crowley14}, it is well established that for single-parameter estimation, the SLD QFI is smaller than the RLD QFI. This relationship ensures that the SLD approach provides the tighter bound for sensitivity \cite{hmz:9, hmz:10}. Furthermore, while the single-parameter SLD bound is asymptotically attainable in the limit of large $\mathcal{M}$, the multiparameter case is more constrained. Recent developments in quantum local asymptotic normality suggest that the bound in Eq. (\ref{M}) is asymptotically saturate if and only if the weak compatibility condition is satisfied \cite{hmz:16}: $\text{Tr}[\rho_{\boldsymbol{\epsilon}} [{\rm L}_{\epsilon_\alpha}, {\rm L}_{\epsilon_\beta}]] = 0.$ In scenarios where this condition is violated, the RLD bound may provide a more stringent limit, thereby assuming greater significance for characterizing the system's precision.

It is therefore essential to identify the conditions under which these inequalities are saturated in multiparameter scenarios, enabling the realization of an optimal measurement strategy. In this regard, it is worth noting that while several studies in quantum multiparameter estimation theory \cite{Ragy16, hmz:24, Vaneph13, Vidrighin14} have focused on the SLD-based framework, they predominantly demonstrate that the Quantum Cram\'er-Rao Bounds (QCRBs) in Eqs. (\ref{M}) and (\ref{BS}) are saturable if and only if:
\begin{equation}
{\rm Tr}\left[{\rho_{\boldsymbol{\epsilon}} \left[ {{\rm L}_{{\epsilon_\alpha }}^{(S)},{\rm L}_{{\epsilon _\beta }}^{(S)}} \right]} \right] = 0 \label{23}.
\end{equation}
It is straightforward to show that condition (\ref{23}) can be reformulated as
\begin{equation}
{\rm{Im}}\left( {{\rm Tr}\left[ {\rho_{\boldsymbol{\epsilon}} {\rm L}_{{\epsilon_\alpha}}^{(S)} {\rm L}_{{\epsilon_\beta}}^{(S)}} \right]} \right) = 0. \label{24}
\end{equation}
It is therefore pertinent to explore the link between the RLD and SLD bounds to identify which is more fundamental. This led to the development of the most informative QCRB ($B_{\rm MI}$) \cite{hmz:27, hmz:28}, defined as follows 
\begin{equation}
B_{\rm MI} = \max\left\{ {{B_R},{B_S}} \right\}. \label{25}
\end{equation}
Accordingly, the most informative QCRB is determined by comparing the SLD and RLD limits. We thus define the following ratio 
\begin{equation}
\mathcal{R} = \frac{{{B_S}}}{{{B_R}}}.
\end{equation}
Depending on the ratio $\mathcal{R}$, $B_{\rm MI}$ takes the following values
$$B_{\rm MI} = \begin{cases} B_S & \text{if } \mathcal{R} > 1, \\ B_R & \text{if } \mathcal{R} < 1,\end{cases}$$
naturally, if $\mathcal{R} = 1$, then $B_{\rm MI} = B_R = B_S$.

In summary, the optimal precision limits in the multiparameter regime can be unified into a single inequality, expressed as
\begin{equation}
\sum^d\limits_\alpha{{\mathop{\rm Var}} \left[ {{\epsilon_\alpha}} \right] \ge \frac{{{B_{\rm MI}}}}{\mathcal{M}}}.
\end{equation}

The SLD and RLD quantum Fisher information matrices are derived in detail in \cite{EPJD2014, hmz:9,Bakmou2020,hmz:28}. The estimation of the parameters: the cavity-magnon coupling ${\rm g_{mc}}$ or dissipation rate of cavity mode $\gamma_{\rm c}$, the RLD quantum Fisher information matrix $\mathbf{J}$, can be calculated as 
\begin{widetext}
\begin{eqnarray}\label{1}
\mathbf{J}&=&\begin{pmatrix}
2\mathtt{vec}{\left[ {{\partial _{{\rm g_{mc}}}}\mathcal{C}} \right]^\dag }{\Sigma ^{ - 1}}\mathtt{vec}\left[ {{\partial_{{\rm g_{mc}}}}\mathcal{C} } \right] + 2{\partial _{{\rm g_{mc}}}}{\mathbf{d}^\top}\hspace{0.1cm}\Gamma ^{-1}\hspace{0.1cm}{\partial _{{\rm g_{mc}}}}\mathbf{d} & 2\mathtt{vec}{\left[ {{\partial _{{\rm g_{mc}}}}\mathcal{C}} \right]^\dag }{\Sigma^{ - 1}}\mathtt{vec}\left[ {{\partial_{{\rm \gamma_{\rm c}}}}\mathcal{C} } \right] + 2{\partial _{{\rm g_{mc}}}}{\mathbf{d}^\top}\hspace{0.1cm}\Gamma ^{-1}\hspace{0.1cm}{\partial _{{\rm \gamma_{c}}}}\mathbf{d} \\
2\mathtt{vec}{\left[ {{\partial _{{\rm \gamma_{c}}}}\mathcal{C}} \right]^\dag }{\Sigma ^{ - 1}}\mathtt{vec}\left[ {{\partial_{{\rm g_{mc}}}}\mathcal{C} } \right] + 2{\partial _{{\rm \gamma_{c}}}}{\mathbf{d}^\top}\hspace{0.1cm}\Gamma ^{-1}\hspace{0.1cm}{\partial _{{\rm g_{mc}}}}\mathbf{d} & 2\mathtt{vec}{\left[ {{\partial _{\gamma_{\rm c}}}\mathcal{C}} \right]^\dag }{\Sigma ^{ - 1}}\mathtt{vec}\left[ {{\partial_{\gamma_{\rm c}}}\mathcal{C} } \right] + 2{\partial _{{\rm \gamma_{c}}}}{\mathbf{d}^\top}\hspace{0.1cm}\Gamma ^{-1}\hspace{0.1cm}{\partial _{{\rm \gamma_{c}}}}\mathbf{d}
\end{pmatrix}, \label{RLDM}
\nonumber\\
&&=\begin{pmatrix}
{\rm J}_{\rm g_{mc}g_{mc}} & {\rm J}_{\rm g_{mc}\gamma_{\rm c}} \\
{\rm J}_{\gamma_{\rm c}{\rm g_{mc}}} & {\rm J}_{\gamma_{\rm c}\gamma_{\rm c}}
\end{pmatrix},\
\end{eqnarray}
\end{widetext}
where $\Gamma = 2\mathcal{C} + i\Omega$, with $\Omega = \begin{bmatrix} 0 & 1 \\ -1 & 0 \end{bmatrix} \oplus \begin{bmatrix} 0 & 1 \\ -1 & 0 \end{bmatrix}$ and $\mathcal{C}$ is the covariance matrix of the cavity-magnon modes. Under this condition, the Moore-Penrose pseudoinverse of $\Gamma$ reduces to its standard inverse, with $\Sigma=\Gamma^{\dag} \otimes\Gamma$, $\mathbf{d}=[\langle \hat{X}_{\rm c}\rangle,\langle \hat{P}_{\rm c}\rangle,\langle \hat{X}_{\rm m}\rangle,\langle \hat{P}_{\rm m}\rangle]^\top$ and the operator $\mathtt{vec}[\cdot]$ denotes the vectorization of a matrix \cite{Safranek2018}.
The SLD quantum Fisher information matrix $\mathbf{H}$ as
\begin{widetext}
\begin{eqnarray}\label{SLDM}
\mathbf{H}&=& \begin{pmatrix}
2\mathtt{vec}{\left[ {{\partial _{{\rm g_{mc}}}}\mathcal{C}} \right]^\dag }{\mathcal{N}^{ - 1}}\mathtt{vec}\left[ {{\partial_{{\rm g_{mc}}}}\mathcal{C} } \right] + {\partial _{{\rm g_{mc}}}}{\mathbf{d}^\top}\hspace{0.1cm}\mathcal{C}^{-1}\hspace{0.1cm}{\partial _{{\rm g_{mc}}}}\mathbf{d} & 2\mathtt{vec}{\left[ {{\partial _{{\rm g_{mc}}}}\mathcal{C}} \right]^\dag }{\mathcal{N}^{ - 1}}\mathtt{vec}\left[ {{\partial_{{\rm \gamma_{c}}}}\mathcal{C} } \right] + {\partial _{{\rm g_{mc}}}}{\mathbf{d}^\top}\hspace{0.1cm}\mathcal{C}^{-1}\hspace{0.1cm}{\partial _{{\rm \gamma_{c}}}}\mathbf{d} \\
2\mathtt{vec}{\left[ {{\partial _{{\rm \gamma_{c}}}}\mathcal{C}} \right]^\dag }{\mathcal{N}^{ - 1}}\mathtt{vec}\left[ {{\partial_{{\rm g_{mc}}}}\mathcal{C} } \right] + {\partial _{{\rm \gamma_{c}}}}{\mathbf{d}^\top}\hspace{0.1cm}\mathcal{C}^{-1}\hspace{0.1cm}{\partial _{{\rm g_{mc}}}}\mathbf{d} & 2\mathtt{vec}{\left[ {{\partial _{\gamma_{\rm c}}}\mathcal{C}} \right]^\dag }{\mathcal{N}^{ - 1}}\mathtt{vec}\left[ {{\partial_{\gamma_{\rm c}}}\mathcal{C} } \right] + {\partial _{{\rm \gamma_{c}}}}{\mathbf{d}^\top}\hspace{0.1cm}\mathcal{C}^{-1}\hspace{0.1cm}{\partial _{{\rm \gamma_{c}}}}\mathbf{d}
\end{pmatrix}
\nonumber\\
&&=\begin{pmatrix}
{\rm H}_{\rm g_{mc}g_{mc}} & {\rm H}_{\rm g_{mc}\gamma_{\rm c}} \\
{\rm H}_{\gamma_{\rm c}{\rm g}_{\rm mc}} & {\rm H}_{\gamma_{\rm c}\gamma_{\rm c}}
\end{pmatrix},\
\end{eqnarray}
\end{widetext}
with $\mathcal{N} = 4{\mathcal{C}^\dagger \otimes \mathcal{C} + \Omega \otimes \Omega}$. Then, $B_S$ can be calculated from Eq. (\ref{SLDM}), and $B_R$ from Eq. (\ref{RLDM}).

Throughout the multiparameter analysis, we consider the SLD and RLD QFIM associated with the parameter vector $\boldsymbol{\epsilon}=(\rm g_{mc},\gamma_c)^\top$, including both its diagonal and off-diagonal elements. The corresponding multiparameter bounds are obtained from the inverse of the SLD and RLD QFIMs. Both parameters $g_{mc}$ and $\gamma_c$ are encoded in the same cavity--magnon Gaussian state. We emphasize that the diagonal quantities $\rm {H}_{{\rm g_{mc}}{\rm g_{mc}}}$ and $\rm {H}_{\gamma_c\gamma_c}$, considered separately later in the comparison with classical Fisher information, are introduced only to quantify the information extracted by specific homodyne and heterodyne measurements for each parameter. Their separate presentation does not replace the full multiparameter analysis.

\subsection{Classical Fisher information} \label{IIIB}

This section details the Classical Fisher Information (CFI) framework. In this context, a quantum measurement is described by a Positive-Operator Valued Measure (POVM) $\{\hat{\mathsf{M}}_{z}\}$, where each operator is positive semi-definite ($\hat{\mathsf{M}}_{z} \geq 0$) and the set satisfies the completeness condition $\sum_{z} \hat{\mathsf{M}}^\dag_{z}\hat{\mathsf{M}}_{z} = \mathbb{I}$ \cite{hmz:30}. We denote the set of all such valid measurement schemes as $\tilde{\Omega}$. The measurement outcomes are characterized by a conditional probability distribution determined by the Born rule
\begin{equation}
\{P(z|\epsilon)=\text{Tr}[{\rho}_{\epsilon}\hat{\mathsf{M}} _{z}]\}, 
\end{equation}%
with $P(z|\epsilon)$ denoting the conditional probability for the measurement result $z$. From this distribution, the value of the parameter $\epsilon$ can be estimated, with the CFI defined as \cite{hmz:11,hmz:31,paris2009}
\begin{equation}
\label{Eq24}
F_{\epsilon}=\int \frac{1}{P(z|\epsilon)}\left[ \frac{\partial P(z|\epsilon)}{\partial \epsilon}\right] ^{2}dz.
\end{equation}
Since the QFI is defined as $\mathsf{H}_{\epsilon} := \max_{\tilde{\Omega}} \{F_{\epsilon}\}$, it inherently accounts for the optimal measurement strategy. Thus, we obtain
\begin{equation}
\text{Var}\left(\hat{\epsilon}\right) \geq \frac{1}{\mathcal{M}F_{\epsilon}}\geq \frac{1}{\mathcal{M}\mathsf{H}_{\epsilon}}.
\end{equation}
The lower bound of the classical CRB is achievable through optimal estimation strategies; for instance, the maximum likelihood estimator is known to be asymptotically efficient. It should be noted that all measurements considered in this work are assumed to be ideal, corresponding to a measurement efficiency of unity. The calculation of the CFI is performed using the second-moment matrix $\mathcal{C}$ and the first-moment vector $\mathbf{d}$.

\noindent{\bf Heterodyne detection.}---An essential Gaussian measurement technique is heterodyne detection, which involves mixing the measured field with a probe field of a different frequency. The corresponding measurement operators are defined by the overcomplete set of coherent state projectors, $\{|\alpha \rangle \langle \alpha |/\pi \}$, which constitute a valid POVM \cite{hmz:10,Meystre2007,Scully1999,hmz:35}. The heterodyne detection scheme typically involves mixing the mode of interest with an ancillary vacuum state via a $50:50$ beam splitter. The $\hat{X}$ and $\hat{P}$ quadratures of the resulting modes are then measured. For the cavity-magnon system, the realization of optimal heterodyne detection on its subsystems is described in \cite{hmz:36}
\begin{equation}
\label{Eq33}
F_{\epsilon,\text{Het}}=\frac{1}{2}\text{Tr}\bigg(\sigma^{-1}\frac{d\sigma}{d\epsilon}\sigma^{-1}\frac{d\sigma}{d\epsilon}\bigg) + \frac{d\mathbf{d}^{\top}}{d\epsilon}\sigma^{-1}\frac{d\mathbf{d}}{d\epsilon},
\end{equation}
where $\sigma = \mathcal{C} + \mathbb{I}_{4\times 4}$, and the subscript “Het”  marks the heterodyne detection. Here, $\mathbf{d}$ and $\mathcal{C}$ are the first and second moments of the bipartite system, while $\mathbb{I}_{4\times 4}$ represents the supplemental noise inherent in the simultaneous measurement of conjugate quadratures $\hat{X}$ and $\hat{P}$.

\noindent{\bf Homodyne detection.}---In homodyne detection, the signal and a strong local oscillator are mixed at the same frequency, resulting in a measurement of a single quadrature. The corresponding POVM elements are $\{|Q\rangle\langle Q|\}$ or $\{|Y\rangle \langle Y|\}$, representing the eigenstates of the $Q$ or $Y$ quadrature \cite{hmz:10,Meystre2007,Scully1999,hmz:35,hmz:36}. Moreover, $Q \in \{\hat{X}_{\rm c}, \hat{X}_{\rm m}\}$ and $Y \in \{\hat{P}_{\rm c}, \hat{P}_{\rm m}\}$, the cavity quadratures are directly measurable, whereas the magnon quadratures are detected via auxiliary fields. This is typically implemented by coupling the YIG sphere to a weakly driven microwave cavity. Under homodyne detection, the CFI with respect to $\epsilon$ can be expressed as follows \cite{Monras2013}
\begin{equation}
\label{Eq31}
F_{\epsilon,\text{Ho}}^{j}=\frac{1}{2\mathcal{C}_{jj}^{2}}\left[ 2\mathcal{C}_{jj}(\partial_{\epsilon} \mathbf{d}_{j})^{2}+\left( \partial_{\epsilon}\mathcal{C}_{jj}\right) ^{2}\right],
\end{equation}%
where, the superscript '$j$' labels the measured quadrature, while the subscript 'Ho' refers to homodyne detection. Specifically, the CFI for parameters such as the cavity-magnon coupling ${\rm g_{mc}}$ or dissipation rate of cavity mode $\gamma_{\rm c}$, utilizing homodyne measurements of $\textbf{d}_{1} = \langle \hat{X}_{\rm c} \rangle$ and $\textbf{d}_{2} = \langle \hat{P}_{\rm c} \rangle$, are expressed as
\begin{align}
\label{Eq31}
F_{\rm g_{mc}g_{mc},\text{Ho}}^{\hat X_{\rm c}}=&\frac{1}{2\mathcal{C}_{11}^{2}}\left[ 2\mathcal{C}_{11}(\partial_{\rm g_{mc}}\langle \hat{X}_{\rm c} \rangle)^{2}+\left( \partial_{\rm g_{mc}}\mathcal{C}_{11}\right) ^{2}\right],\\ \nonumber
 F_{\gamma_{\rm c}\gamma_{\rm c},\text{Ho}}^{\hat X_{\rm c}}=&\frac{1}{2\mathcal{C}_{11}^{2}}\left[ 2\mathcal{C}_{11}(\partial_{\gamma_{\rm c}}\langle \hat{X}_{\rm c} \rangle)^{2}+\left( \partial_{\gamma_{\rm c}}\mathcal{C}_{11}\right) ^{2}\right],
\end{align}
and 
\begin{align}
\label{Eq31}
F_{\rm g_{mc}g_{mc},\text{Ho}}^{\hat P_{\rm c}}=&\frac{1}{2\mathcal{C}_{22}^{2}}\left[ 2\mathcal{C}_{22}(\partial_{\rm g_{mc}}\langle \hat{P}_{\rm c} \rangle)^{2}+\left( \partial_{\rm g_{mc}}\mathcal{C}_{22}\right) ^{2}\right],\\ \nonumber
 F_{\gamma_{\rm c}\gamma_{\rm c},\text{Ho}}^{\hat P_{\rm c}}=&\frac{1}{2\mathcal{C}_{22}^{2}}\left[ 2\mathcal{C}_{22}(\partial_{\gamma_{\rm c}}\langle \hat{P}_{\rm c} \rangle)^{2}+\left( \partial_{\gamma_{\rm c}}\mathcal{C}_{22}\right) ^{2}\right].
\end{align}

\section{Results and discussions \label{sec5}}

We now examine the sensitivity of the estimation precision for the coupling strength ${\rm g_{mc}}$ and cavity decay $\gamma_{\rm c}$ to several system parameters in steady and dynamical state. These include the environmental temperature $\rm T$, driving power $\rm P$, and the respective dissipation rates and detunings of the cavity and magnon modes ($\gamma_{\rm c,m}$ and $\Delta_{\rm c,m}$). Moreover, the effects of the OPA parameters, namely the gain $\lambda$ and squeeze phase $\theta$, are considered. Additionally, we compare the SLD-optimized quantum bound with the CFI results achieved under heterodyne and homodyne detection. Besides, the feasible value of the parameters reported in table \ref{table}: 
\begin{widetext}
\begin{table}[!h]
\caption{Parameters used in numerical simulations \cite{Deng24,Zhang2014,Ebrahimi21,Peng2025}.}
\label{table}
\begin{tabular*}{\textwidth}{@{\extracolsep{\fill}} l c c c}
\hline\hline
Parameter & Symbol & Value & Unit \\
\hline\hline
Drive power & $\rm P$ & $500\times10^{-3}$ & W \\
Drive field frequency & $\omega_{0}/2\pi$ & $10\times10^{9}$ & Hz \\
Temperature & $\rm T$ & $10\times10^{-3}$ & K \\
Cavity decay rate & $\gamma_{\rm c}/2\pi$ & $5\times10^{6}$ & Hz \\
Magnon decay rate & $\gamma_{\rm m}/2\pi$ & $40\times10^{6}$ & Hz \\
Cavity detuning & $\Delta_{\rm c}/2\pi$ & $40\times10^{6}$ & Hz \\
Magnon detuning & $\Delta_{\rm m}/2\pi$ & $0$ & Hz \\
Coupling magnon-photon & $\rm g_{\rm mc}/2\pi$ & $41\times10^{6}$ & Hz \\
\hline\hline
\end{tabular*}
\end{table}
\end{widetext}

\subsection{Steady-state}

In the following, $B_{\rm MI}$ and $\mathcal{R}$ characterize the joint estimation of ${\rm g_{mc}}$ and $\gamma_c$, and are evaluated using the SLD and RLD QFIMs rather than independent single-parameter Fisher information matrices.
\begin{figure}[!htb]
\minipage{0.5\textwidth}
\includegraphics[width=1\linewidth]{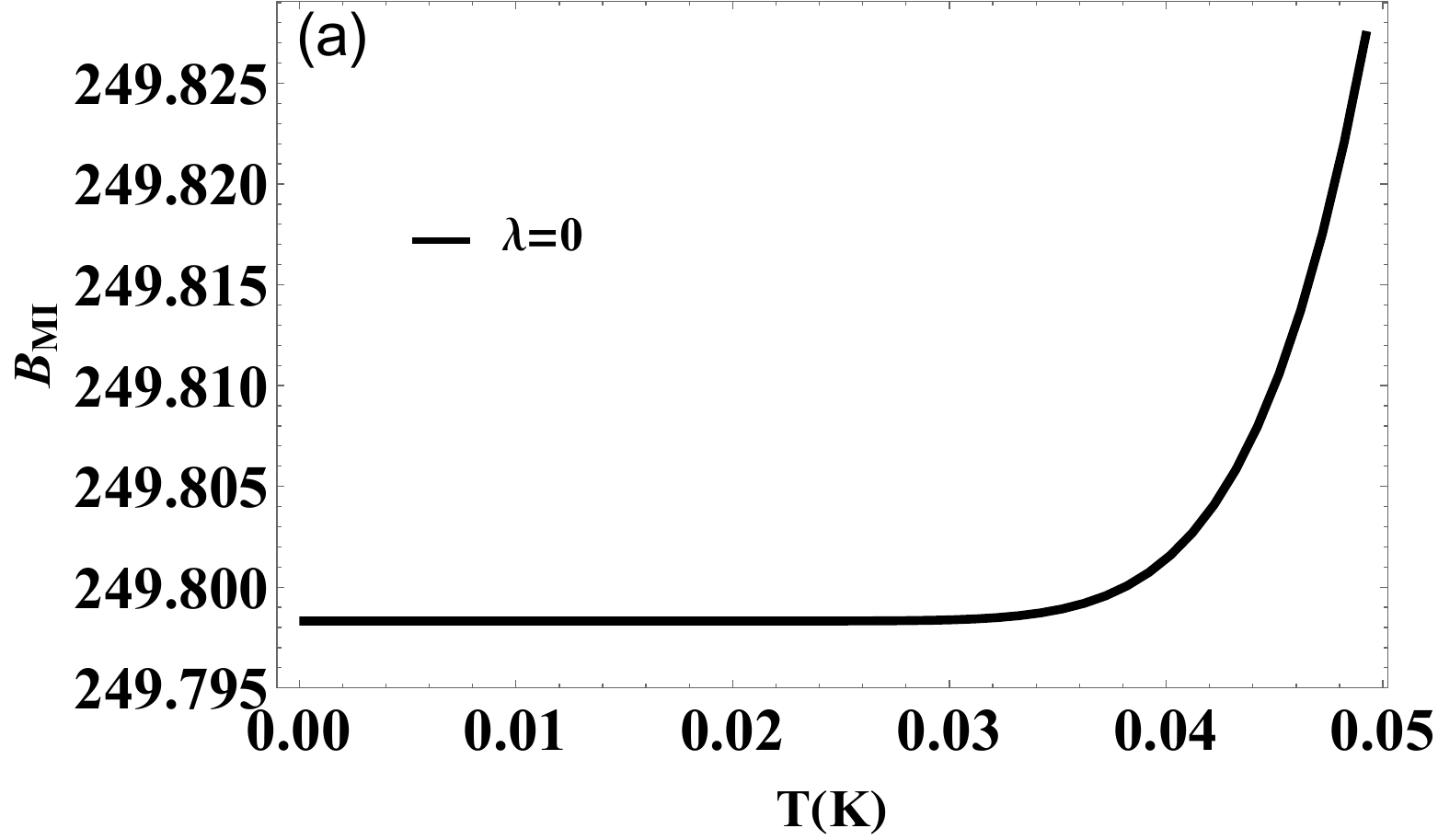}
\endminipage\hfill
\minipage{0.5\textwidth}
\includegraphics[width=1\linewidth]{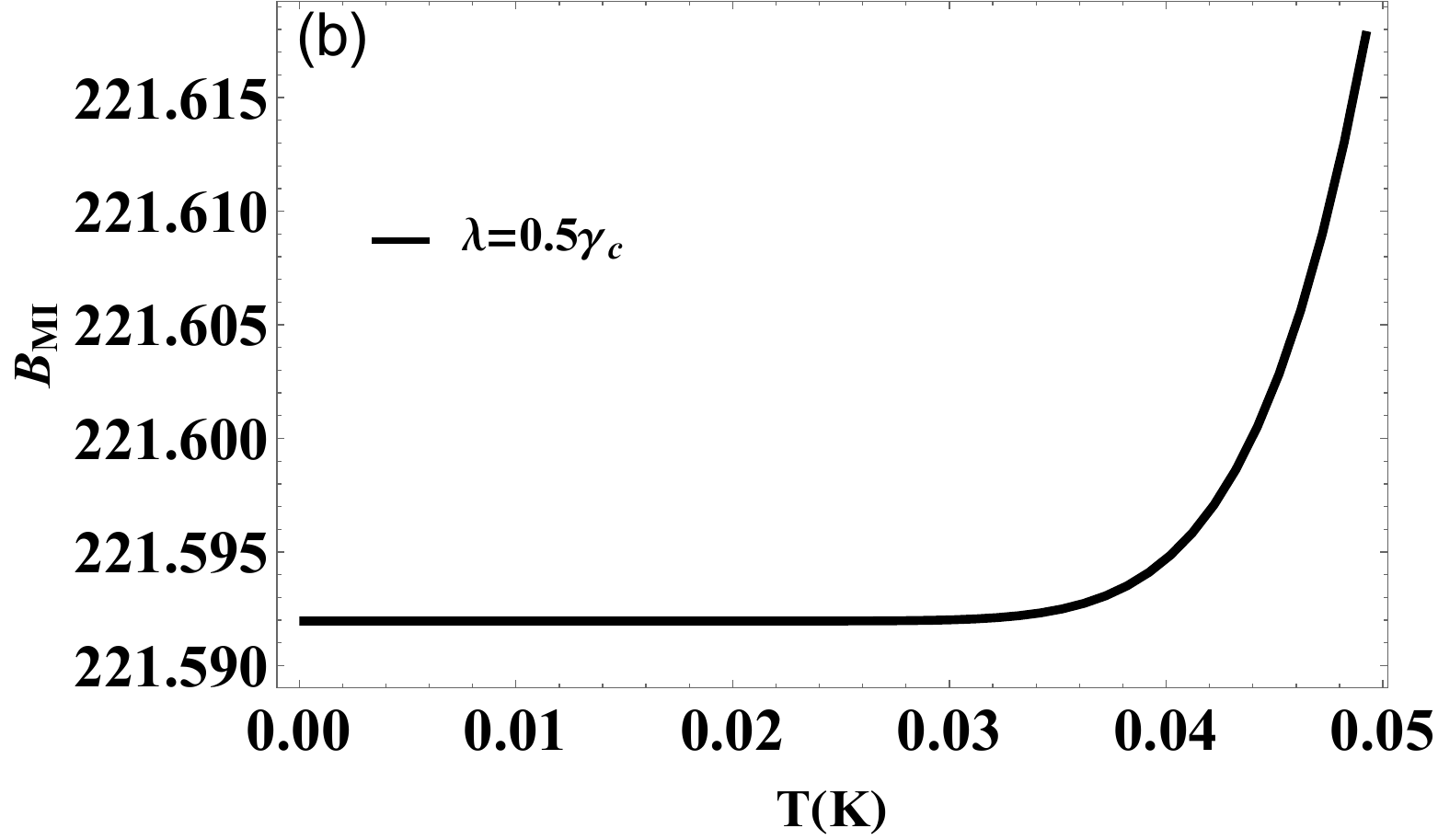}
\endminipage\hfill
\minipage{0.5\textwidth}
\includegraphics[width=1\linewidth]{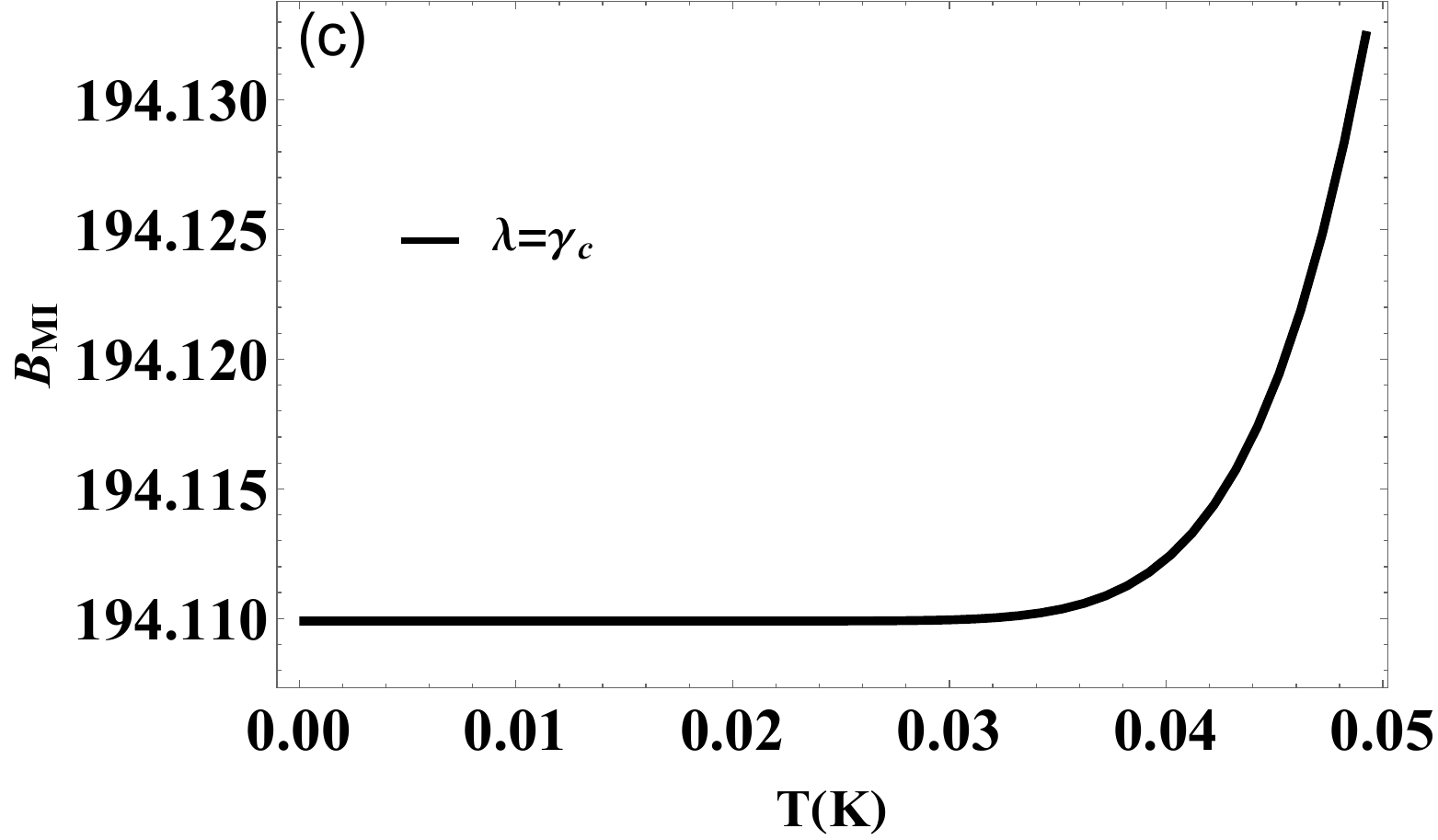}
\endminipage\hfill
\minipage{0.5\textwidth}
\includegraphics[width=1\linewidth]{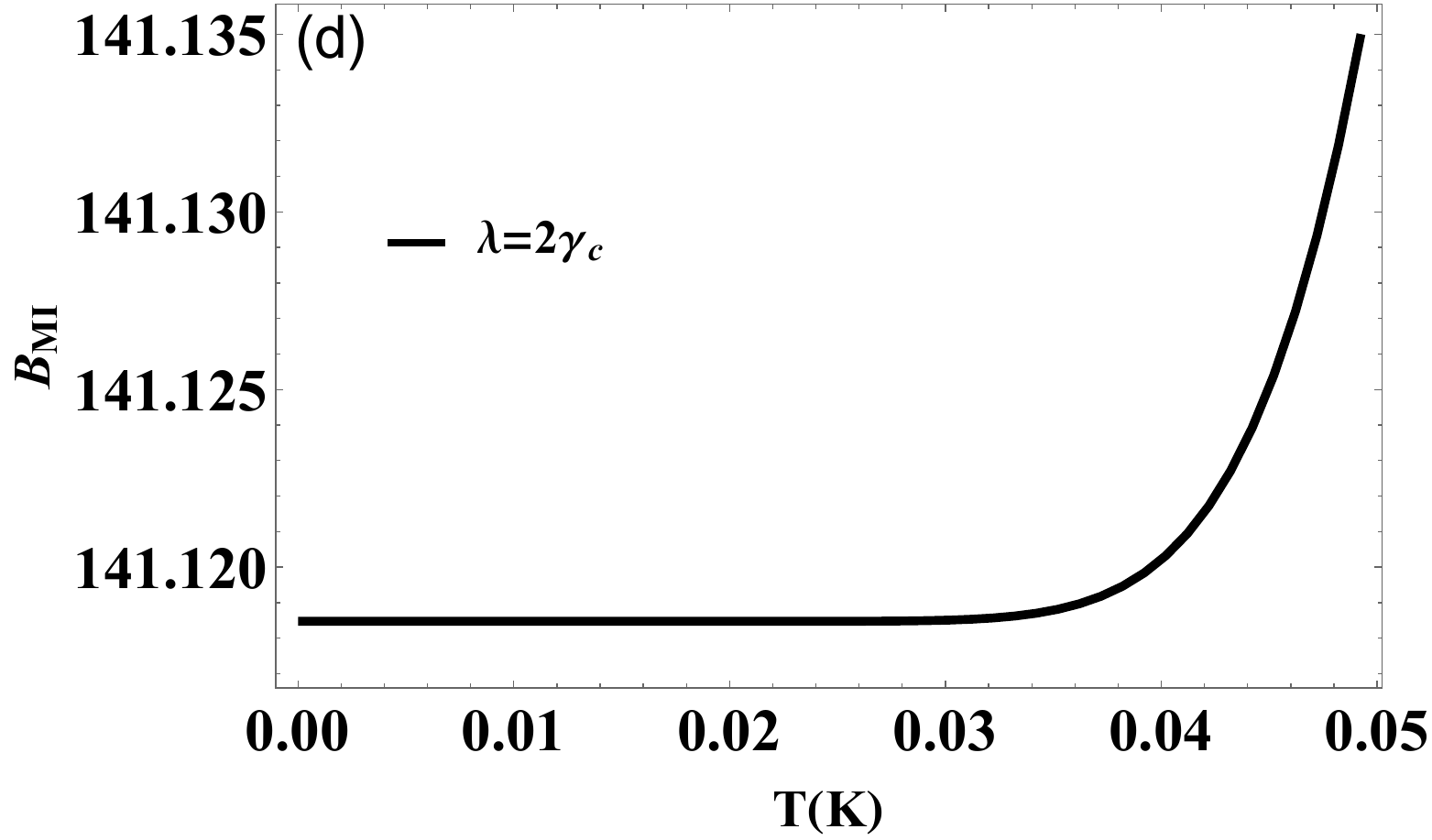}
\endminipage\hfill
\caption{Plot of the most informative quantum Cram\'er-Rao bound $B_{\rm MI}$ as function of the temperature $\rm T$ for different values of the gain $\lambda$, with $\theta=3\pi/2$. The other parameters are the same as in the table \ref{table}.}
\label{LT}
\end{figure}
Fig.~\ref{LT} explores the $B_{\rm MI}$ as a function of the temperature $\rm T$ for different values of the OPA gain $\lambda$. One can observe that the $B_{\rm MI}$ increases monotonically with increasing temperature for all considered values of $\lambda$, indicating a progressive degradation of the multiparameter estimation precision due to thermal noise. Furthermore, increasing the gain $\lambda$ significantly reduces the $B_{\rm MI}$ over a broad temperature range, as shown in Figs.~\ref{LT}(a-d). This behavior demonstrates that the presence of the OPA effectively enhances the precision of multiparameter estimation. Moreover, the degradation of the estimation precision is mitigated due to the suppression of thermal fluctuations through the OPA. This means that the squeezed field generated by the OPA modifies the distribution of quantum fluctuations and enhances the sensitivity of the Gaussian steady-state to small parameter variations, thereby reducing the $B_{\rm MI}$ and improving the attainable precision. Nevertheless, despite this improvement, the $B_{\rm MI}$ still increases monotonically with the temperature $\rm T$. This increase becomes more pronounced once the temperature exceeds approximately $0.03\,\mathrm{K}$, confirming that thermal effects continue to negatively affect the estimation performance even in the presence of the OPA.%

\begin{figure}[!htb]
\minipage{0.5\textwidth}
\includegraphics[width=1\linewidth]{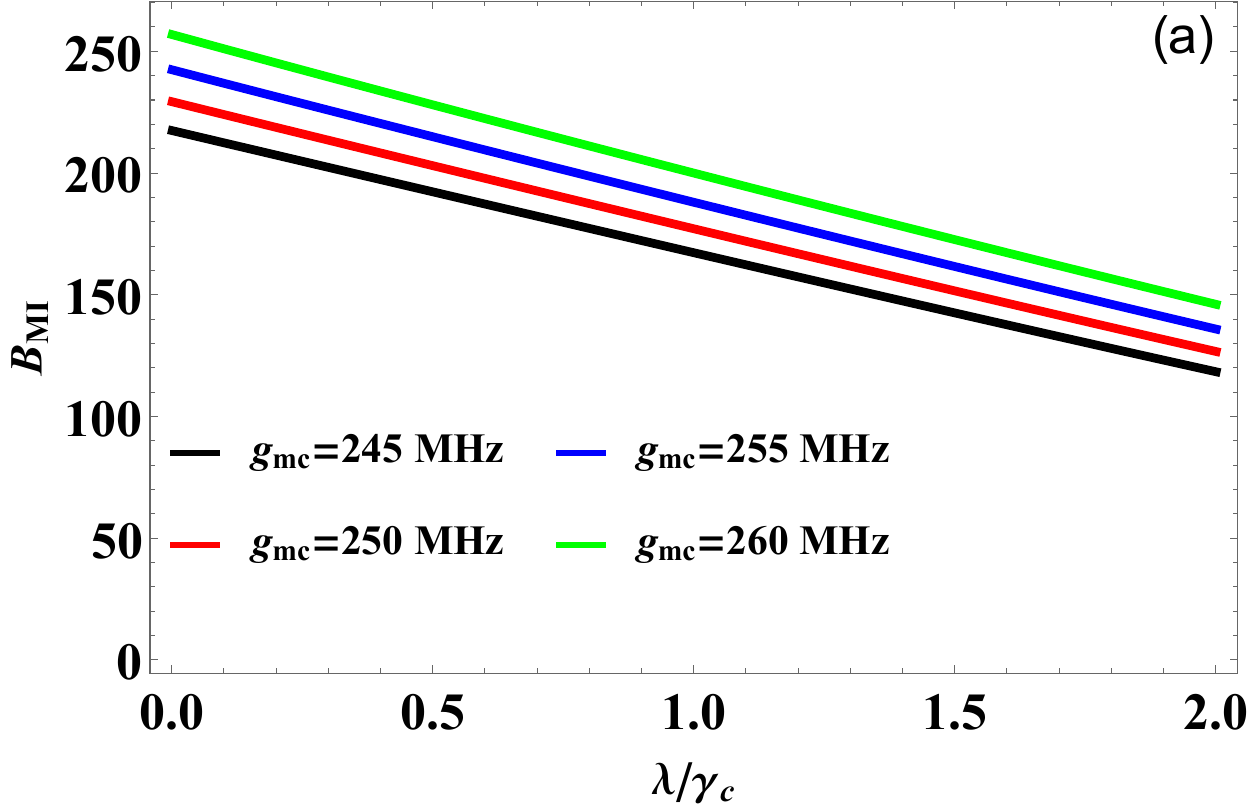}
\endminipage\hfill
\minipage{0.5\textwidth}
\includegraphics[width=1\linewidth]{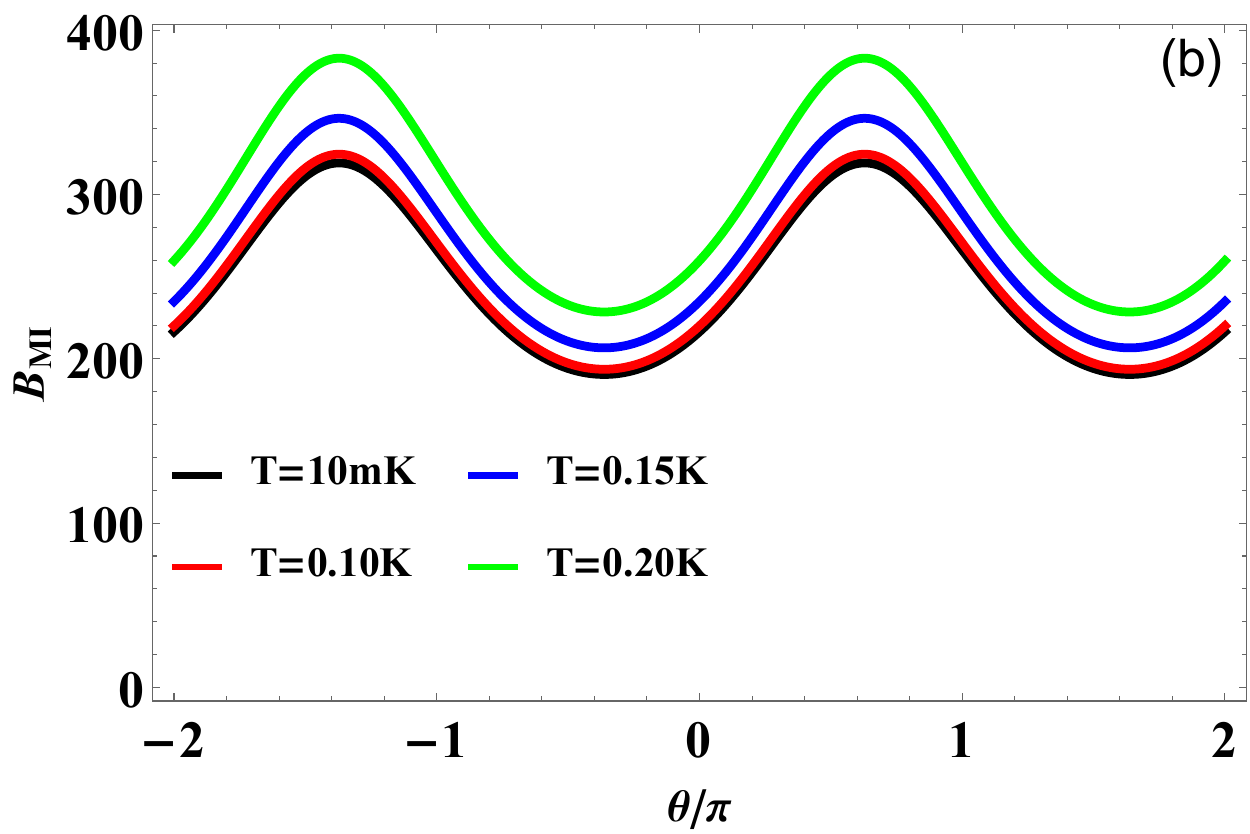}
\endminipage\hfill
\caption{Plot of the variation of the $B_{\rm MI}$ versus: (a) $\lambda/\gamma_{\rm c}$ for different values of the cavity-magnon coupling ${\rm g_{mc}}$ with $\theta=3\pi/2$; (b) $\theta/\pi$ for different values of temperature $\rm T$, with $\lambda=\gamma_{\rm c}$. The other parameters are the same as in the table \ref{table}.}
\label{STSAB3}
\end{figure}

Figure \ref{STSAB3} presents the behavior of the $B_{\rm MI}$ as a function of (a) the normalized gain $\lambda/\gamma_{\rm c}$ and (b) the phase $\theta/\pi$ for various values of the cavity-magnon coupling strength ${\rm g_{mc}}$. In Fig.~\ref{STSAB3}(a), the $B_{\rm MI}$ decreases as the OPA gain $\lambda$ increases, which indicates an enhancement of the multiparameter estimation precision. This means that the OPA reduces the quantum fluctuations of the cavity field and enhances the useful quantum correlations inside the system, leading to a larger amount of extractable information and consequently smaller $B_{\rm MI}$ values. Moreover, the reduction of the $B_{\rm MI}$ remains significant over a broad range of $\lambda$, demonstrating the important role of the OPA in improving the estimation precision. In addition, increasing the cavity-magnon coupling strength ${\rm g}_{\rm mc}$ leads to larger $B_{\rm MI}$ values, as depicted in Fig.~\ref{STSAB3}(a). This behavior indicates that strong hybrid coupling progressively degrades the estimation precision. This stronger coupling increases the redistribution of quantum fluctuations between the cavity and magnon modes, which reduces the sensitivity of the steady Gaussian state to small parameter variations.

Figure~\ref{STSAB3}(b) shows that the $B_{\rm MI}$ exhibits a periodic dependence on the phase $\theta/\pi$, highlighting the phase-sensitive nature of the OPA. The alternating maxima and minima indicate that the estimation precision can be controlled through the phase of the parametric interaction. In particular, the minima of the $B_{\rm MI}$ correspond to regions of enhanced estimation precision, whereas the maxima correspond to regimes with lower precision. Furthermore, increasing the temperature $\rm T$ increases the minimum values of the $B_{\rm MI}$, indicating that thermal effects progressively reduce the estimation precision. This means that the thermal effects weakens the quantum correlations generated by the OPA and increases the fluctuations inside the cavity-magnon system.

\begin{figure}[!htb]
\minipage{0.5\textwidth}
\includegraphics[width=1\linewidth]{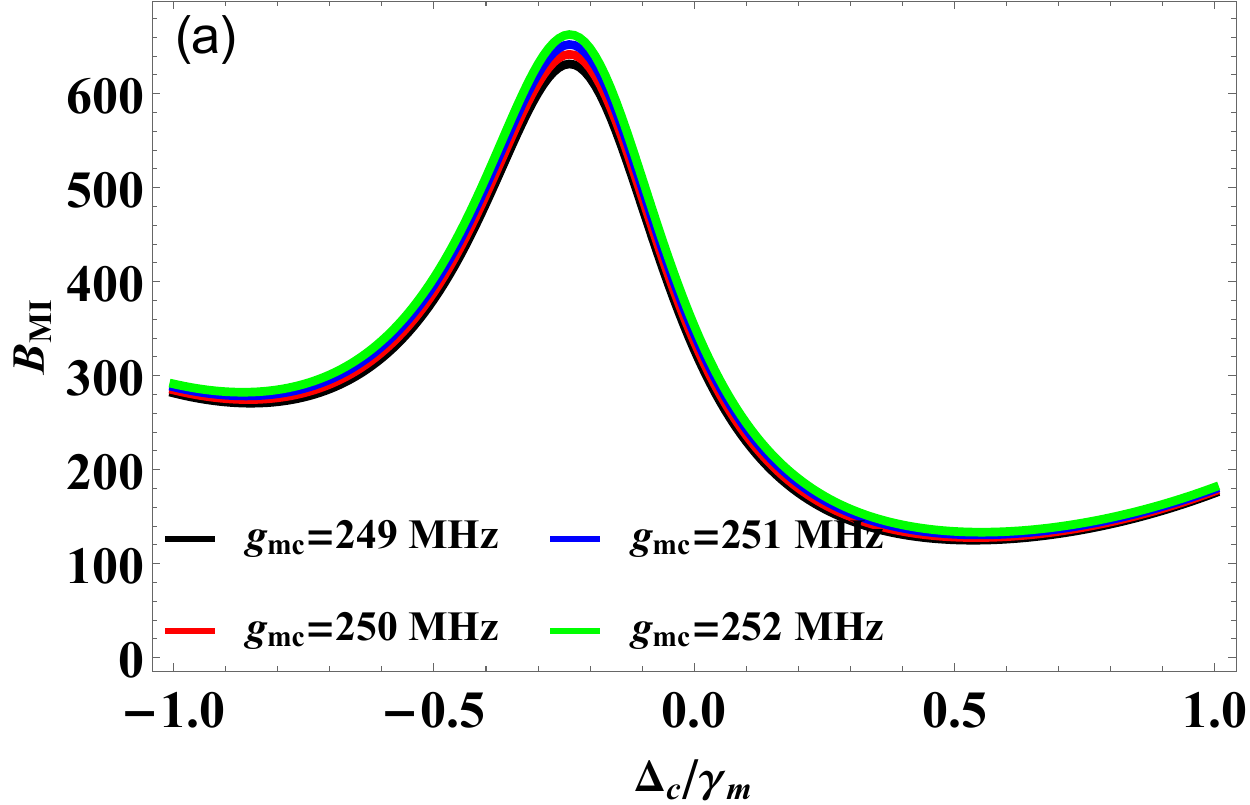}
\endminipage\hfill
\minipage{0.5\textwidth}
\includegraphics[width=1\linewidth]{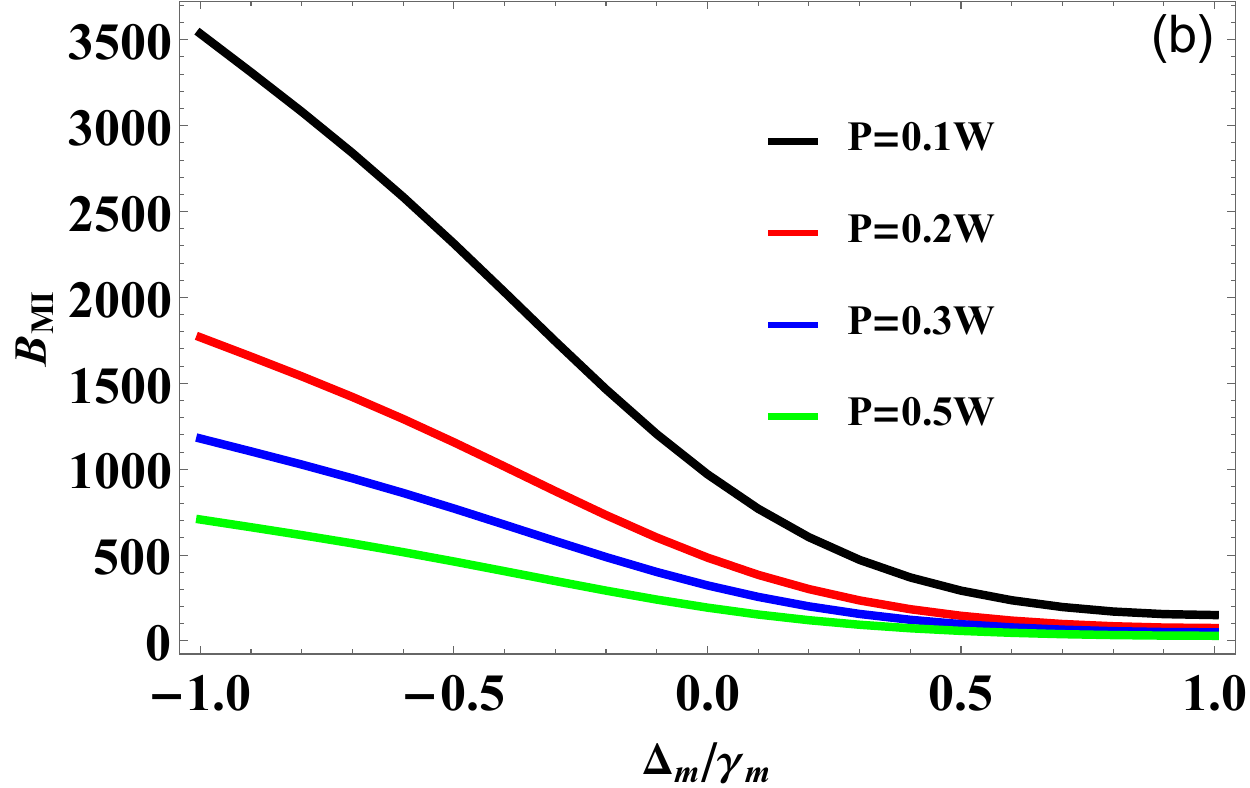}
\endminipage\hfill
\caption{Plot of the variation of the $B_{\rm MI}$ as function of: (a) cavity detuning $\Delta_{\rm c}/\gamma_{\rm m}$ for different value of the cavity-magnon coupling ${\rm g}_{\rm mc}$ and (b) magnon detuning $\Delta_{\rm m}/\gamma_{\rm m}$ for different value of the power $\rm P$. The other parameters are the same as in the table \ref{table}, with $\lambda = \gamma_{\rm c}$ and $\theta=3\pi/2$.}
\label{STSAB4}
\end{figure}

We plot in Fig~\ref{STSAB4}(a) the $B_{\rm MI}$ as a function of the normalized cavity detuning $\Delta_{\rm c}/\gamma_{\rm m}$ for different values of the cavity-magnon coupling strength ${\rm g}_{\rm mc}$. One can observe that the $B_{\rm MI}$ initially increases as the cavity detuning varies from  $\Delta_{\rm c} / \omega_{\rm m} \approx -0.75$ toward the resonance region around $\Delta_{\rm c} / \omega_{\rm m} \approx -0.25$, where the $B_{\rm MI}$ reaches its maximum value. This behavior indicates a degradation of the multiparameter estimation precision near the resonance region, which can due to the enhancement of fluctuation transfer between the cavity and magnon modes. For $\Delta_{\rm c}/\gamma_{\rm m}>0$, the $B_{\rm MI}$ decreases rapidly and reaches its minimum value around $\Delta_{\rm c}/\gamma_{\rm m}\approx0.5$, corresponding to a region of enhanced estimation precision. Moreover, large positive cavity detuning weakens the effective interaction and suppresses the thermal effects between the subsystems, thereby reducing the $B_{\rm MI}$. Furthermore, increasing the coupling strength ${\rm g}_{\rm mc}$ slightly increases the $B_{\rm MI}$ over the entire parameter region, indicating that stronger hybrid coupling reduces the amount of extractable information associated with the estimated parameters.

Figure~\ref{STSAB4}(b) presents the $B_{\rm MI}$ as a function of the normalized magnon detuning $\Delta_{\rm m}/\gamma_{\rm m}$ for different values of the driving power $\rm P$. It can be observed that the $B_{\rm MI}$ decreases monotonically as $\Delta_{\rm m}/\gamma_{\rm m}$ increases from $-1$ to $1$ for all considered driving powers. A similar trend is observed for higher driving powers, although the overall $B_{\rm MI}$ values become significantly smaller. This behavior indicates a continuous improvement of the multiparameter estimation precision as the magnon detuning increases within the considered interval. Besides, no pronounced resonance feature is observed near the magnon condition $\Delta_{\rm m}=0$. This result indicates that the estimation precision is not solely determined by the magnitude of the magnon response but rather by the sensitivity of the hybrid Gaussian state to variations of the estimated parameters. Moreover, increasing $\Delta_{\rm m}$ modifies the balance between the coherent cavity-magnon interaction and dissipation, leading to a progressive reduction of the $B_{\rm MI}$. Furthermore, increasing the driving power from $0.05~{\rm W}$ to $0.2~{\rm W}$ significantly reduces the $B_{\rm MI}$ throughout the entire detuning range. These results demonstrate that stronger driving enhances the amount of accessible information encoded in the cavity-magnon system and consequently improves the achievable multiparameter estimation precision.

One can calculate the Wigner function of the optical mode to verify the presence of optical squeezing 
\begin{equation}
\mathcal{W}(u_{\rm c}) = \frac{\exp\left[ \frac{-1}{2} u_{\rm c}^T \mathcal{C}_{\rm c}^{-1} u_{\rm c} \right]}{2\pi \sqrt{\det \mathcal{C}_{\rm c}}},
\end{equation}
where $u_{\rm c} = (\delta {X}_{\rm c}, \delta {P}_{\rm c})$ denotes the phase-space variables associated with the fluctuations of the optical position and momentum, and $\mathcal{C}_{\rm c}$ is the CM of the optical mode which can be achieved in steady-state from Eq. (\ref{eq:Sigmamm}).
\begin{figure}[!htb]
  \centering
  \begin{minipage}{0.48\columnwidth}
    \centering
    \includegraphics[width=\linewidth]{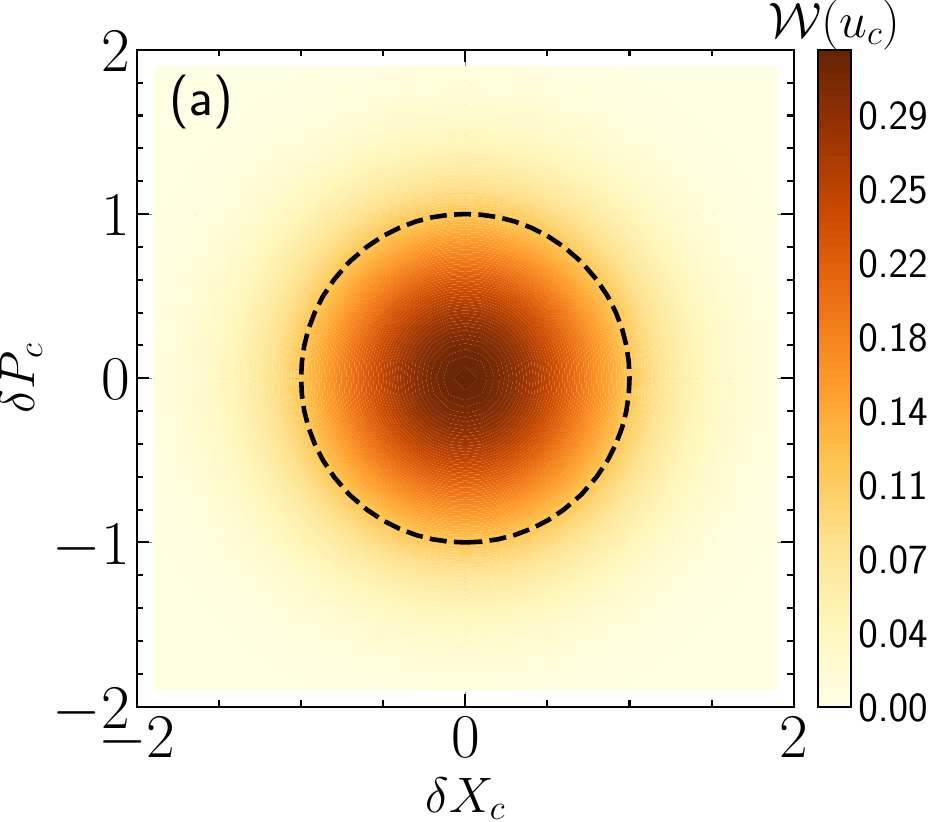}
  \end{minipage}\hfill
  \begin{minipage}{0.48\columnwidth}
    \centering
    \includegraphics[width=\linewidth]{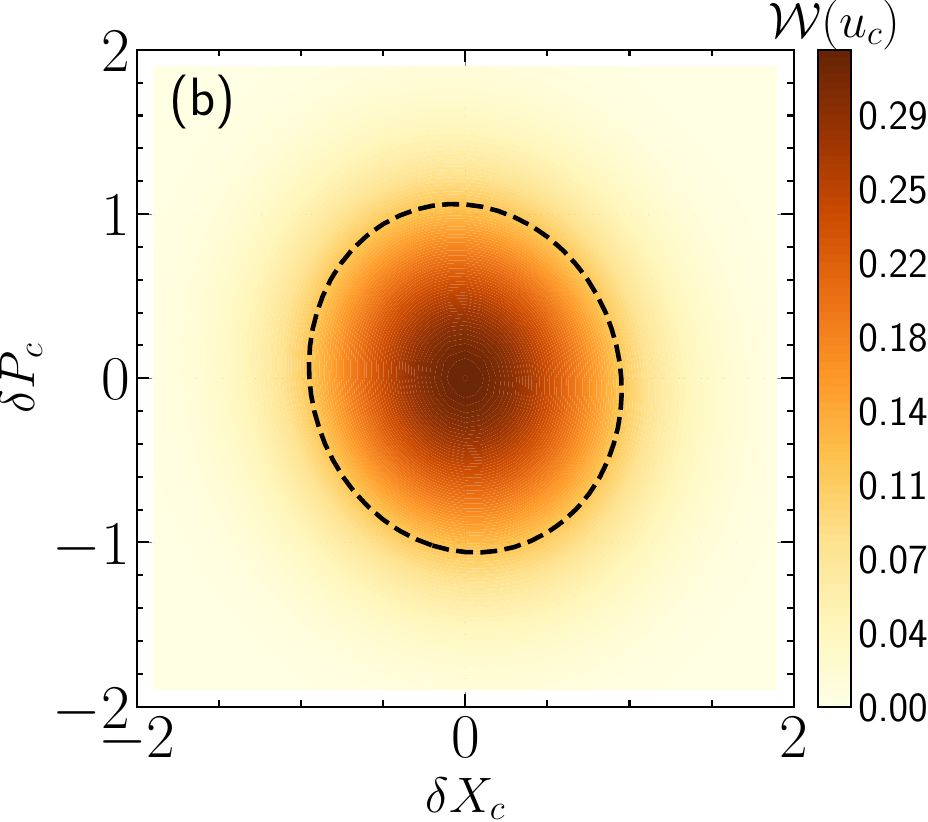}
  \end{minipage}

  \vspace{1em}

  \begin{minipage}{0.48\columnwidth}
    \centering
    \includegraphics[width=\linewidth]{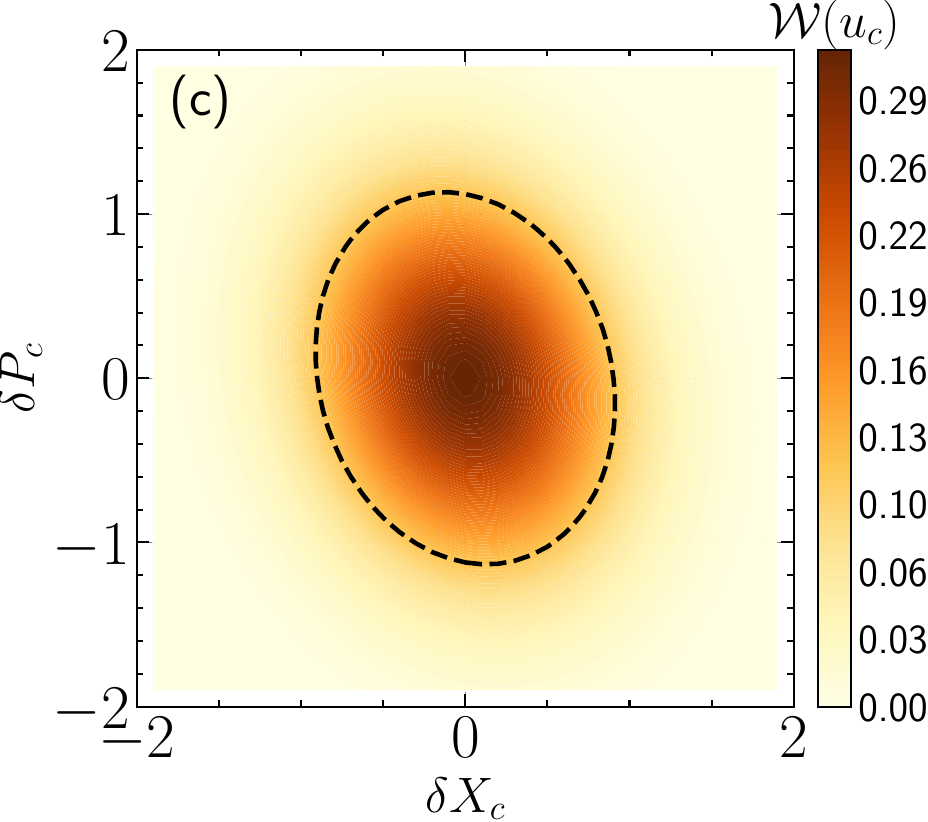}
  \end{minipage}\hfill
  \begin{minipage}{0.48\columnwidth}
    \centering
    \includegraphics[width=\linewidth]{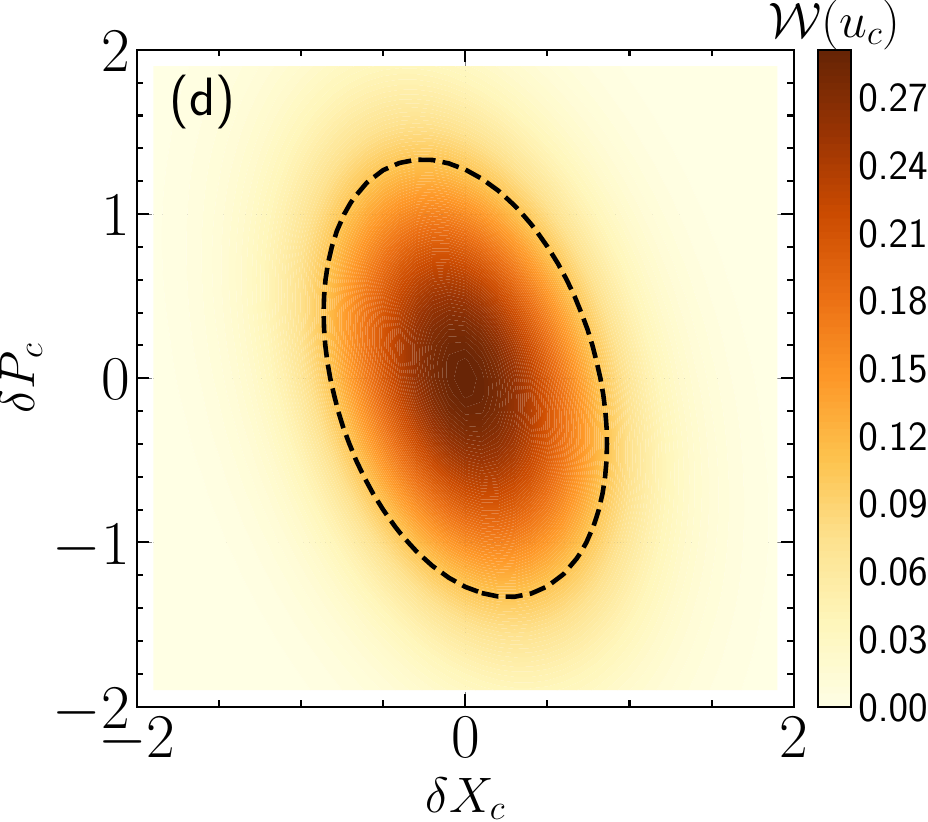}
  \end{minipage}

  \caption{Plot of the Wigner function $\mathcal{W}(u_{\rm c})$ of the cavity mode as a function of the phase c-number variables of the microwave mode $\delta {X}_{\rm c}$ and $\delta {P}_{\rm c}$ for different values of the gain $\lambda$: (a) $\lambda=0$, (b) $\lambda=0.5\gamma_{\rm c}$, (c) $\lambda=\gamma_{\rm c}$, and (d) $\lambda=2\gamma_{\rm c}$ with $\theta=3\pi/2$. The dashed ellipse (circle) marks $1/e$ of the maximum value of $\mathcal{W}(u_{\text{c}})$. The other parameters are the same as in Table~\ref{table}.}
  \label{W}
\end{figure}

We display in Fig. \ref{W} the Wigner function of the cavity mode for different values of the OPA gain $\lambda$. In the absence of the OPA ($\lambda = 0$), the Wigner distribution exhibits a symmetric Gaussian profile centered around the origin, which corresponds to a coherent Gaussian state, i.e., without squeezing. As the gain $\lambda$ increases, the Wigner function progressively becomes elongated along one axis and compressed along the orthogonal direction. The dashed ellipse (circle) marks $1/e$ of the maximum value of $\mathcal{W}(u_{\text{c}})$, i.e., the dashed ellipse and circle signify a reduction to $1/e$ of the maximum value of $\mathcal{W}(u_{\text{c}})$ for the steady state of the cavity mode. This deformation constitutes a direct signature of squeezing generated by the nonlinear parametric process. The squeezing becomes more pronounced for larger values of $\lambda$, as observed in panels (c) and (d), where the elliptical structure becomes increasingly narrow. The rotation of the ellipse additionally indicates the presence of phase-dependent correlations between the phase c-number variables. Importantly, the reduction of fluctuations along the phase c-number variables explains the enhancement of the $B_{\rm MI}$ by increasing $\lambda$ as observed in previous figures. Hence, Fig. \ref{W} clearly demonstrates the physical origin of the enhanced multiparameter quantum estimation.\\

\begin{figure}[!htb]
\minipage{0.5\textwidth}
\includegraphics[width=1\linewidth]{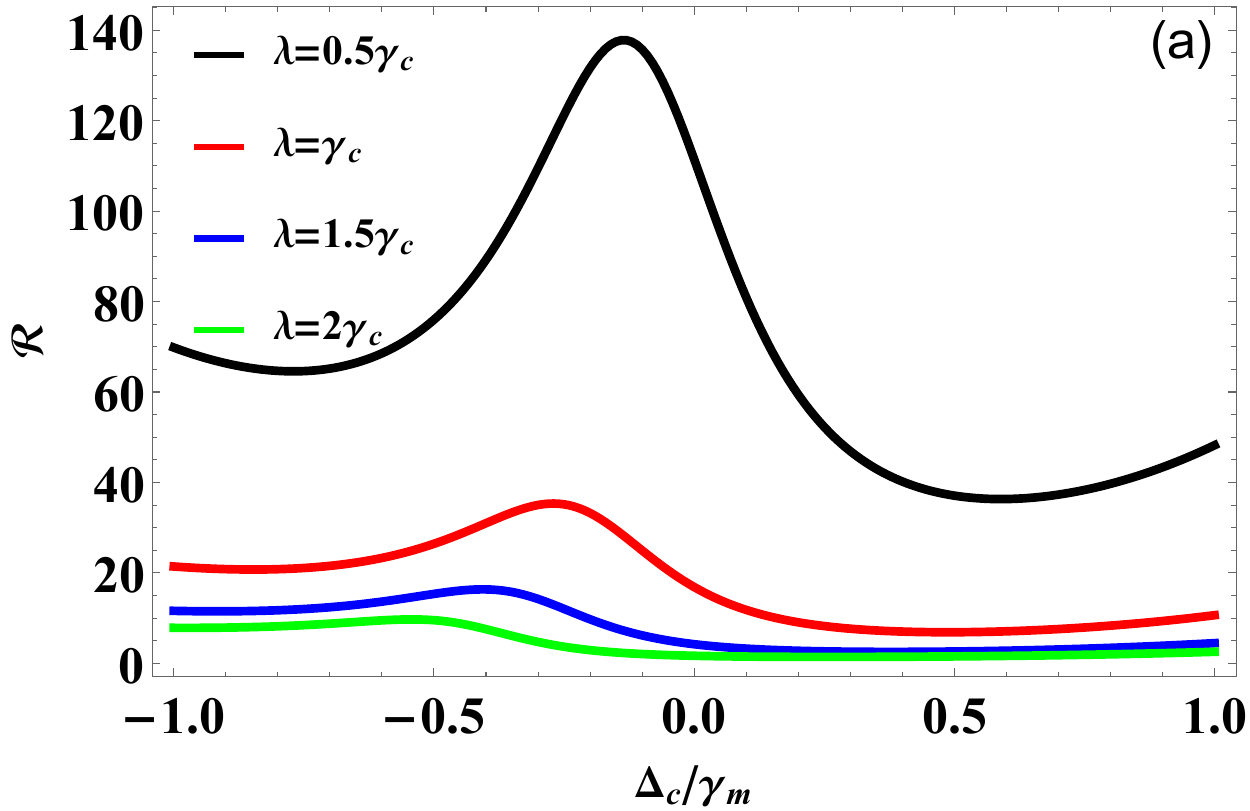}
\endminipage\hfill
\minipage{0.5\textwidth}
\includegraphics[width=1\linewidth]{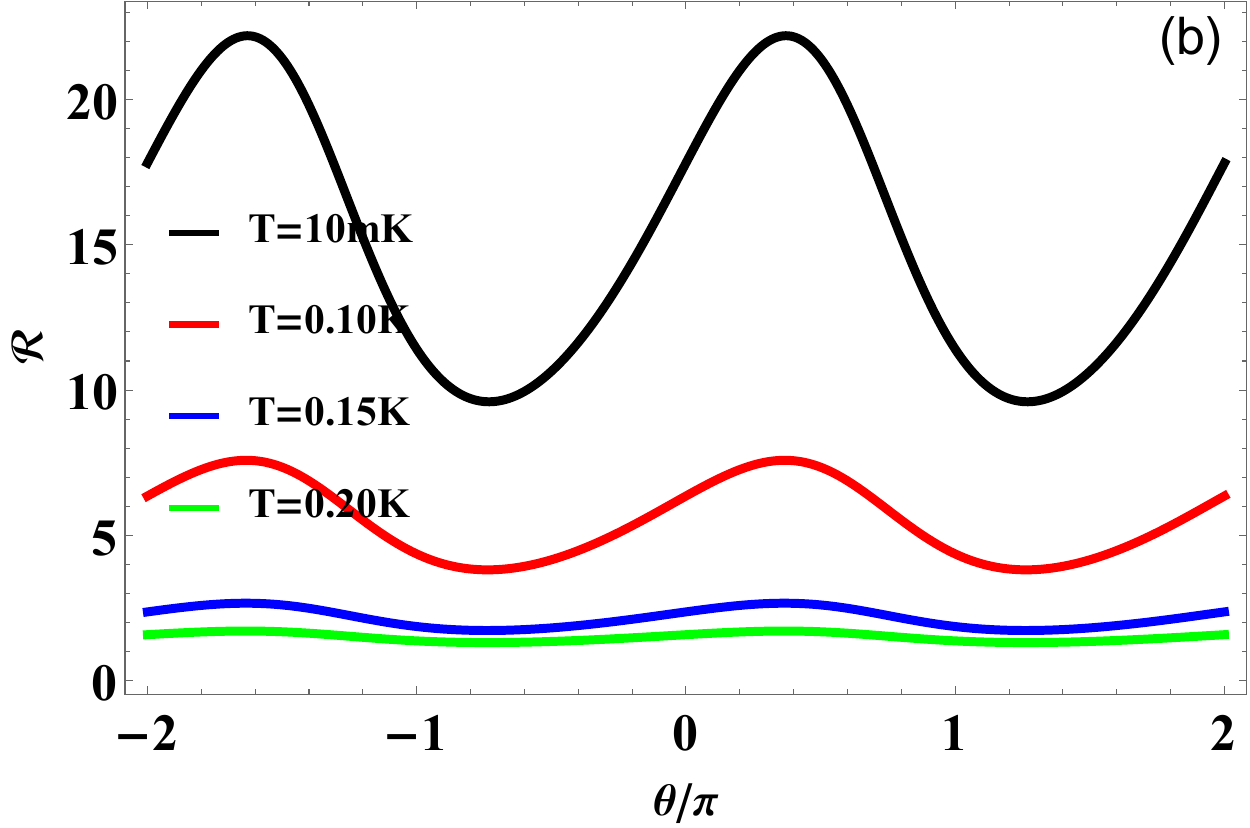}
\endminipage\hfill
\caption{Plot of the ratio $\mathcal{R}=B_S/B_R$ as a function of the: (a) the detuning $\Delta_{\rm c}/\gamma_{\rm m}$ for different value of the OPA gain $\lambda$ with $\theta=3\pi/2$; (b) the phase $\theta$ for various value of the temperature $\rm T$ with $\lambda =\gamma_{\rm c}$. The other parameters are the same as in the table \ref{table}.}
\label{Rg}
\end{figure}

Figure~\ref{Rg} presents the ratio $\mathcal{R}=B_S/B_R$, which quantifies the relative importance of the SLD and RLD quantum Cram\'er-Rao bounds in the multiparameter estimation protocol. We plot in Fig.~\ref{Rg}(a) the ratio $\mathcal{R}$ as a function of the normalized cavity detuning $\Delta_{\rm c}/\gamma_{\rm m}$ for different values of the OPA gain $\lambda$. One can see that $\mathcal{R}>1$ throughout the entire detuning range, implying that $B_S>B_R$. Consequently, the $B_{\rm MI}$ is always determined by the SLD bound in the present system. The ratio exhibits a pronounced maximum around $\Delta_{\rm c}/\gamma_{\rm m}\approx-0.2$, where $\mathcal{R}$ reaches values exceeding $10^2$ for $\lambda=0.5\gamma_{\rm c}$. As the gain $\lambda$ increases, the ratio decreases significantly and the peak becomes less pronounced. This behavior indicates that the OPA reduces the difference between the SLD and RLD bounds and therefore improves the compatibility of the simultaneous estimation strategy. This means that the squeezed field generated by the OPA suppresses part of the quantum fluctuations responsible for the discrepancy between the two bounds. Comparing Fig.~\ref{Rg}(a) with Fig.~\ref{STSAB4}(a), the largest values of $\mathcal{R}$ occur near the detuning region where the $B_{\rm MI}$ is also maximal, showing that the deterioration of the estimation precision is accompanied by a stronger separation between the SLD and RLD limits.

Figure~\ref{Rg}(b) shows the dependence of $\mathcal{R}$ on the squeezing phase $\theta$ for different temperatures $\rm T$. The ratio $\mathcal{R}$ displays a clear periodic behavior, demonstrating the phase-sensitive nature of the OPA. For all considered temperatures, $\mathcal{R}$ remains larger than unity, confirming that the SLD bound continues to determine the $B_{\rm MI}$. The oscillation amplitude decreases markedly as the temperature increases. In particular, the maximum value of $\mathcal{R}$ decreases from approximately $22$ at $\rm T=10\,{\rm mK}$ to values close to unity at $\rm T=0.20\,{\rm K}$. This behavior indicates that thermal fluctuations reduce the distinction between the SLD and RLD bounds by progressively destroying the phase-sensitive quantum correlations generated by the OPA. The minima of $\mathcal{R}$ correspond to regions where the two bounds become closer, whereas the maxima indicate stronger incompatibility between the optimal measurements associated with the estimated parameters. Comparing Fig.~\ref{Rg}(b) with Fig.~\ref{STSAB3}(b), it is evident that the periodic modulation of the $B_{\rm MI}$ and the ratio $\mathcal{R}$ originates from the same phase-sensitive squeezing mechanism generated by the OPA. While the squeezing phase strongly modifies the estimation precision, the condition $\mathcal{R}>1$ is preserved for all values of $\theta$, implying that the SLD bound remains the dominant contribution to the $B_{\rm MI}$. Consequently, the improvement of the multiparameter estimation precision achieved at the optimal phases is associated with an increase in the SLD quantum Fisher information, while the fundamental estimation regime remains unchanged.

\begin{figure}[!htb]
\minipage{0.5\textwidth}
\includegraphics[width=1\linewidth]{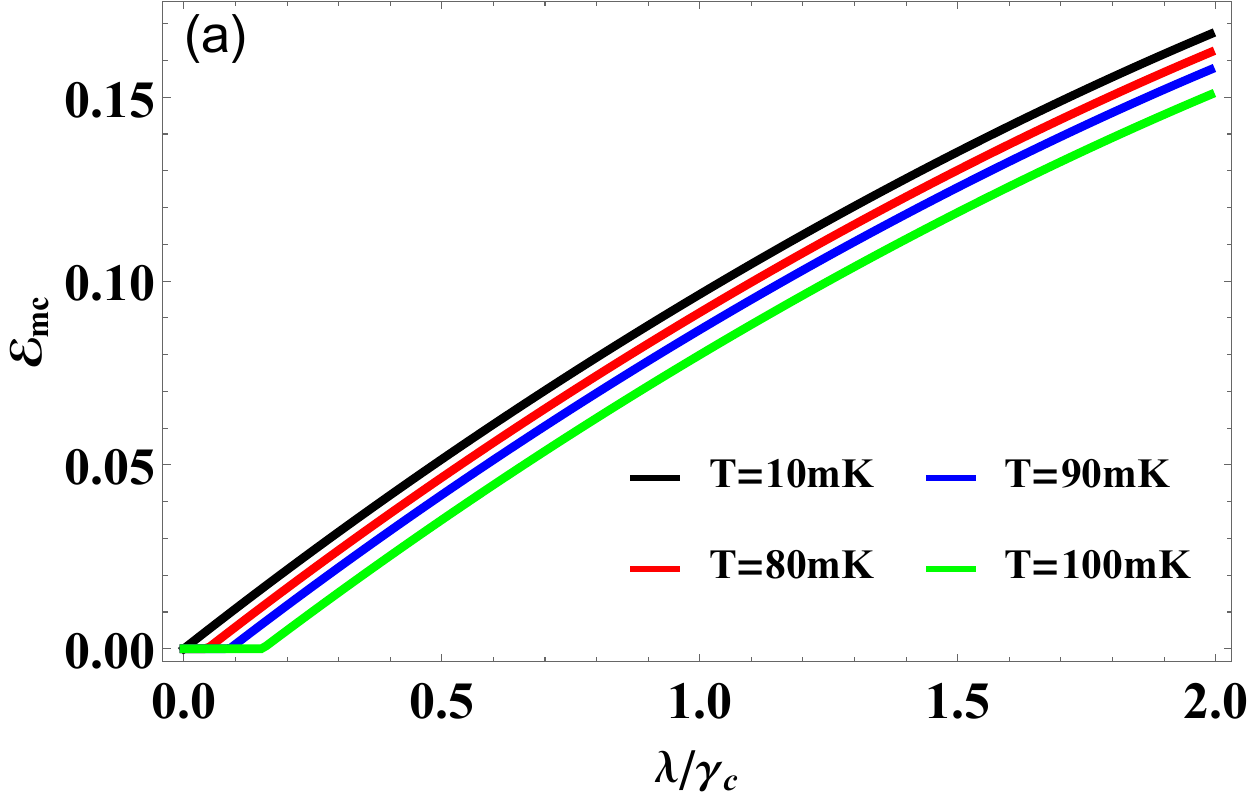}
\endminipage\hfill
\caption{Plot of the logarithmic negativity $\mathcal{E}_{\rm mc}$ versus the normalized OPA gain $\lambda/\gamma_{\rm c}$ for various values of the temperature $\rm T$ with $\theta=3\pi/2$. The other parameters are the same as in the table \ref{table}.}
\label{EN}
\end{figure}

Figure \ref{EN}(a) presents the logarithmic negativity $\mathcal{E}_{\rm mc}$ versus the normalized OPA gain $\lambda/\gamma_{\rm c}$ various values of the temperature $\rm T$. We remark that the entanglement between the cavity and magnon modes increases monotonically as the OPA gain $\lambda$ becomes larger. This means that the parametric amplification process enhances the generation of correlated photon pairs inside the cavity, which strengthens the quantum correlations transferred to the magnon subsystem through the cavity-magnon interaction. Besides, the hybrid cavity-magnon state becomes more entangled for stronger nonlinear gain. Moreover, the temperature strongly affects the amount of generated entanglement, i.e., increasing the temperature progressively reduces the logarithmic negativity over the entire range of $\lambda/\gamma_{\rm c}$, as depicted in Fig.~\ref{EN}(a). This behavior originates from thermal effects (decoherence phenomenon), which destroy quantum correlations between photon and magnon. Nevertheless, even at relatively high temperatures, the entanglement remains nonzero for sufficiently large values of the gain $\lambda$, demonstrating the robustness of the quantum correlations against thermal noises. Interestingly, the almost linear increase of $\mathcal{E}_{\rm mc}$ with $\lambda$ indicates that the nonlinear parametric interaction provides an efficient mechanism for continuously enhancing $\mathcal{E}_{\rm mc}$. Since quantum entanglement is directly connected to the sensitivity of Gaussian quantum states, the enhancement of $\mathcal{E}_{\rm mc}$ explains the reduction of the $B_{\rm MI}$ observed in previous figures. Therefore, Fig.~\ref{EN}(a) establishes a direct connection between OPA-enhanced entanglement generation and the improvement of multiparameter quantum estimation precision in the cavity-magnon system.

\subsection{Dynamical state}%

\begin{figure}[!htb]
\minipage{0.5\textwidth}
\includegraphics[width=1\linewidth]{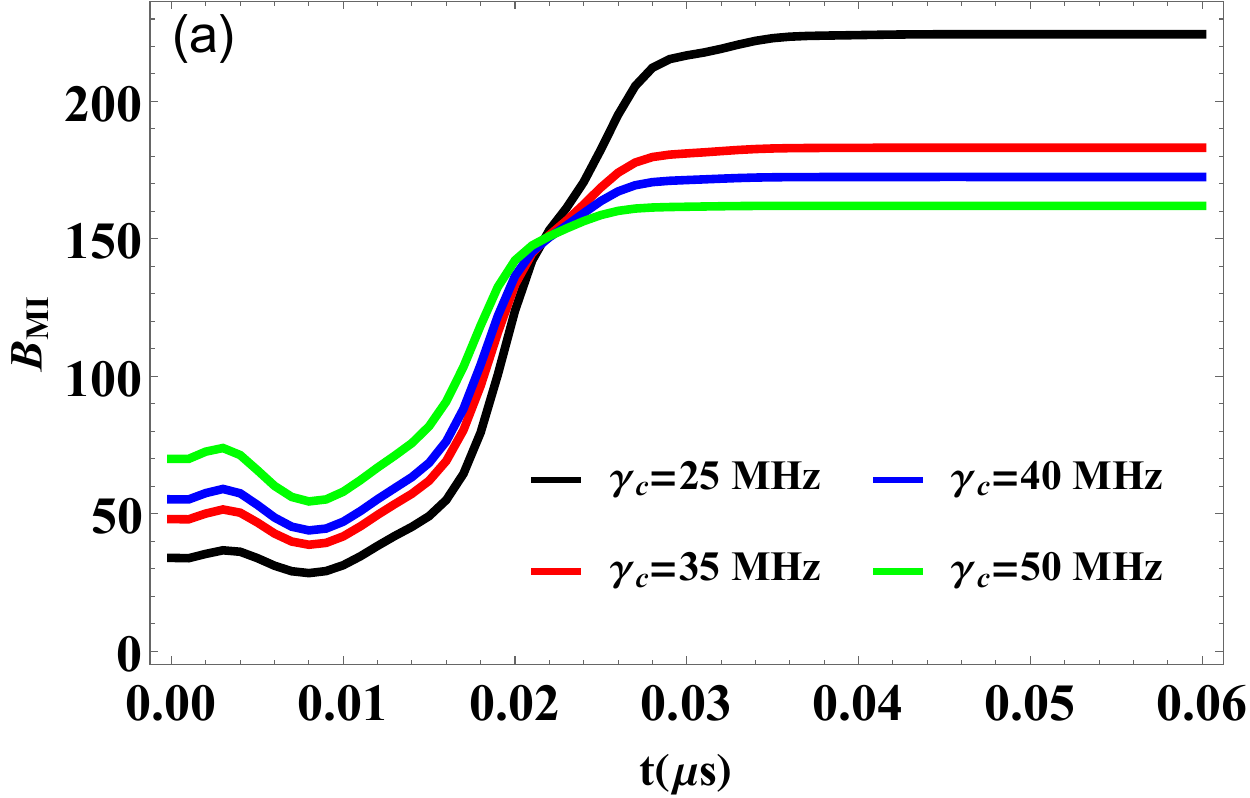}
\endminipage\hfill
\minipage{0.5\textwidth}
\includegraphics[width=1\linewidth]{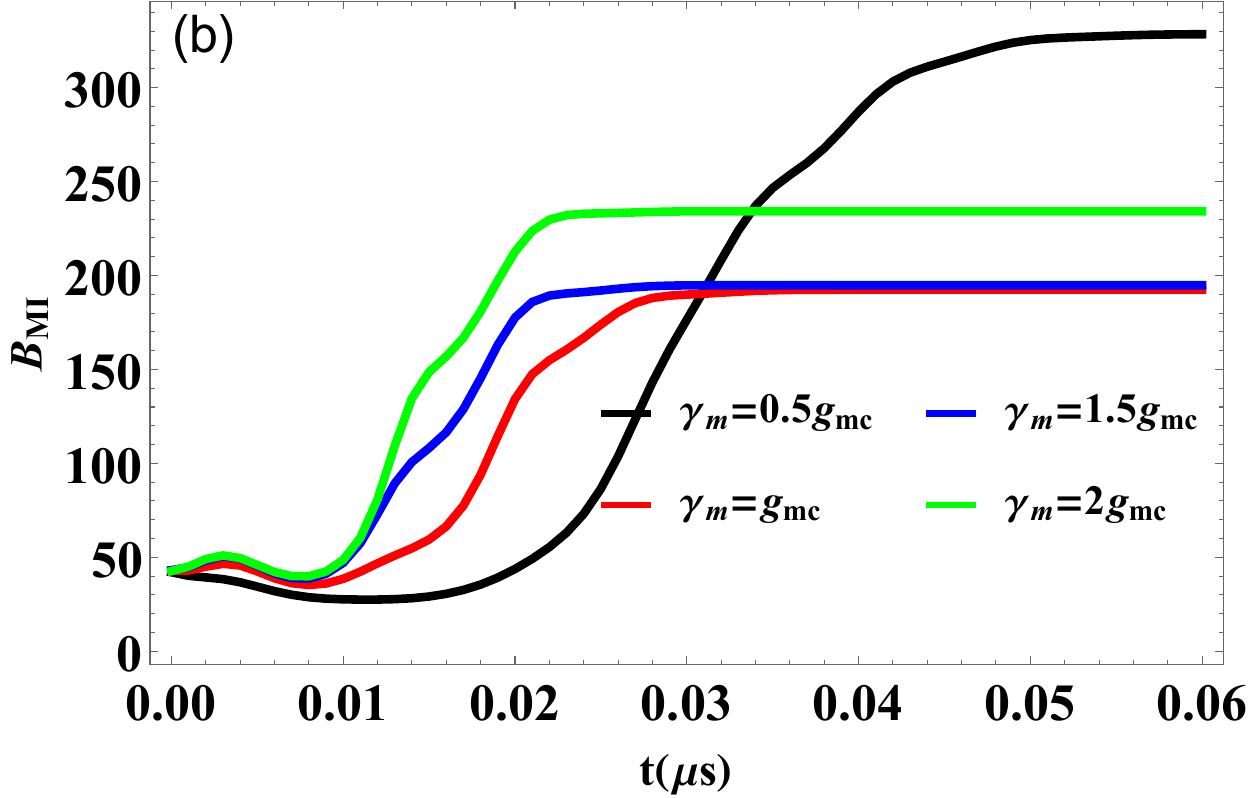}
\endminipage\hfill
\caption{Plot of the dynamical evolution of the $B_{\rm MI}$ for various value of (a) $\gamma_{\rm c}$ and (b) $\gamma_{\rm m}$ for $\rm g_{\rm mc}=2\pi\times 41$ MHz with $\lambda =10\pi$ MHz and $\theta=3\pi/2$. The other parameters are the same as in the table \ref{table}.}
\label{STSAB5}
\end{figure}

We present in Fig.~\ref{STSAB5}(a) the dynamical evolution of the multiparameter quantum Cram\'er--Rao bound $B_{\rm MI}$ for different values of the cavity decay rate $\gamma_{\rm c}$. At short times ($t \lesssim 0.01~\mu{\rm s}$), the system exhibits a transient regime where $B_{\rm MI}$ remains relatively small, indicating that the quantum state still retains a high degree of coherence and hence a large amount of accessible quantum information for parameter estimation. As time increases, $B_{\rm MI}$ grows and eventually saturates to a steady-state value, reflecting the relaxation of the system toward its open-system equilibrium under the combined effects of dissipation and continuous quantum noise.
It is observed that increasing the cavity decay rate $\gamma_{\rm c}$ enhances $B_{\rm MI}$ during the transient regime and modifies its approach to the steady state. Physically, stronger cavity losses reduce the photon lifetime inside the resonator, thereby weakening the build-up of intracavity squeezing and suppressing the OPA that are essential for enhancing the quantum Fisher information. As a result, the distinguishability between neighboring quantum states is reduced, which leads to a degradation of the multiparameter estimation precision. In contrast, smaller values of $\gamma_{\rm c}$ allow the system to maintain coherent intracavity dynamics for a longer time, thereby improving the estimation performance, particularly in the early-time regime ($t < 0.02~\mu{\rm s}$).
Figure~\ref{STSAB5}(b) shows the time evolution of $B_{\rm MI}$ for different magnon damping rates $\gamma_{\rm m}$. Similarly, $B_{\rm MI}$ increases from its initial value and approaches a stationary regime after a finite evolution time, consistent with the establishment of a dissipative steady state. However, the steady-state value of $B_{\rm MI}$ increases with increasing $\gamma_{\rm m}$, indicating a reduction in the achievable estimation precision. From a physical perspective, magnon damping suppresses the coherent exchange of excitations between the cavity and magnon subsystems, thereby weakening the hybrid quantum correlations that encode information about the parameters. This leads to a reduction of the quantum Fisher information and consequently to larger values of $B_{\rm MI}$.
Comparing Figs.~\ref{STSAB5}(a) and \ref{STSAB5}(b), it is evident that both cavity and magnon dissipation play a detrimental role in the multiparameter estimation protocol. In both cases, dissipation channels act as sources of decoherence that degrade quantum correlations and limit the amount of extractable information. Nevertheless, despite these losses, the system eventually converges to a well-defined steady state in which the competition between coherent dynamics, OPA-induced squeezing, and dissipation determines the ultimate precision bound.

\begin{figure}[!htb]
\minipage{0.5\textwidth}
\includegraphics[width=1\linewidth]{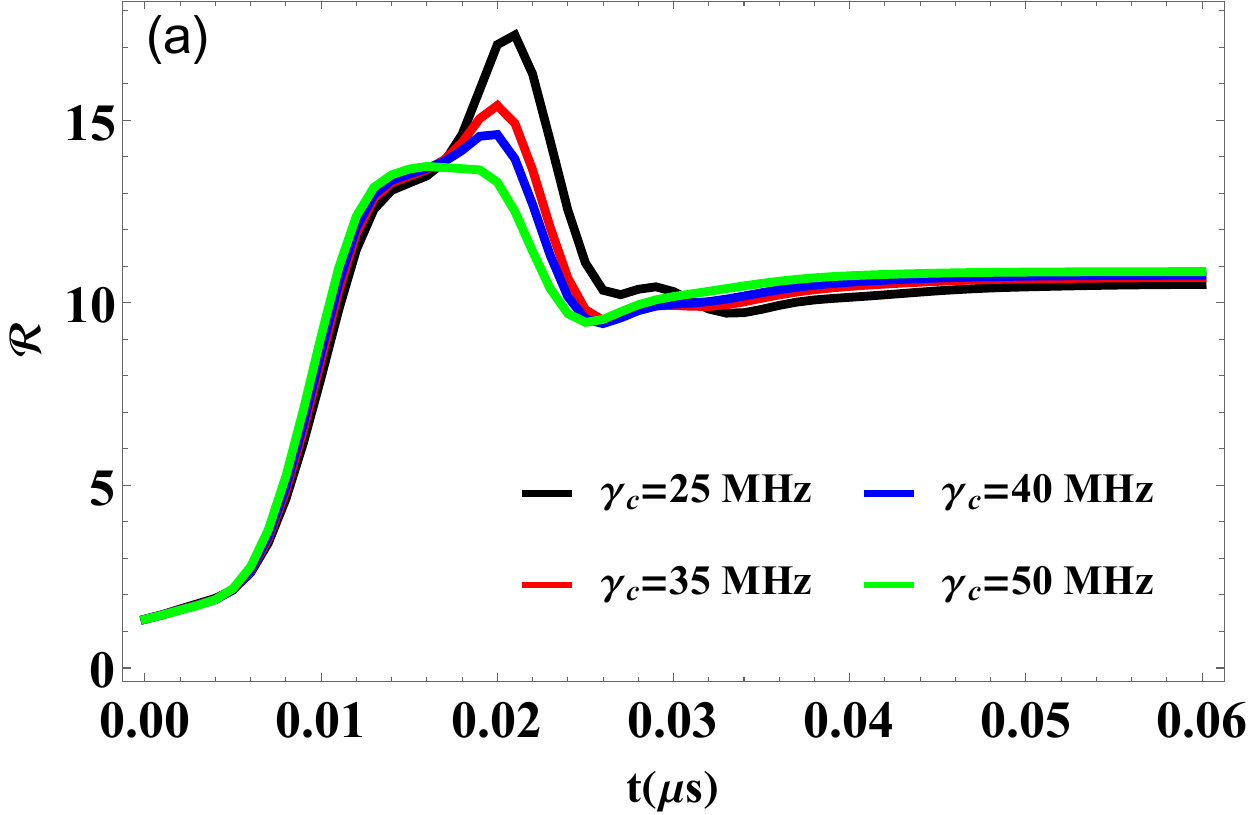}
\endminipage\hfill
\minipage{0.5\textwidth}
\includegraphics[width=1\linewidth]{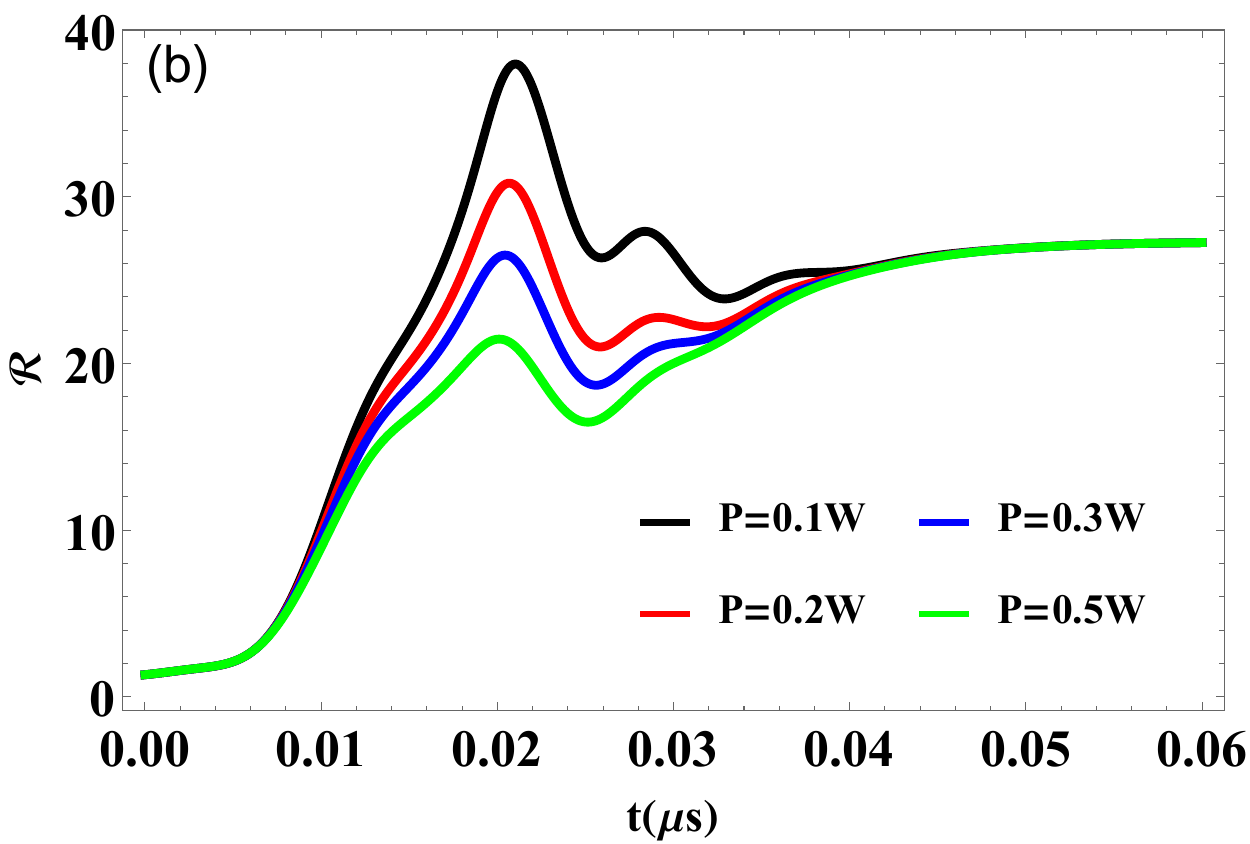}
\endminipage\hfill
\caption{Plot of the dynamical evolution of the ratio $\mathcal{R}=B_S/B_R$ for various value of: (a) the cavity damping rate $\gamma_{\rm c}$ and (b) the power $\rm P$ with $\lambda =10\pi$ MHz and $\theta=3\pi/2$. The other parameters are the same as in the table \ref{table}.}
\label{Rd}
\end{figure}

We plot in Fig.~\ref{Rd} dynamical evolution of the ratio $\mathcal{R}$ for different cavity decay rates $\gamma_c$ and driving powers $\rm P$. The ratio evolves nontrivially over time, indicating that the relative performance of the SLD and RLD bounds changes during the transient dynamics of the system. In Fig. \ref{Rd}(a), increasing the cavity decay rate modifies both the amplitude and saturation value of the $\mathcal{R}$. At short times, the ratio $\mathcal{R}$ remains relatively small because coherent quantum correlations dominate the dynamics. However, as dissipation increases, the ratio becomes larger ($\mathcal{R}\gg 1$), suggesting that incompatibility effects between estimated parameters become more significant in dissipative regimes. This means that the cavity losses introduce additional noise channels that modify the structure of the quantum correlations encoded in the covariance matrix $\mathcal{C}$.\\
In Fig. \ref{Rd}(b), increasing the driving power strongly enhances the ratio $\mathcal{R}$. Stronger driving increases the intracavity photon population and amplifies the nonlinear squeezing process generated by the OPA. Consequently, the difference between the SLD and RLD bounds becomes more pronounced. The oscillation-free saturation toward stationary values indicates that the system ultimately reaches a steady state. These results demonstrate that both dissipation and driving strength critically influence the attainability of optimal multiparameter quantum estimation bounds.

\begin{figure}[!htb]
\minipage{0.5\textwidth}
\includegraphics[width=1\linewidth]{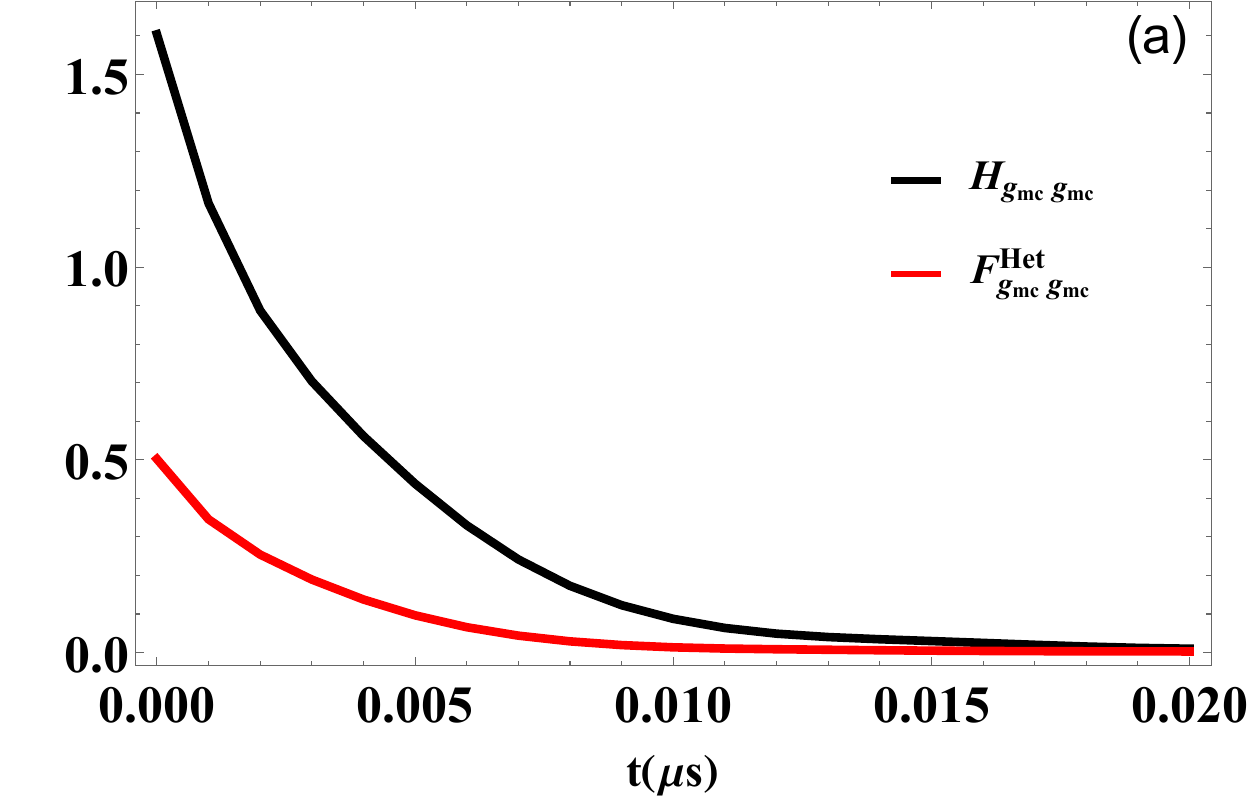}
\endminipage\hfill
\minipage{0.5\textwidth}
\includegraphics[width=1\linewidth]{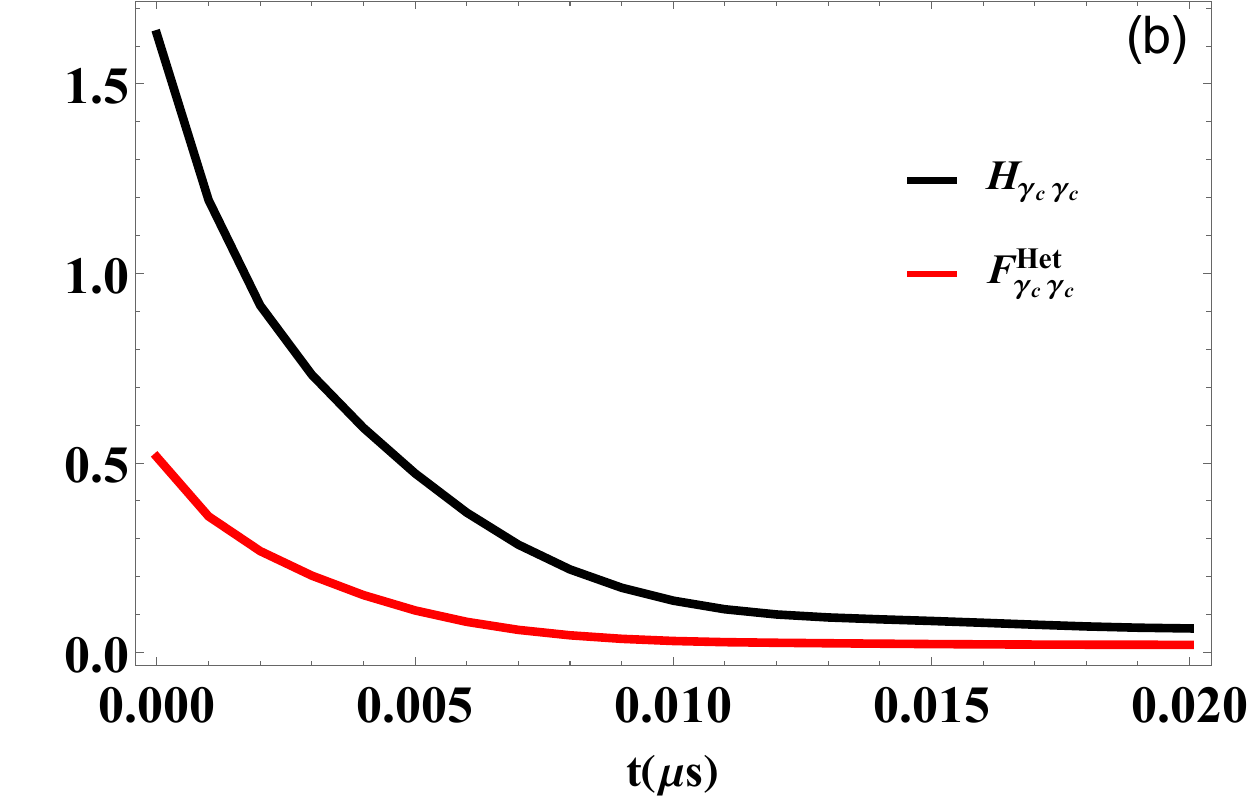}
\endminipage\hfill
\caption{Dynamical evolution of the SLD QFI (Black curve) and the CFI in heterodyne detection (Red curve), in (a) $\rm \textbf{H}_{ g_{mc}g_{mc}}$ and $F_{\rm g_{mc}g_{mc}}^{\rm Het}$; in (b) $\rm \textbf{H}_{\gamma_{\rm c}\gamma_{\rm c}}$ and $F_{\gamma_{\rm c}\gamma_{\rm c}}^{\rm Het}$, with $\lambda =\gamma_{\rm c}$ and $\theta=3\pi/2$. The other parameters are the same as in the table \ref{table}.}
\label{HFd}
\end{figure}

To assess the sensitivity of $\rm g_{mc}$ and $\gamma_{\rm c}$ over time, we compare the ultimate quantum estimation (QFI) with the classical information (CFI) accessible via heterodyne detection. One can see that as the beginning of the evolution, the SLD QFI is significantly larger than the CFI, indicating that the quantum state contains considerably more information than what can be accessed through heterodyne detection, as depicted in \ref{HFd}(a)-(b). Moreover, it is evident that as time progresses, both the SLD QFI and the CFI decrease. This indicates a degradation of the quantum estimation precision for ${\rm g}_{\rm mc}$ and $\gamma_{\rm c}$. Furthermore, the decline in the CFI reflects a corresponding loss of estimation accuracy for heterodyne detection. For times exceeding $t > 0.02\,\mu\text{s}$, the QFI $\rm H_{g_{mc} g_{mc}}$ and CFI $F_{g_{\rm mc}g_{\rm mc}}^{\rm Het}$ become equal and converge to zero, as shown in \ref{HFd}(a). This indicates that no quantum advantage remains and that precise parameter estimation becomes effectively impossible in this region. However, in Fig. \ref{HFd}(b), quantum estimation precision for $\gamma_{\rm c}$ still exists, even if it is very small compared to its initial value of $\rm \textbf{H}_{\gamma_{\rm c}\gamma_{\rm c}} \approx 1.7$ at $t \approx 0$ s; eventually, $\rm \textbf{H}_{\gamma_{\rm c}\gamma_{\rm c}}$ reaches a steady state of approximately $0.1$. Apparently, the classical information $F_{\gamma_{\rm c}\gamma_{\rm c}}^{\rm Het}$ is very small with respect to $\rm \textbf{H}_{\gamma_{\rm c}\gamma_{\rm c}}$ when $t > 0.01\,\mu\text{s}$.%

\begin{figure}[!htb]
\minipage{0.5\textwidth}
\includegraphics[width=1\linewidth]{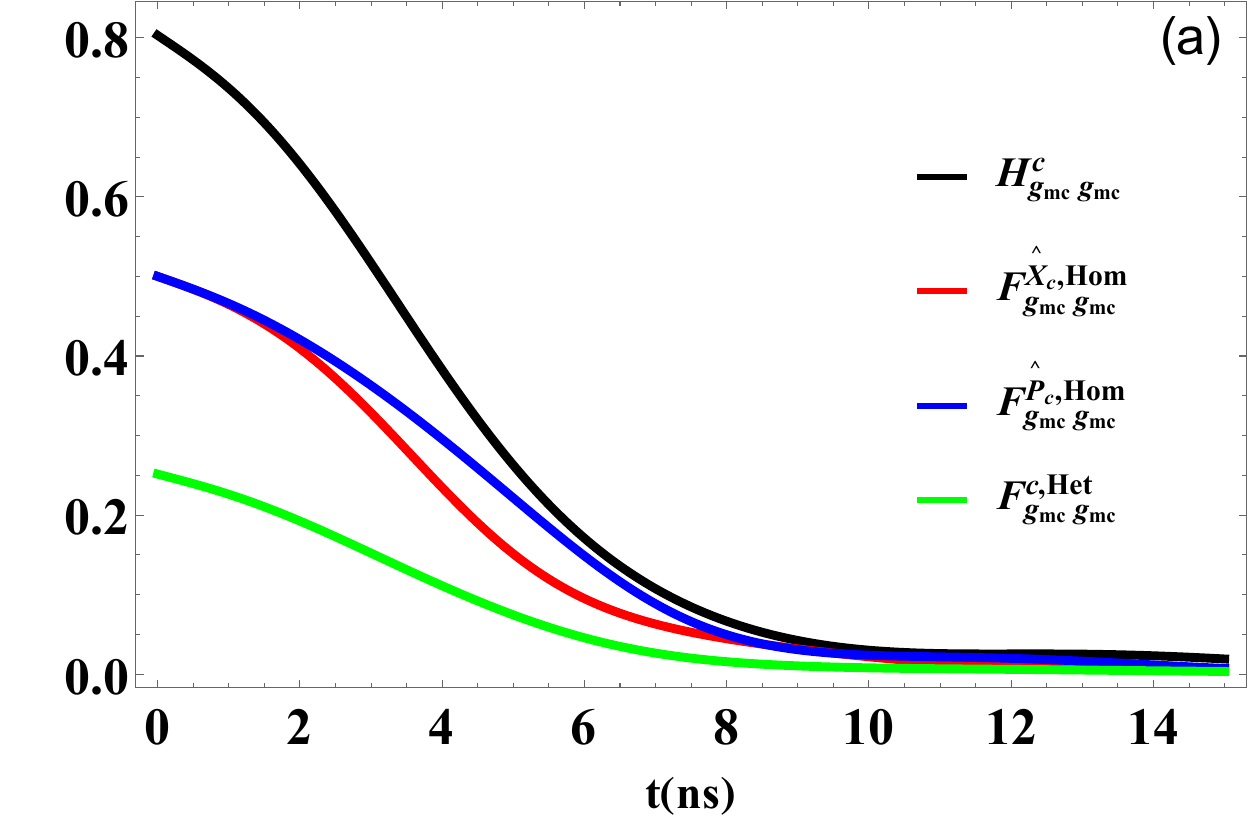}
\endminipage\hfill
\minipage{0.5\textwidth}
\includegraphics[width=1\linewidth]{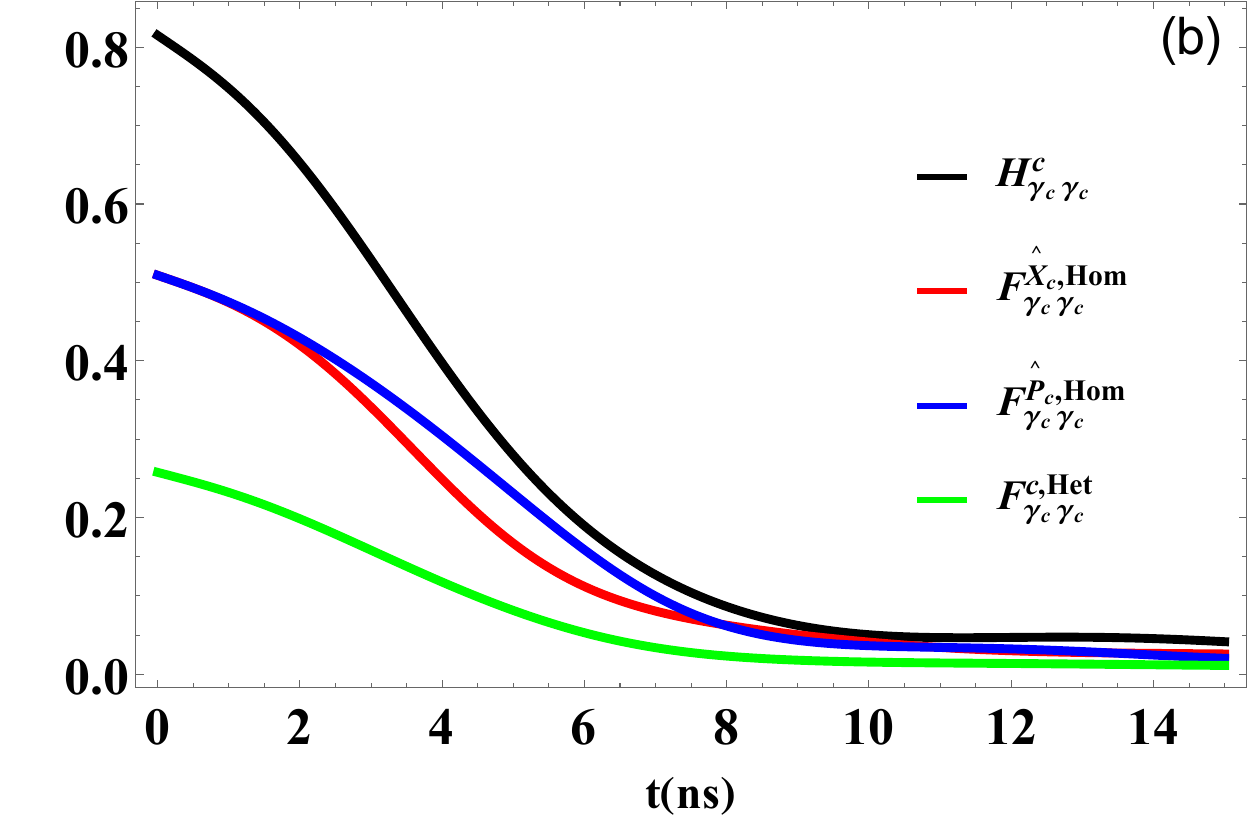}
\endminipage\hfill
\caption{Time evolution of the SLD QFI for the cavity mode ($\rm c$) $\rm \textbf{H}^{\rm c}_{u}$ and the CFI in homodyne detection $F^{\hat{X}_{\rm c},{\rm Hom}}_{u}$ and $F^{\hat{P}_{\rm c},{\rm Hom}}_{u}$, for quadrature operators $\hat{X}_{\rm c}$ and $\hat{P}_{\rm c}$, and in heterodyne detection $F^{\rm c}_{u}$, with $u={\rm g_{mc}\rm g_{mc}}$ in (a) and $u=\gamma_{\rm c}\gamma_{\rm c}$ in (b), using $\lambda =\gamma_{\rm c}$ and $\theta=3\pi/2$. The other parameters are the same as in the table \ref{table}.}
\label{STSAB7}
\end{figure}

Let us now compare the SLD QFI and the CFI for both homodyne and heterodyne detection of the cavity mode ($\rm c$). As seen in Figs. \ref{STSAB7}(a) and (b), the CFI is bounded by the SLD QFI, indicating that the QFI provides higher precision than the CFI. Moreover, as shown in these figures, the QFI decreases monotonically over time, indicating that the quantum precision for estimating $\rm g_{\rm mc}$ and $\gamma_{\rm c}$ degrades. We observe that for approximately $t \in [0, 2\,\text{ns}]$, the two homodyne CFIs are equal, satisfying $F^{\hat{X}_{\rm c},\mathrm{Hom}}_{\rm g_{\rm mc}g_{\rm mc}} = F^{\hat{P}_{\rm c},\mathrm{Hom}}_{\rm g_{\rm mc}g_{\rm mc}}$. In the intermediate regime  ($2\,\text{ns} < t < 8\,\text{ns}$), the hierarchy $F^{\hat{P}_{\rm c},\mathrm{Hom}}_{\rm g_{\rm mc}g_{\rm mc}} > F^{\hat{X}_{\rm c},\mathrm{Hom}}_{g_{\rm mc}g_{\rm mc}}$ emerges, showing that the $\hat{P}_{\rm c}$-quadrature becomes more informative than the $\hat{X}_{\rm c}$-quadrature. Furthermore, for sufficiently long times ($t > 12\,\text{ns}$), both the SLD QFI and the CFI decay to zero, indicating a complete loss of information regarding the parameter $g_{\rm mc}$, as depicted in Fig. \ref{STSAB7}(a).

A similar behavior is observed for the SLD QFI and CFI in the estimation of the cavity decay rate $\gamma_{\rm c}$, as shown in Fig. \ref{STSAB7}(b). However, at long times, only the heterodyne CFI $F^{{\rm c},\mathrm{Het}}_{\gamma_{\rm c}\gamma_{\rm c}}$ vanishes completely, whereas the homodyne CFIs retain small but nonzero values; this highlights the lower sensitivity of heterodyne detection to cavity dissipation in the long-time limit. Furthermore, comparing Figs.~\ref{STSAB7}(a) and \ref{STSAB7}(b), one finds that
$F^{\rm c,Het}_{\rm g_{\rm mc}g_{\rm mc}} = F^{\rm c,Het}_{\gamma_{\rm c}\gamma_{\rm c}}$
for $t\in[0,2]\,\mathrm{ns}$, indicating equal estimation precision in the transient regime. For $t>2\,\mathrm{ns}$, the inequality
$F^{\rm c,Het}_{\rm g_{\rm mc}g_{\rm mc}} < F^{\rm c,Het}_{\gamma_{\rm c}\gamma_{\rm c}}$
shows that the cavity decay rate $\gamma_{\rm c}$ can be estimated more precisely than the coupling $\rm g_{\rm mc}$. Besides, for $t>8\,\mathrm{ns}$, the SLD QFI and CFIs decrease to vanish, implying a complete loss of information and making reliable estimation of both parameters impossible in the long-time regime. Moreover, one can explain the gap between the SLD-QFI and the CFI quantifies the amount of information that remains inaccessible due to measurement constraints, as depicted in Figs. \ref{STSAB7}(a-b). These results therefore establish a direct connection between the fundamental quantum precision limits and experimentally realizable detection protocols, showing that homodyne measurements provide the most suitable route for approaching optimal multiparameter estimation in cavity-magnonic system.

\section{\label{sec6}FEASIBILITY}
\begin{figure*}[!htb]
  \centering
  \includegraphics[width=15cm]{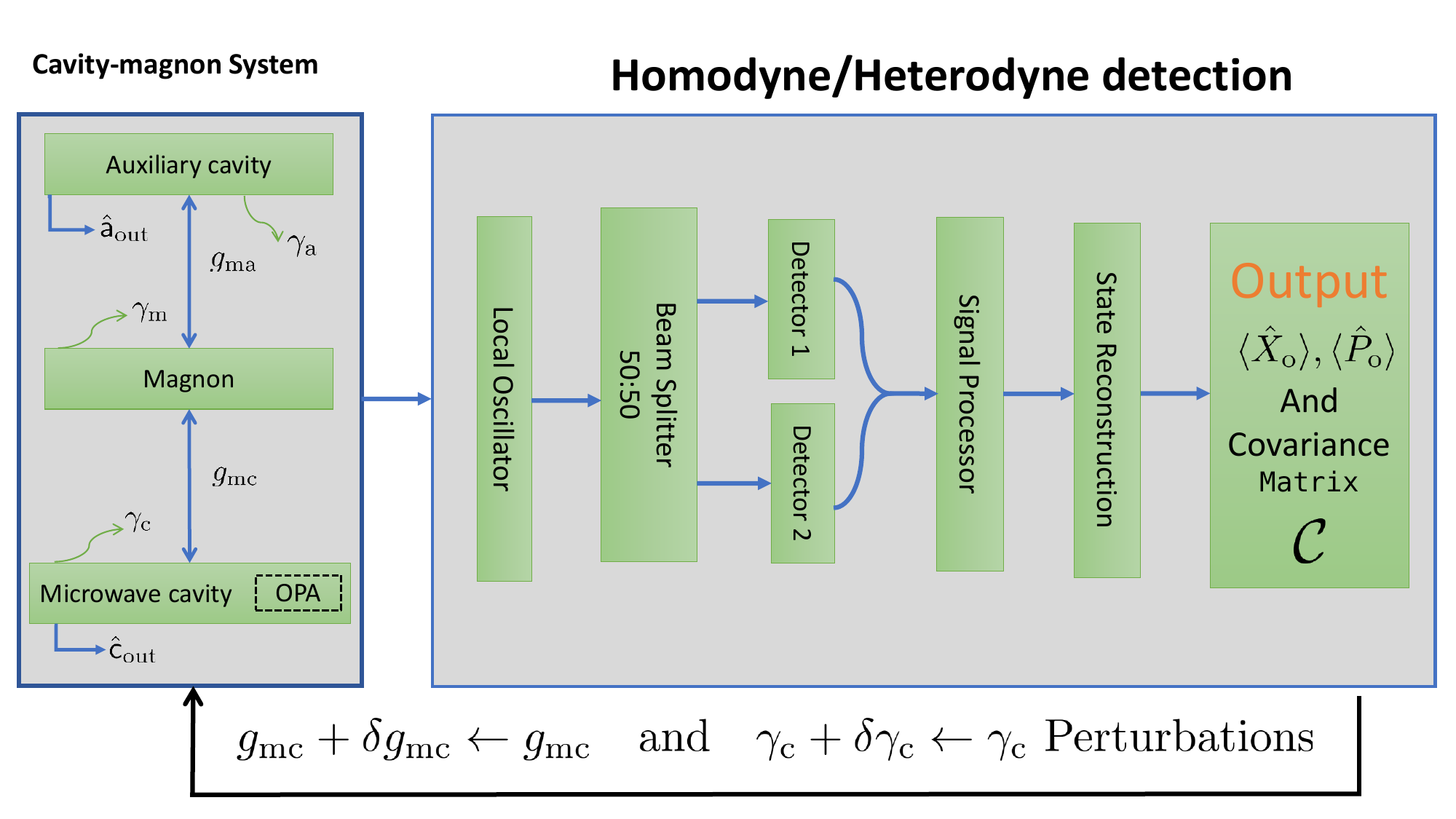}
  \caption{Schematic diagram illustrating the reconstruction of the first and second order moments of a Gaussian state via homodyne and heterodyne detection, where $g_{\rm ma}$ represents the auxiliary cavity-magnon coupling. The reconstruction is based on the input-output relations $\hat{\rm c}_{\rm out} = \hat{\rm c}_{\rm in} - \sqrt{2\gamma_{\rm c}}\,\hat{\rm c}$ and $\hat{\rm a}_{\rm out} = \hat{\rm a}_{\rm in} - \sqrt{2\gamma_{\rm a}}\,\hat{\rm a}$, where $\hat{\rm a}_{\rm in}$ and $\gamma_{\rm a}$ denote the input field and the decay rate of the auxiliary cavity, respectively.}
  \label{Fig8}
\end{figure*}
The experimental reconstruction of the first-order moments and the covariance matrix of a Gaussian state can be achieved using homodyne and heterodyne detection-techniques that are well-established in continuous-variable quantum systems \cite{Vitali07,hmz:10,Peng2025}. Since direct experimental access to the magnon mode is relatively challenging, a YIG sphere supporting magnon excitations is coupled to a microwave cavity with a coupling strength ${\rm g}_{\rm mc}$ and dissipation rate $\gamma_{\rm c}$. This cavity incorporates a degenerate optical parametric amplifier (OPA) that acts as a phase-sensitive pre-amplifier to enhance the intracavity field via parametric amplification. This process strengthens the imprint of magnon-induced fluctuations on the cavity output field, thereby improving measurement sensitivity. Furthermore, the magnon mode is coupled to an auxiliary readout cavity with a coupling strength ${\rm g}_{\rm ma}$ and dissipation rate $\gamma_{\rm a}$, driven by a weak red-detuned laser field. The output field, carrying the magnon quadratures, is then amplified in a phase-sensitive manner, effectively serving as a pre-amplifier for the readout.%

Measurement of the cavity output quadratures yields stationary time traces used to reconstruct steady-state probability distributions. From these, the first-order moments and the covariance matrix are extracted \cite{aspelmeyer2014}. The covariance matrix is determined by computing the time-averaged products of the recorded quadrature components. This procedure utilizes input-output theory and spectral filtering \cite{aspelmeyer2014}, employing a causal filter function of the form
\begin{equation}
\vartheta(t)=\frac{\varrho(t)-\varrho(t-h)}{h^{1/2}}\exp(-i\omega t),
\end{equation}
where $\varrho(t)$ is the Heaviside step function, while $h$ and $\omega$ denote the filter bandwidth and central frequency, respectively \cite{yan2019}. The inclusion of the auxiliary OPA significantly enhances the feasibility of the detection scheme by amplifying the relevant quadratures prior to measurement, thereby compensating for finite detector efficiency and technical noise. This enhancement is particularly effective for homodyne detection \cite{saheked2018,racz2023,frascella2021}, as it measures a single quadrature aligned with the OPA-amplified direction, enabling near quantum-limited sensitivity. 

In heterodyne detection, which simultaneously measures both conjugate quadratures \cite{feng2016}, the amplified quadrature benefits from the OPA; however, the conjugate quadrature remains subject to an intrinsic vacuum-noise contribution, limiting the overall improvement. Furthermore, the OPA increases the signal strength and improves robustness against losses, rendering both homodyne and heterodyne detection schemes experimentally viable. The reconstructed first-order moments and covariance matrix fully characterize the Gaussian state, providing all the information required to determine the QFI experimentally. Finally, by implementing the OPA-assisted homodyne and heterodyne detection schemes illustrated in Fig.~\ref{Fig8} and repeating the measurement procedure across various system parameters, the first-order moments and the full covariance matrix of the Gaussian state can be reconstructed experimentally. A quantitative assessment of the proposed metrological scheme is then achieved by comparing the classical estimation precisions for homodyne and heterodyne detection with the quantum estimation precision. Consequently, our OPA-assisted cavity–magnon quantum metrology protocol is experimentally feasible and can be realized using current state-of-the-art technology \cite{weedbrook2012}.

\section{\label{sec7} CONCLUSION}

In summary, we have proposed a theoretical scheme to enhance multiparameter quantum estimation in a feasible cavity-magnon system by incorporating a degenerate optical parametric amplifier (OPA). We have utilized the most informative quantum Cram\'er-Rao bound to evaluate the estimation performance, employing both the RLD and SLD operators. Besides, we have shown that the presence of the optical parametric amplifier significantly enhances estimation precision by reducing the $B_{\rm MI}$. This improvement arises from noise suppression facilitated by the OPA.  Furthermore, we have discussed the effects of various physical parameters on the $B_{\rm MI}$ in the steady state. We have used the Gaussian measurement schemes that can be realized experimentally. Moreover, we compared the CFI for heterodyne and homodyne detection with the SLD QFI, showing that these measurement schemes offer efficient and experimentally accessible strategies for parameter estimation.


\begin{thebibliography}{99} 
\bibitem{Paris05} A. Ferraro,  S. Olivares, and M. G. Paris, Gaussian states in continuous variable quantum information.  arXiv preprint quant-ph/0503237 (2005).

\bibitem{Van05} S. L. Braunstein, P. Van Loock, Quantum information with continuous variables. Rev. Mod. Phys. 77, 513 (2005).

\bibitem{Kim02} M. S. Kim, J. Lee, W. Munro, Experimentally realizable characterizations of continuous-variable Gaussian states. Phys. Rev. A, 66, 030301 (2002).

\bibitem{Vitali07} D. Vitali, S. Gigan, A. Ferreira, H. R. B\"ohm, P. Tombesi, A. Guerreiro, V. Vedral, A. Zeilinger, and M. Aspelmeyer, Optomechanical entanglement between a movable mirror and a cavity field. Phys. Rev. Lett. 98, 030405 (2007).

\bibitem{Mauro07} M. Paternostro, D. Vitali, S. Gigan, M. S. Kim, C. Brukner, J. Eisert, and M. Aspelmeyer, Creating and probing multipartite macroscopic entanglement with light. Phys. Rev. Lett. 99, 250401 (2007).

\bibitem{Tian10} L. Tian, H. Wang, Optical wavelength conversion of quantum states with optomechanics. Rev. Phys. A, 82, 053806 (2010).

\bibitem{Groblasher09} S. Gr\"ob­lacher, K. Hammerer, M. R. Vanner, and M. Aspelmeyer, Observation of strong coupling between a micromechanical resonator and an optical cavity field.  Nature, 460, 724-727 (2009).

\bibitem{Teufel11} J. D. Teufel, T. Donner, D. Li, J. W. Harlow, M. S. Allman, K. Cicak, A. J. Sirois, J. D. Whittaker, K. W. Lehnert, and R. W. Simmonds, Sideband cooling of micromechanical motion to the quantum ground state. Nature, 475, 359 (2011).

\bibitem{Chan11} J. Chan,  T. Alegre,  A. Safavi-Naeini, J. T. Hill,  A. Krause,  S. Gr\"ob­lacher,  M. Aspelmeyer and O. Painter, Laser cooling of a nanomechanical oscillator into its quantum ground state.  Nature, 478, 89 (2011).

\bibitem{Paris03} S. Olivares, M. G. Paris, and A. R. Rossi, Optimized teleportation in Gaussian noisy channels. Phys. Lett. A, 319(1-2), 32 (2003).

\bibitem{Zhang16} X. Zhang, C. L. Zou, L. Jiang, H. X. Tang, Cavity magnomechanics. Sci. Adv. 2, e1501286 (2016).

\bibitem{Li18} J. Li, S. Y. Zhu, and G. S. Agarwal, Magnon-photon-phonon entanglement in cavity magnomechanics. Phys. Rev. Lett. 121(20) 203601 (2018).

\bibitem{Kevrekidis03} P. G. Kevrekidis, G. Theocharis, D. J. Frantzeskakis, and  B. A. Malomed, Feshbach resonance management for Bose-Einstein condensates. Phys. Rev. Lett. 90, 230401 (2003).

\bibitem{Gross11} C. Gross, H. Strobel, E. Nicklas, T. Zibold, N. Bar-Gill, G. Kurizki, and M. K. Oberthaler, Atomic homodyne detection of continuous-variable entangled twin-atom states. Nature, 480, 219 (2011).

\bibitem{Wade16} A. C. J. Wade, J. F. Sherson, and K. M{\o}lmer, Manipulation of collective quantum states in Bose-Einstein condensates by continuous imaging. Phys. Rev. A, 93, 023610 (2016).

\bibitem{Aasi13} J. Aasi, et al. Enhanced sensitivity of the LIGO gravitational wave detector by using squeezed states of light. Nat. Photonics, 7, 613 (2013).

\bibitem{Wagle2024} D. Wagle, A. Rai, M. T. Kaffash, and M. B. Jungfleisch, Controlling magnon-photon coupling in a planar geometry.  J. Phys. Mater. 7, 025005 (2024).

\bibitem{Tabuchi2014} Y. Tabuchi, S. Ishino, T. Ishikawa, R. Yamazaki, K. Usami, and Y. Nakamura, Hybridizing ferromagnetic magnons and microwave photons in the quantum limit. Phys. Rev. Lett. 113, 083603 (2014).

\bibitem{Zhang2014} X. Zhang, C.-L. Zou, L. Jiang, and H. X. Tang, Strongly coupled magnons and cavity microwave photons. Phys. Rev. Lett. 113, 156401 (2014).

\bibitem{Andrade2025} I. L. S. Andrade, K. Pirota,  A. O. Caldeira,  and  F. Rouxinol, Observation of Purcell effect in electrically coupled cavity-magnet system. arXiv:2508.02946 (2025).

\bibitem{Ebrahimi21} M. S. Ebrahimi, A. Motazedifard, and M. Bagheri Harouni, Single-quadrature quantum magnetometry in cavity electromagnonics. Phys. Rev. A, 103, 062605 (2021).

\bibitem{Xu2023} D. Xu, X.-K. Gu, H.-K. Li, Y.-C. Weng, Y.-P. Wang, J. Li, H. Wang, S.-Y. Zhu, and J. Q. You, Quantum control of a single magnon in a macroscopic spin system. Phys. Rev. Lett. 130, 193603 (2023).

\bibitem{Peng2025} J.-X. Peng, N. Akhtar, X. Yang, Y.-J. Zhu, Z.-W. Yu, and J.-F. Li, Nonreciprocal quantum metrology in cavity magnonics. Chaos Solitons Fractals, 200, 117118 (2025).

\bibitem{yuan} H. Y. Yuan, Y. Cao, A. Kamra, R. A. Duine, and P. Yan, Quantum magnonics: When magnon spintronics meets quantum information science. Phys. Rep. 965, 1 (2022).

\bibitem{Goryachev2014} M. Goryachev, W. G. Farr, D. L. Creedon, Y. Reshitnyk, M. Kostylev, and M. E. Tobar, High-cooperativity cavity QED with magnons at microwave frequencies. Phys. Rev. Applied, 2, 054002 (2014).

\bibitem{19} T. Liu, X. Zhang, H. X. Tang, and M. E. Flatt\'e, Optomagnonics in magnetic solids. Phys. Rev. B, 94, 060405(R) (2016).

\bibitem{20} S. Viola Kusminskiy, H. X. Tang, and F. Marquardt, Coupled spin-light dynamics in cavity optomagnonics. Phys. Rev. A, 94, 033821 (2016).

\bibitem{21} S. Sharma, Y. M. Blanter, and G. E. Bauer, Light scattering by magnons in whispering gallery mode cavities. Phys. Rev. B, 96, 094412 (2017).

\bibitem{22} X. Zhang, N. Zhu, C.-L. Zou, and H. X. Tang, Optomagnonic whispering gallery microresonators. Phys. Rev. Lett. 117, 123605 (2016).

\bibitem{23} A. Osada, R. Hisatomi, A. Noguchi, Y. Tabuchi, R. Yamazaki, K. Usami, M. Sadgrove, R. Yalla, T. Nomura, and Y. Nakamura, Cavity optomagnonics with spin-orbit coupled photons. Phys. Rev. Lett. 116, 223601 (2016).

\bibitem{Flaig17} H. Maier-Flaig, S. Klingler, C. Dubs, O. Surzhenko, R. Gross, M. Weiler, H. Huebl, and S. T. B. Goennenwein, Temperature-dependent magnetic damping of yttrium iron garnet spheres. Phys. Rev. B, 95, 214423 (2017).

\bibitem{25} A. V. Chumak, V. I. Vasyuchka, A. A. Serga, and B. Hillebrands, Magnon spintronics. Nat. Phys. 11, 453 (2015).

\bibitem{magnon sensing} S. P. Wolski, D. Lachance-Quirion, Y. Tabuchi, S. Kono, A. Noguchi, K. Usami, and Y. Nakamura, Dissipation-based quantum sensing of magnons with a superconducting qubit. Phys. Rev. Lett. 125, 117701 (2020).

\bibitem{Nori} D. Zhang, X. Wang, T. Li, X. Luo, W. Wu, F. Nori, and J. You, Cavity quantum electrodynamics with ferromagnetic magnons in a small yttrium-iron-garnet sphere. npj Quantum Inf. 1, 15014 (2015).

\bibitem{Highcooperativity} H. Huebl, C. W. Zollitsch, J. Lotze, F. Hocke, M. Greifenstein, A. Marx, R. Gross, and S. T. B. Goennenwein, High cooperativity in coupled microwave resonator ferrimagnetic insulator hybrids. Phys. Rev. Lett. 111, 127003 (2013).

\bibitem{Dynamical Backaction Magnomechanics} C. A. Potts, E. Varga, V. A. Bittencourt, S. V. Kusminskiy, and J. P. Davis, Dynamical backaction magnomechanics. Phys. Rev. X, 11, 031053 (2021).

\bibitem{30} R. Hisatomi, A. Osada, Y. Tabuchi, T. Ishikawa, A. Noguchi, R. Yamazaki, K. Usami, and Y. Nakamura, Bidirectional conversion between microwave and light via ferromagnetic magnons. Phys. Rev. B, 93, 174427 (2016).

\bibitem{32} M. Yu, H. Shen, and J. Li, Magnetostrictively induced stationary entanglement between two microwave fields. Phys. Rev. Lett. 124, 213604 (2020).

\bibitem{33} M. Yu, S.-Y. Zhu, and J. Li, Macroscopic entanglement of two magnon modes via quantum correlated microwave fields. J. Phys. B: At. Mol. Opt. Phys. 53, 065402 (2020).

\bibitem{34} J. Li and S. Gr\"oblacher, Entangling the vibrational modes of two massive ferromagnetic spheres using cavity magnomechanics. Quantum Sci. Technol. 6, 024005 (2021).

\bibitem{35} J. Li and S.-Y. Zhu, Entangling two magnon modes via magnetostrictive interaction. New J. Phys. 21, 085001 (2019).

\bibitem{36} J. M. P. Nair and G. S. Agarwal, Quantum drives produce strong entanglement between YIG samples without using intrinsic nonlinearities. arXiv:1905.07884 (2019).

\bibitem{last} Z. Zhang, M. O. Scully, and G. S. Agarwal, Quantum entanglement between two magnon modes via Kerr nonlinearity driven far from equilibrium. Phys. Rev. Res. 1, 023021 (2019).

\bibitem{electromagnonics-optomechanics} H. Tan and J. Li, Einstein-Podolsky-Rosen entanglement and asymmetric steering between distant macroscopic mechanical and magnonic systems. Phys. Rev. Res. 3, 013192 (2021).

\bibitem{foroudcrystalentanglement} F. Bemani, R. Roknizadeh, A. Motazedifard, M. H. Naderi, and D. Vitali, Quantum correlations in optomechanical crystals. Phys. Rev. A, 99, 063814 (2019).

\bibitem{hussian2022} B. Hussain, S. Qamar, and M. Irfan, Entanglement enhancement in cavity magnomechanics by an optical parametric amplifier. Phys. Rev. A, 105, 063704 (2022).

\bibitem{Amazioug2023} M. Amazioug, S. K. Singh, B. Teklu, and M. Asjad, Feedback control of quantum correlations in a cavity magnomechanical system with magnon squeezing. Entropy, 25, 1462 (2023).

\bibitem{Amazioug2025} M. Amazioug, S. Abdel-Khalek, and M. Asjad, Controlled nonreciprocal magnon and photon blockade using Barnett effects in a spinning microwave magnomechanical system. Ann. Phys. 537, e00289 (2025).

\bibitem{Deng24} X. Deng, K.-K. Zhang, T. Shui, and W.-X. Yang, Nonreciprocal unconventional magnon blockade via the Sagnac-Fizeau shift in an optomagnonic system. Phys. Rev. A, 110, 063711 (2024).

\bibitem{37} J. Li, Y.-P. Wang, J.-Q. You, and S.-Y. Zhu, Squeezing microwaves by magnetostriction. Natl. Sci. Rev. 10, nwac247 (2023).

\bibitem{microwace field squeezing} J. Li, S.-Y. Zhu, and G. S. Agarwal, Squeezed states of magnons and phonons in cavity magnomechanics. Phys. Rev. A, 99, 021801(R) (2019).

\bibitem{38} M.-S. Ding, L. Zheng, and C. Liu, Phonon laser in a cavity magnomechanical system. Sci. Rep. 9, 1 (2019).

\bibitem{39} C. A. Potts, V. A. Bittencourt, S. Viola Kusminskiy, and J. P. Davis, Magnon-phonon quantum correlation thermometry. Phys. Rev. Applied, 13, 064001 (2020).

\bibitem{one} N. Crescini, C. Braggio, G. Carugno, A. Ortolan, and G. Ruoso, Cavity magnon polariton based precision magnetometry. Appl. Phys. Lett. 117, 144001 (2020).

\bibitem{phasemodulatedmagnetometry} N. Crescini, G. Carugno, and G. Ruoso, Phase-modulated cavity magnon polaritons as a precise magnetic field probe. Phys. Rev. Applied, 16, 034036 (2021).

\bibitem{8} Y. Cao and P. Yan, Exceptional magnetic sensitivity of $\mathcal{PT}$-symmetric cavity magnon polaritons. Phys. Rev. B, 99, 214415 (2019).

\bibitem{magnonblockade1} Z.-X. Liu, H. Xiong, and Y. Wu, Magnon blockade in a hybrid ferromagnet-superconductor quantum system. Phys. Rev. B, 100, 134421 (2019).

\bibitem{magnonblockade2} J.-K. Xie, S.-L. Ma, and F.-L. Li, Quantum-interference-enhanced magnon blockade in an yttrium-iron-garnet sphere coupled to superconducting circuits. Phys. Rev. A, 101, 042331 (2020).

\bibitem{Sharma} S. Sharma, V. A. Bittencourt, A. D. Karenowska, and S. V. Kusminskiy, Spin cat states in ferromagnetic insulators. Phys. Rev. B, 103, L100403 (2021).

\bibitem{quantumillumination} Q. Cai, J. Liao, B. Shen, G. Guo, and Q. Zhou, Microwave quantum illumination via cavity magnonics. Phys. Rev. A, 103, 052419 (2021).

\bibitem{photon-phononconversion} S.-F. Qi and J. Jing, Magnon-assisted photon-phonon conversion in the presence of structured environments. Phys. Rev. A, 103, 043704 (2021).

\bibitem{Cavity magnomechanical storage and retrieval of quantum states} B. Sarma, T. Busch, and J. Twamley, Cavity magnomechanical storage and retrieval of quantum states. New J. Phys. 23, 043041 (2021).

\bibitem{Zhu2025} Y. J. Zhu, X. Han, H. F. Wang, and S. Zhang, Enhanced Force Sensitivity Based on Quadratic Optomechanical Coupling and an Optical Parametric Amplifier. Adv. Quantum Technol. 8, e2500148 (2025).

\bibitem{Holevo11} A. S. Holevo, Probabilistic and Statistical Aspects of Quantum Theory. (Edizioni della Normale, Pisa, 2011).

\bibitem{Helstrom67} C. W. Helstrom, Minimum mean-squared error of estimates in quantum statistics. Phys. Lett. A, 25, 101 (1967).

\bibitem{Helstrom76} C. W. Helstrom, Quantum Detection and Estimation Theory. Vol. 123. (Academic Press, New York, 1976).

\bibitem{Bures69} D. Bures, An extension of Kakutani's theorem on infinite product measures to the tensor product of semifinite $w^*$-algebras. Trans. Am. Math. Soc. 135, 199 (1969).

\bibitem{Braunstein94} S. L. Braunstein and C. M. Caves, Statistical distance and the geometry of quantum states. Phys. Rev. Lett. 72, 3439 (1994).

\bibitem{paris2009} M. G. A. Paris, Quantum estimation for quantum technology. Int. J. Quantum Inf. 7, 125 (2009).

\bibitem{Szczykulska2016} M. Szczykulska, T. Baumgratz, and A. Datta, Multi-parameter quantum metrology. Adv. Phys.: X, 1, 621--639 (2016).

\bibitem{Giovannetti04} V. Giovannetti, S. Lloyd, and L. Maccone, Quantum-enhanced measurements: Beating the standard quantum limit. Scien. 306, 1330 (2004).

\bibitem{Zwierz10} M. Zwierz, C. A. Pérez-Delgado, and P. Kok, General Optimality of the Heisenberg Limit for Quantum Metrology. Phys. Rev. Lett. 105, 180402 (2010).

\bibitem{Giovannetti11} V. Giovannetti, S. Lloyd, and L. Maccone, Advances in quantum metrology. Nat. Photonics, 5, 222 (2011).

\bibitem{Demkowicz12} R. Demkowicz-Dobrzański, J. Kołodyński, and M. Guţă, Elusive Heisenberg limit in quantum-enhanced metrology. Nat. Commun. 3, 1063 (2012).

\bibitem{Abbott16} B. P. Abbott et al. (LIGO Scientific Collaboration and Virgo Collaboration), Observation of gravitational waves from a binary black hole merger. Phys. Rev. Lett. 116, 061102 (2016).

\bibitem{Correa15} L. A. Correa, M. Mehboudi, G. Adesso, and A. Sanpera, Individual quantum probes for optimal thermometry. Phys. Rev. Lett. 114, 220405 (2015).

\bibitem{Hofer17} P. P .Hofer, J. B. Brask, M. Perarnau-Llobet, N. Brunner, Quantum Thermal Machine as a Thermometer. Phys. Rev. Lett. 119, 090603 (2017).

\bibitem{Monras11} A. Monras and F. Illuminati, Measurement of damping and temperature: Precision bounds in Gaussian quantum systems. Phys. Rev. A: At. Mol. Opt. Phys. 83, 012315 (2011).

\bibitem{Mirkhalaf24} S. Mirkhalaf, M. Mehboudi, Z. Nafari Qaleh, and S. Rahimi-Keshari, Operational significance of nonclassicality in nonequilibrium Gaussian quantum thermometry. New J. Phys. 26, 023046 (2024).
\bibitem{Khan22} M. M. Khan, M. Mehboudi, H. Terças, M. Lewenstein, and M. A. Garcia-March, Subnanokelvin thermometry of an interacting $d$-dimensional homogeneous Bose gas. Phys. Rev. Res. 4, 023191 (2022).

\bibitem{Ballester04} M. A. Ballester, Estimation of unitary quantum operations. Phys. Rev. A, 69, 022303 (2004).

\bibitem{Gaiba09} R. Gaiba and M. G. A. Paris, Squeezed vacuum as a universal quantum probe. Phys. Lett. A, 373, 934 (2009).

\bibitem{Frowis14} F. Fr\"owis, M. Skotiniotis, B. Kraus, and W. D\"ur, Optimal quantum states for frequency estimation. New J. Phys. 16, 083010 (2014).

\bibitem{Zhang13} Y. L. Zhang, Y. R. Zhang,  L. Z. Mu, and H. Fan, Criterion for remote clock synchronization with Heisenberg-scaling accuracy. Phys. Rev. A, 88, 052314 (2013).

\bibitem{Matsumoto02} K. Matsumoto, A new approach to the Cram\'er--Rao-type bound of the pure-state model. J. Phys. A: Math. Gen. 35, 3111 (2002).

\bibitem{Vaneph13} C. Vaneph, T. Tufarelli, and M. G. Genoni, Quantum estimation of a two-phase spin rotation. Quantum Meas. Quantum Metrol. 1, 12 (2013).

\bibitem{Vidrighin14} M. D. Vidrighin, G. Donati, M. G. Genoni, X.-M. Jin, W. S. Kolthammer, M. Kim, A. Datta, M. Barbieri, and I. A. Walmsley, Joint estimation of phase and phase diffusion for quantum metrology. Nat. Commun. 5, 3532 (2014).

\bibitem{Crowley14} P. J. Crowley, A. Datta, M. Barbieri, and I. A. Walmsley, Tradeoff in simultaneous quantum-limited phase and loss estimation in interferometry. Phys. Rev. A, 89(2), 023845 (2014).

\bibitem{Ragy16} S. Ragy, M. Jarzyna, and R. Demkowicz-Dobrza\'nski, Compatibility in multiparameter quantum metrology. Phys. Rev. A, 94, 052108 (2016).

\bibitem{Seveso19} L. Seveso, F. Albarelli, M. G. Genoni, and M. G. A. Paris, On the discontinuity of the quantum Fisher information for quantum statistical models with parameter dependent rank. J. Phys. A: Math. Theor. 53, 02LT01 (2020).

\bibitem{Hauke16} P. Hauke, M. Heyl, L. Tagliacozzo, and P. Zoller, Measuring multipartite entanglement through dynamic susceptibilities. Nat. Phys. 12, 778 (2016).

\bibitem{Liu17} C.-C. Liu, D. Wang, W.-Y. Sun, and L. Ye, Quantum Fisher information, quantum entanglement and correlation close to quantum critical phenomena. Quantum Inf. Process. 16, 219 (2017).

\bibitem{Naimy2025} A. Naimy, A. Slaoui, A. Lakhfif, and R. A. Laamara, Optimal multiparameter quantum estimation of magnonic couplings in a magnomechanical cavity. arXiv preprint arXiv:2511.12352 (2025).

\bibitem{Asjad2023} M. Asjad, B. Teklu, and M. G. A. Paris, Joint quantum estimation of loss and nonlinearity in driven-dissipative Kerr resonators. Phys. Rev. Res. 5, 013185 (2023).

\bibitem{Meystre2007} P. Meystre and M. Sargent III, Elements of Quantum Optics. 4th ed. (Springer, Berlin, Heidelberg, 2007).

\bibitem{Scully1999} M. O. Scully and M. S. Zubairy,  Quantum Optics. (Cambridge University Press, Cambridge, 1997).

\bibitem{Wolf17} R. Wolf, I. Breunig, H. Zappe, and K. Buse, Cascaded second-order optical nonlinearities in on-chip micro rings. Opt. Express, 25, 29927 (2017).

\bibitem{Szabados20} J. Szabados, D. N. Puzyrev, Y. Minet, L. Reis, K. Buse, A. Villois, D. V. Skryabin, and I. Breunig, Frequency comb generation via cascaded second-order nonlinearities in microresonators. Phys. Rev. Lett. 124, 203902 (2020).

\bibitem{Wang20} M. Wang, N. Yao, R. Wu, Z. Fang, S. Lv, J. Zhang, J. Lin, W. Fang, and Y. Cheng, Strong nonlinear optics in on-chip coupled lithium niobate microdisk photonic molecules. New J. Phys. 22, 073030 (2020).

\bibitem{28} D. Lachance-Quirion, Y. Tabuchi, A. Gloppe, K. Usami, and Y. Nakamura, Hybrid quantum systems based on magnonics. Appl. Phys. Express, 12, 070101 (2019).

\bibitem{27} Y. Tabuchi, S. Ishino, A. Noguchi, T. Ishikawa, R. Yamazaki, K. Usami, and Y. Nakamura, Quantum magnonics: The magnon meets the superconducting qubit. C. R. Phys. 17, 729 (2016).

\bibitem{1} M. Tsang, and C. M. Caves, Coherent quantum-noise cancellation for optomechanical sensors. Phys. Rev. Lett. 105, 123601 (2010).

\bibitem{2} X. Xu, and J. M. Taylor, Squeezing in a coupled two-mode optomechanical system for force sensing below the standard quantum limit. Phys. Rev. A, 90, 043848 (2014).

\bibitem{3} B. A. Levitan, A. Metelmann, and A. A. Clerk, Optomechanics with two-phonon driving. New J. Phys. 18, 093014 (2016).

\bibitem{4} A. Motazedifard, A. Dalafi, F. Bemani, and  M. H. Naderi, Force sensing in hybrid Bose-Einstein-condensate optomechanics based on parametric amplification. Phys. Rev. A, 100, 023815 (2019). 

\bibitem{Dejesus1987} E. X. DeJesus, and C. Kaufman, Routh-Hurwitz criterion in the examination of eigenvalues of a system of nonlinear ordinary differential equations. Phys. Rev. A, 35, 5288 (1987).

\bibitem{ref:31} P. C. Parks, and V. Hahn, Stability Theory (Prentice Hall, 1993).

\bibitem{adesso} G. Adesso, A. Serafini, and F. Illuminati, Extremal entanglement and mixedness in continuous variable systems. Phys. Rev. A: At. Mol. Opt. Phys. 70, 022318 (2004).

\bibitem{simon} R. Simon, Peres-Horodecki separability criterion for continuous variable systems. Phys. Rev. Lett. 84, 2726 (2000).

\bibitem{Vidal} G. Vidal, and R. F. Werner, Computable measure of entanglement. Phys. Rev. A, 65, 032314 (2002).

\bibitem{Plenio} M. B. Plenio, Logarithmic negativity: A full entanglement monotone that is not convex. Phys. Rev. Lett. 95, 090503 (2005).

\bibitem{EPJD2014} Q. Zheng, Y. Yao, and Y. Li, Optimal quantum channel estimation of two interacting qubits subject to decoherence. Eur. Phys. J. D, 68, 170 (2014). 

\bibitem{hmz:9} H. P. Yuen, and M. Lax, Multiple-parameter quantum estimation and measurement of nonselfadjoint observables. IEEE Trans. Inf. Theory, 19, 740 (1973). 

\bibitem{Safranek2018} D. Safr\'anek, Estimation of Gaussian quantum states. J. Phys. A: Math. Theor. 52, 035304 (2019).

\bibitem{Bakmou2020} L. Bakmou, M. Daoud, and R. A. Laamara, Multiparameter quantum estimation theory in quantum Gaussian states. J. Phys. A: Math. Theor. 53, 385301 (2020). 

\bibitem{Razhin2017} R. Yousefjani, R. Nichols, S. Salimi, and G. Adesso, Estimating phase with a random generator: Strategies and resources in multiparameter quantum metrology. Phys. Rev. A, 95, 062307 (2017).

\bibitem{hmz:27} M. G. Genoni, M. G. A. Paris, G. Adesso, H. Nha, P. L. Knight, and M. S. Kim, Optimal estimation of joint parameters in phase space. Phys. Rev. A: At. Mol. Opt. Phys. 87, 012107 (2013).

\bibitem{hmz:11} C. W. Helstrom, Quantum detection and estimation theory. J. Stat. Phys. 1, 231 (1969).

\bibitem{hmz:12} Y. H. Yan,  Y. P. Tan,  Y. Y. Lu,  S. J. Xiong, and Z. Sun, Impact of decoherence on the metrological advantage of weak-value amplification.  Chin. Phys. B, 35, 010307 (2026).

\bibitem{hmz:13} V. P. Belavkin, Nondemolition principle of quantum measurement theory. Found. Phys. 24, 685 (1994).

\bibitem{hmz:14} A. S. Holevo, Estimation of shift parameters of a quantum state. Rep. Math. Phys. 13, 379 (1978).

\bibitem{hmz:15} V. Katariya, and M. M. Wilde, RLD Fisher information bound for multiparameter estimation of quantum channels. New J. Phys. 23, 073040 (2021).

\bibitem{hmz:16} R. Demkowicz-Dobrza\'nski, W. G\'orecki, and M. Gu\c{t}\u{a}, Multi-parameter estimation beyond quantum Fisher information. J. Phys. A: Math. Theor. 53, 363001 (2020).

\bibitem{hmz:17} J. Suzuki, Parameter estimation of qubit states with unknown phase parameter. Int. J. Quantum Inf. 13, 1450044 (2015). 

\bibitem{hmz:18} J. Suzuki, Y. Yang, and M. Hayashi, Quantum state estimation with nuisance parameters. J. Phys. A: Math. Theor. 53, 453001 (2020). 

\bibitem{hmz:19} H. Yuan, and C.-H. F. Fung, Quantum parameter estimation with general dynamics. npj Quantum Inf. 3, 14 (2017). 

\bibitem{hmz:21} A. Monras, and F. Illuminati, Information geometry of Gaussian channels. Phys. Rev. A: At. Mol. Opt. Phys. 81, 062326 (2010).

\bibitem{hmz:10} A. Serafini, Quantum Continuous Variables: A Primer of Theoretical Methods. (CRC Press, 2023).

\bibitem{hmz:24} J. Miyazaki, and K. Matsumoto, Imaginarity-free quantum multiparameter estimation. Quantum, 6, 665 (2022).

\bibitem{hmz:28} Y. Gao, and H. Lee, Bounds on quantum multiple-parameter estimation with Gaussian state. Eur. Phys. J. D, 68, 347 (2014). 

\bibitem{hmz:30} C. L. Degen, F. Reinhard, and P. Cappellaro, Quantum sensing. Rev. Mod. Phys. 89, 035002 (2017).

\bibitem{hmz:31} J. Liu, H. Yuan, X.-M. Lu, and X. Wang, Quantum Fisher information matrix and multiparameter estimation. J. Phys. A: Math. Theor. 53, 023001 (2020).

\bibitem{hmz:35} C. Oh, C. Lee,  C. Rockstuhl,  H. Jeong,  J. Kim,  H. Nha,  and S. Y. Lee, Optimal Gaussian measurements for phase estimation in single-mode Gaussian metrology. npj Quantum Inf. 5, 10 (2019).

\bibitem{hmz:36} J. X. Peng,  B. Zhu, W. Zhang, and K. Zhang, Estimation theory of photon-magnon coupling strength in a driven-dissipative double-cavity-magnon system. Phys. Rev. A, 109, 022601 (2024).

\bibitem{Monras2013} A. Monras, Phase space formalism for quantum estimation of Gaussian states. arXiv:1303.3682 (2013).

\bibitem{aspelmeyer2014} M. Aspelmeyer, T. J. Kippenberg, and F. Marquardt, Cavity optomechanics. Rev. Mod. Phys. 86, 1391 (2014).

\bibitem{yan2019} X. B. Yan, Z. J. Deng, X. D. Tian, and J. H. Wu, Entanglement optimization of filtered output fields in cavity optomechanics. Opt. Express, 27, 24393 (2019).

\bibitem{saheked2018} Y. Shaked, Y. Michael, R. Z. Vered, L. Bello, M. Rosenbluh, and A. Pe\'er, Lifting the bandwidth limit of optical homodyne measurement with broadband parametric amplification. Nat. Commun. 9, 609 (2018). 

\bibitem{racz2023} \'E. R\'acz, L. Ruppert, and R. Filip, Efficient OPA tomography of non-Gaussian states of light. arXiv:2310.18821 (2023).

\bibitem{frascella2021} G. Frascella, S. Agne, F. Y. Khalili, and M. V. Chekhova, Overcoming detection loss and noise in squeezing-based optical sensing. npj Quantum Inf. 7, 72 (2021). 

\bibitem{feng2016} S. Feng, D. He, and B. Xie, Quantum theory of phase-sensitive heterodyne detection. J. Opt. Soc. Am. B, 33, 1365 (2016).

\bibitem{weedbrook2012} C. Weedbrook, S. Pirandola, R. Garc\'ia-Patr\'on, N. J. Cerf, T. C. Ralph, J. H. Shapiro, and S. Lloyd, Gaussian quantum information. Rev. Mod. Phys. 84, 621 (2012).

\end{thebibliography}
\end{document}